\documentclass[twocolumn,twocolappendix]{aastex631}

\usepackage{natbib}
\usepackage{graphicx}
\usepackage{multirow}
\usepackage{mathtools}
\usepackage{makecell}
\usepackage{rotating}

\usepackage{hyperref}
\hypersetup{
    colorlinks=true,
    citecolor=blue,
    linkcolor=blue,
    filecolor=magenta,      
    urlcolor=cyan,
}

\newcommand{\msun}{\mbox{M}_{\odot}}
\newcommand{\mstar}{$M_*$}
\newcommand{\mmol}{$M_{\text{mol}}$}
\newcommand{\mhi}{$M_{\text{HI}}$}
\newcommand{\mumol}{$\mu_{\text{mol}}$}
\newcommand{\mdyn}{$M_{\text{dyn}}$}

\newcommand{\hii}{H~\textsc{ii}}
\newcommand{\one}{~\textsc{i}}
\newcommand{\ii}{~\textsc{ii}}
\newcommand{\iii}{~\textsc{iii}}

\newcommand{\ebvgas}{E(B-V)$_{\text{gas}}$}

\newcommand{\mugas}{$\mu_{\text{gas}}$}
\newcommand{\mgas}{$M_{\text{gas}}$}

\newcommand{\zism}{$Z_{\text{ISM}}$}
\newcommand{\mzism}{$M_{\text{Z,ISM}}$}
\newcommand{\mzstars}{$M_{\text{Z,stars}}$}
\newcommand{\mzdust}{$M_{\text{Z,dust}}$}

\newcommand{\etaout}{$\eta_{\text{out}}$}
\newcommand{\mdotin}{$\dot{M}_{\text{in}}$}
\newcommand{\mdotout}{$\dot{M}_{\text{out}}$}

\newcommand{\cott}{CO(3$-$2)}
\newcommand{\coto}{CO(2$-$1)}
\newcommand{\cooz}{CO(1$-$0)}
\newcommand{\sigsfr}{$\Sigma_{\text{SFR}}$}
\newcommand{\sigmol}{$\Sigma_{\text{mol}}$}
\newcommand{\siggas}{$\Sigma_{\text{gas}}$}
\newcommand{\aco}{$\alpha_{\text{CO}}$}
\newcommand{\acoks}{$\alpha_{\text{CO}}^{\text{KS}}$}
\newcommand{\acodyn}{$\alpha_{\text{CO}}^{\text{dyn}}$}
\newcommand{\acounits}{M$_{\odot}$~(K km s$^{-1}$ pc$^2$)$^{-1}$}
\newcommand{\lcott}{L$^{\prime}_{\text{CO}(3-2)}$}

\newcommand{\lcooz}{L$^{\prime}_{\text{CO}(1-0)}$}
\newcommand{\scott}{S$_{\text{CO}(3-2)}$}
\newcommand{\delsfms}{$\Delta$log(SFR)$_{\text{MS}}$}
\newcommand{\delmzr}{$\Delta$log(O/H)$_{\text{MZR}}$}
\newcommand{\delmmol}{$\Delta\log(M_{\text{mol}})_{\text{T18,MS}}$}
\newcommand{\htwo}{H$_2$}
\newcommand{\tdepl}{$t_{\mathrm{depl}}$}

\shorttitle{Molecular gas and metallicity at $z\sim2.3$}
\shortauthors{Sanders et al.}

\begin{document}

\title{CO Emission, Molecular Gas, and Metallicity in Main-Sequence Star-Forming Galaxies at $z\sim2.3$
\footnote{
The data presented herein were obtained at the W. M. Keck Observatory, which is operated as a scientific partnership among the California Institute of Technology, the University of California and the National Aeronautics and Space Administration. The Observatory was made possible by the generous financial support of the W. M. Keck Foundation.
}
}

\author[0000-0003-4792-9119]{Ryan L. Sanders}\altaffiliation{NHFP Hubble Fellow}\affiliation{Department of Physics and Astronomy, University of California, Davis, One Shields Ave, Davis, CA 95616, USA}

\author[0000-0003-3509-4855]{Alice E. Shapley}\affiliation{Department of Physics \& Astronomy, University of California, Los Angeles, 430 Portola Plaza, Los Angeles, CA 90095, USA}

\author[0000-0001-5860-3419]{Tucker Jones}\affiliation{Department of Physics and Astronomy, University of California, Davis, One Shields Ave, Davis, CA 95616, USA}

\author[0000-0003-4702-7561]{Irene Shivaei}\affiliation{Steward Observatory, University of Arizona, Tucson, AZ 85721, USA}

\author[0000-0003-1151-4659]{Gerg{\"o} Popping}\affiliation{European Southern Observatory, Karl-Schwarzschild-Str. 2, D-85748, Garching, Germany}

\author[0000-0001-9687-4973]{Naveen A. Reddy}\affiliation{Department of Physics \& Astronomy, University of California, Riverside, 900 University Avenue, Riverside, CA 92521, USA}

\author[0000-0003-2842-9434]{Romeel Dav\'{e}}\affiliation{Institute for Astronomy, Unversity of Edinburgh, James Clerk Maxwell Building, Peter Guthrie Tait Road, Edinburgh, EH9 3FD, United Kingdom}

\author[0000-0002-0108-4176]{Sedona H. Price}\affiliation{Max-Planck-Institut f{\"u}r Extraterrestrische Physik, Postfach 1312, Garching, D-85741, Germany}

\author{Bahram Mobasher}\affiliation{Department of Physics \& Astronomy, University of California, Riverside, 900 University Avenue, Riverside, CA 92521, USA}

\author[0000-0002-7613-9872]{Mariska Kriek}\affiliation{Leiden Observatory, Leiden University, P.O.\ Box 9513, NL-2300 AA Leiden, The Netherlands}

\author[0000-0002-2583-5894]{Alison L. Coil}\affiliation{Center for Astrophysics and Space Sciences, University of California, San Diego, 9500 Gilman Dr., La Jolla, CA 92093-0424, USA}

\author[0000-0002-4935-9511]{Brian Siana}\affiliation{Department of Physics \& Astronomy, University of California, Riverside, 900 University Avenue, Riverside, CA 92521, USA}

\email{email: rlsand@ucdavis.edu}

\begin{abstract}
We present observations of \cott\ in 13 main-sequence $z=2.0-2.5$ star-forming galaxies at $\log(M_*/\msun)=10.2-10.6$
 that span a wide range in metallicity (O/H) based on rest-optical spectroscopy.
We find that \lcott/SFR decreases with decreasing metallicity, implying that the CO luminosity per unit gas mass
 is lower in low-metallicity galaxies at $z\sim2$.
We constrain the CO-to-\htwo\ conversion factor (\aco) and find that \aco\ inversely correlates with metallicity at $z\sim2$.
We derive molecular gas masses (\mmol) and characterize the relations among \mstar, SFR, \mmol, and metallicity.
At $z\sim2$, \mmol\ increases and molecular gas fraction (\mmol/\mstar) decrease with increasing \mstar,
 with a significant secondary dependence on SFR.
Galaxies at $z\sim2$ lie on a near-linear molecular KS law that is well-described by a constant depletion time of 700~Myr.
We find that the scatter about the mean SFR$-$\mstar, O/H$-$\mstar, and \mmol$-$\mstar\ relations is correlated such that,
 at fixed \mstar, $z\sim2$ galaxies with larger \mmol\ have higher SFR and lower O/H.
We thus confirm the existence of a fundamental metallicity relation at $z\sim2$ where O/H is inversely correlated with
 both SFR and \mmol\ at fixed \mstar.
These results suggest that the scatter of the $z\sim2$ star-forming main sequence, mass-metallicity relation,
 and \mmol-\mstar\ relation are primarily driven by stochastic variations in gas inflow rates.
We place constraints on the mass loading of galactic outflows and perform a metal budget analysis, finding that
 massive $z\sim2$ star-forming galaxies retain only 30\% of metals produced, implying that a large mass of metals resides
 in the circumgalactic medium.
\end{abstract}




\section{Introduction}\label{sec:intro}

The cold gas mass and abundance of heavy elements (i.e., metallicity) are fundamental
 properties of galaxies that are key to understanding galaxy formation and evolution.
Molecular gas clouds are the sites of star formation such that galaxy star-formation rates (SFRs)
 depend on the mass of molecular gas (\mmol) available in the interstellar medium (ISM).
This close tie is manifested in a tight correlation between the surface densities of SFR (\sigsfr)
 and molecular gas mass (\sigmol) known as the molecular Kennicutt-Schmidt relation \cite[.e.g,][]{ken98,big08,ler08,rey19,ken21}.
Likewise, metallicity is connected to star formation as new metals are produced via nucleosynthesis and returned to
 the ISM through the processes of stellar evolution and death,
 such that the trend of increasing metallicity over time traces the buildup of galaxy stellar mass (\mstar).

Galaxy-integrated SFR, metallicity, and \mmol\ have been found to depend strongly on \mstar\ in star-forming populations,
 resulting in a number of scaling relations:
 the star-forming main sequence (MS) in which SFR increases with increasing \mstar\ \citep[e.g.,][]{bri04,noe07};
 the mass-metallicity relation (MZR) in which ISM metallicity, traced by the gas-phase oxygen abundance,
 increases with increasing \mstar\ \citep[e.g.,][]{tre04,and13,cur20b};
 and the \mmol-\mstar\ relation in which more massive galaxies have larger molecular gas reservoirs \citep[e.g.,][]{bot14,sai17}.
At $z\sim0$, these scaling relations display secondary dependences that connect these properties in
 3-dimensional parameter spaces.
At fixed \mstar, O/H decreases with increasing SFR, forming the SFR$-$O/H$-$\mstar\ ``Fundamental Metallicity Relation"
 (SFR-FMR; e.g., \citealt{man10,cre19,cur20b}).
Similarly, O/H and \mmol\ are inversely related at fixed \mstar in the \mmol$-$O/H$-$\mstar\ relation, or ``Gas-FMR"
 \citep{bot16a,bot16b}.
A Gas-FMR has also been found at $z\sim0$ using atomic hydrogen gas masses \citep{bot13,hug13,lar13,bro18}.
Since SFR is determined by the amount of cold gas present, it is likely that the Gas-FMR is a more fundamental
 relation from which the SFR-FMR emerges.
These higher-order relations are realized as correlated scatter around pairs of scaling relations:
 the SFR-FMR is represented by an anti-correlation between residuals around the MS
 and residuals around the MZR, while a signature of the Gas-FMR is that residuals around the
 \mmol-\mstar\ relation are anti-correlated with those about the MZR.

Characterizing the interrelation between \mstar, SFR, \mmol, and metallicity is of central
 importance to understanding the cycle of baryons that governs galaxy growth.
The class of gas-regulator, equilibrium, or bathtub models of galaxy formation indicate that
 the gas fraction (\mgas/\mstar) and ISM metallicity of a galaxy is governed by the rates of gas
 accretion (\mdotin) and outflow (\mdotout) relative to the SFR \citep[e.g.,][]{pee11,dav12,lil13,pen14}, such that
 the mass-loading factor of outflows (\etaout=\mdotin/SFR) can be constrained using measurements of
 both metallicity and gas fraction.
Such models suggest that the SFR-FMR and Gas-FMR arise as a direct result of baryon cycling.
Freshly accreted unenriched gas will expand the gas reservoir leading to a larger SFR, while
 simultaneously diluting metals in the ISM leading to lower O/H.
The SFR-FMR and Gas-FMR are also ubiquitous features of numerical simulations of galaxy formation
 in a cosmological context that include feedback \citep[e.g.,][]{dav17,dav19,der17,tor18,tor19}.
The existence of both a SFR-FMR and Gas-FMR, and the associated correlated residuals around scaling relations,
 is thus a signature of a self-regulated baryon cycle governing galaxy growth.

Searching for the SFR-FMR and Gas-FMR at high redshift presents an opportunity to understand baryon cycling
 during an epoch when galaxy formation was proceeding rapidly,
 when gas inflows and outflows are expected to be more intense than locally on average.
There has been great progress in characterizing
 metallicity scaling relations at high redshift in the past several years
 thanks to extensive rest-optical spectroscopic surveys of representative star-forming galaxies
 at $z\sim1-3$ \citep[e.g.,][]{ste14,kri15,mom16,kas19}.
The MZR has been found to exist out to $z\sim3.5$, evolving toward lower O/H
 at fixed \mstar\ with increasing redshift \citep[e.g.,][]{erb06,mai08,tro14,cul14,cul21,san15,san21,top21}.
The existence of a SFR-FMR has also been confirmed at $z=1.5-2.5$
 \citep{zah14b,san18,san21,hen21}.

\mmol\ is commonly inferred from indirect tracers including CO line emission and dust continuum emission.
The former requires a conversion factor between CO luminosity and \mmol\ \citep[\aco; e.g., ][]{wol10,sch12,bol13,acc17},
 while the latter requires an assumed dust-to-gas ratio \citep[e.g.,][]{san13,dev19}
 or an empirical calibration between Rayleigh-Jeans tail dust emission and \mmol\ \citep{sco16}.
These conversion factors have been found to depend strongly on ISM metallicity in the local universe,
 such that knowledge of the metallicity is key to accurate inferences of \mmol.
Progress in understanding the cold gas properties of high-redshift galaxies has been challenging due to the difficulty of
 detecting these tracers at cosmological redshifts.

CO line emission is the most common tracer of molecular gas in the nearby universe,
 but has proven difficult to measure in high-redshift main-sequence galaxies.
Early work at $z>1$ was limited to extreme sources including submillimeter galaxies (SMGs) and
 ultraluminous infrared galaxies (ULIRGs) \citep[e.g.,][]{gre05,tac08},
 and more typical galaxies could only be reached with strong gravitational lensing \citep{bak04,cop07,dan11,sai13}.
The sensitive IRAM NOEMA interferometer enabled targeted surveys to reach main-sequence galaxies at $z\sim1-2$
 \citep{dad10,dad15,magn12}, culminating with the PHIBSS survey that measured CO for $\sim50$ main-sequence
 galaxies at $z=1-2.5$ \citep{tac10,tac13,tac18}.

More recently, the Atacama Large Millimeter Array (ALMA) has provided further improvement in sensitivity
 at millimeter wavelengths.
Using ALMA, the ASPECS blind spectral scan survey has detected \cott\ and \coto\ for dozens of main-sequence galaxies at
 $z=1-3$ in the Hubble Ultra Deep Field \citep{dec19,gon19,boo19}, while targeted campaigns measured CO for a number of other
 $z>1$ galaxies \citep[e.g.,][]{sil15,sil18}.
Main-sequence \cooz\ detections at $z=2-3$ have also been obtained with deep VLA observations \citep{pav18,rie20}.
However, these existing samples of main-sequence CO-detected galaxies at high redshift still only probe the 
 most massive main-sequence galaxies, with nearly all targets at $z>2$ having stellar masses above $10^{10.7}~\msun$.
This fact highlights an important disconnect between high-redshift galaxies with cold gas information and the
 large samples at $z\sim1-3$ with detailed rest-optical spectroscopy spanning $\log(M_*/\msun)=9.0-10.5$, for which
 robust metallicity constraints are available.
It is necessary to obtain CO observations of lower-mass galaxies that are more typical of high-redshift star-forming
 populations and critically have metallicity measurements in order to search for a Gas-FMR at high redshift,
 improve constraints on baryon cycling in early galaxies, and understand how to reliably estimate \mmol\ in main-sequence
 high-redshift galaxies when CO and dust tracers are not available.

In this work, we analyze \cott\ observations obtained with ALMA for 13 near main-sequence galaxies
 at $z\sim2.3$ with a mean stellar mass of $10^{10.4}~\msun$.
These targets also have deep rest-optical spectroscopy from the MOSDEF survey \citep{kri15},
 providing robust determinations of gas-phase metallicity and SFR.
This unique combination of measurements allows us to carry out a systematic analysis of the
 relations among \mstar, SFR, \mmol, and metallicity for typical star-forming galaxies at $z\sim2$,
 and place constraints on gas flows and baryon cycling in these early galaxies.

This paper is organized as follows.
In Section~\ref{sec:obs}, we present the observations and describe how derived properties are calculated.
The results are presented in Section~\ref{sec:results}.
The sample properties are described in Sec.~\ref{sec:sample}.
Empirical relations between \cott\ luminosity and galaxy properties are explored in Sec.~\ref{sec:lco32}.
We place constraints on the relation between \aco\ and O/H at $z\sim2$ in Sec.~\ref{sec:aco}.
In Sec.~\ref{sec:mgas}, we derive molecular gas masses and characterize the relationships between
 \mstar, SFR, \mmol, and metallicity.
We discuss these results in Section~\ref{sec:discussion}, and summarize our conclusions in Section~\ref{sec:summary}.
 
Throughout, we assume a standard $\Lambda$CDM cosmology with H$_0$=70~km~s$^{-1}$~Mpc$^{-1}$, $\Omega_{\text{m}}$=0.3, and
 $\Omega_{\Lambda}$=0.7.
Magnitudes are in the AB system \citep{oke83} and wavelengths are given in air.
Stellar masses and SFRs are on the \citet{cha03} initial mass function (IMF) scale.
The term metallicity refers to the gas-phase oxygen abundance unless otherwise stated,
 where solar metallicity is 12+log(O/H$)_{\odot}=8.69$ \citep{asp21}.
Molecular gas masses include a 36\% contribution from helium and metals.

\section{Observations}\label{sec:obs}

\subsection{Sample Selection and Rest-Optical Spectroscopic Observations}

Our sample was selected from the MOSFIRE Deep Evolution Field (MOSDEF) survey \citep{kri15},
 a deep near-infrared (rest-frame optical) spectroscopic survey of $\sim$1500 galaxies at $1.4\le z\le3.8$
 using the MOSFIRE instrument \citep{mcl12} on the 10~m Keck~I telescope.
Survey targets were selected in the five CANDELS extragalactic legacy fields \citep{gro11,koe11}
 from the photometric catalogs of the 3D-HST survey \citep{bra12a,ske14,mom16}
 down to fixed $H$-band magnitude measured from \textit{HST}/WFC3 F160W imaging in three redshift intervals,
 with limits of $H_{\text{AB}}\le24.0$, 24.5, and 25.0 for targets at $1.37\le z\le1.70$, $2.09\le z\le2.61$,
 and $2.95\le z\le3.80$, respectively.
These redshift intervals are chosen such that the strong rest-frame optical emission lines [O\ii]$\lambda\lambda$3726,3729,
 H$\beta$, [O\iii]$\lambda\lambda$4959,5007, H$\alpha$, [N\ii]$\lambda$6584, and [S\ii]$\lambda\lambda$6716,6731 fall
 in windows of atmospheric transmission in the near-infrared.
Targets were selected based on pre-existing spectroscopic or \textit{HST} grism redshifts when available, and
 photometric redshifts otherwise.
The completed MOSDEF survey measured robust redshifts for $\sim1300$ targeted galaxies.
See \citet{kri15} for a full description of the survey design, observations, and data reduction.

Using the MOSDEF catalogs, we selected a sample of 14 star-forming galaxies in the middle redshift interval
 at $z\sim2.3$ where the \cott\ emission line can be observed in Band~3 with ALMA.
Targets were required to be in the COSMOS field for accessibility with ALMA,
 have spectroscopic redshifts in the range $2.0\le z\le2.6$,
 have detections at S/N$\ge$3 for the H$\beta$, [O\iii]$\lambda$5007, H$\alpha$, and [N\ii]$\lambda$6584 lines
 to ensure robust metallicity and SFR constraints,
 and have stellar masses within $\pm$0.1~dex of log($M_*/\msun)=10.5$\footnote{The stellar masses used to select the sample observed with ALMA were not
 corrected for the contribution of strong emission lines to the broadband photometry.  The stellar masses used in this paper
 have been derived from emission-line corrected photometry \citep{san21}.
Accordingly, the final stellar mass range differs slightly from that originally used in target selection.}.
Active galactic nuclei (AGN) were removed from the sample, having been identified based on their
 X-ray and IR properties \citep{coi15,aza17,aza18,leu19} and when log([N\ii]/H$\alpha)>-0.3$.
This selection yielded a sample of 14 star-forming galaxies at $z=2.08-2.47$ for followup ALMA observations,
 the properties of which are presented in Table~\ref{tab:mosdefalma}.

\subsection{ALMA Observations and \cott\ Measurements}\label{sec:almaobs}

These 14 targets were observed in ALMA Cycle 6 Program 2018.1.01128.S (PI: R.~Sanders)
 with the Band~3 receiver in 21 scheduling blocks over 7 Nov.\ to 29 Dec.\ 2018.
The observations were designed such that the \cott\ line ($\nu_{\text{rest}}=345.796$~GHz) fell within one spectral window
 in Band~3, between 99 and 113~GHz at the redshifts of our targets.
The requested spectral configuration provided a bandpass of 1.875~GHz per spectral window with
 a native resolution of 7.8125~MHz, corresponding to $\approx22$~km~s$^{-1}$ at $z\sim2.3$.
Observations were carried out with 43 12~m antennas in the C43-3 to C43-5 configurations,
 providing beam major axes of 0.7\arcsec$-$2.5\arcsec with a mean beam size of 1.5\arcsec.
On-source integration times were 10 to 170~minutes per target
 reaching sensitivies of 86 to 456~$\mu$Jy/beam integrated over 50~km~s$^{-1}$.

\begin{figure*}
 \includegraphics[width=\textwidth]{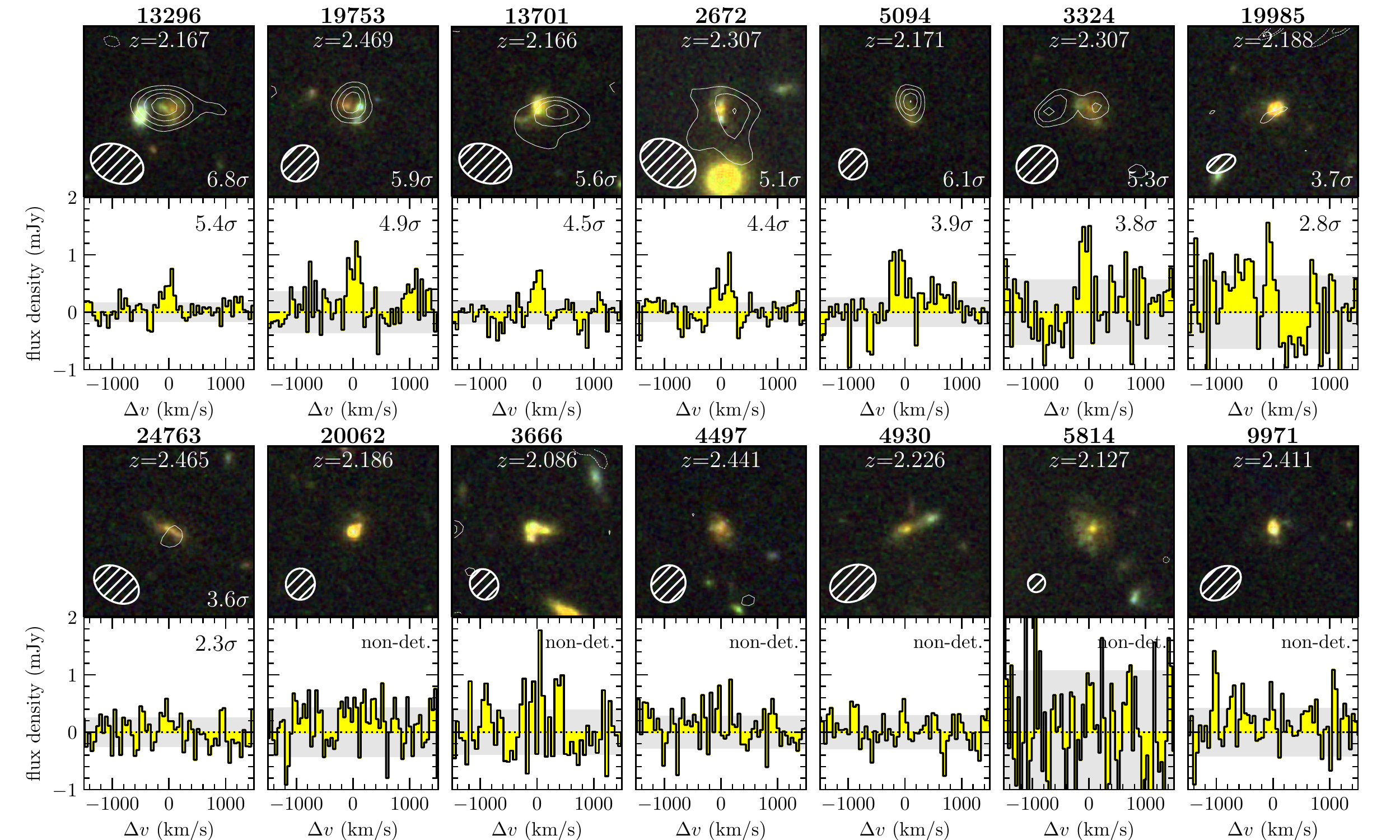}
 \centering
 \caption{
ALMA \cott\ observations for the 14 sources in our sample at $z=2.08-2.47$, sorted by decreasing integrated \cott\ S/N.
The top panel of each subplot displays a \textit{HST} false-color image (red=F160W; green=F125W; blue=F814W)
 overlaid with \cott\ S/N contours (white), with solid lines denoting positive values starting at +3 in steps of 1
 and dashed lines showing negative values starting at $-$3 in steps of $-$1.
The beam shape at line center is shown in the lower left corner, and the peak \cott\ S/N is given in the lower right corner.
The bottom panel presents the extracted 1D spectrum, with the RMS noise displayed in gray and the integrated \cott\ S/N
 given in the upper right corner for robust and tentative detections.
}\label{fig:obs}
\end{figure*}

The ALMA data were calibrated and imaged using the Common Astronomy Software Applications (CASA) package \citep{casa}.
We utlized flux, phase, and bandpass calibrated measurement sets provided by the North American ALMA Regional Center at NRAO.
The CASA task \textsc{tclean} was used to produce clean imaged datacubes with a velocity resolution of 50~km~s$^{-1}$
 ($\approx1/3$ of the rest-optical line FWHM; \citealt{pri16}) using natural weighting.
We searched for detections by collapsing the channels within $\pm$200~km~s$^{-1}$ of the systemic redshift determined from
 the rest-optical lines and searching for positive peaks in these moment-0 maps exceeding 3$\sigma$ at the spatial
 location of each target galaxy.
Figure~\ref{fig:obs} shows \cott\ contours superposed on \textit{HST} false-color images for our sample
 (top panel of each subplot).
We found \cott\ emission is robustly detected in 6 targets with peak S/N$>$5 in the moment-0 maps, tentatively detected in
 two targets with peak S/N=$3-4$, and not detected in the remaining 6 targets.

For all detected sources, we perform optimal extractions to produce one-dimensional science spectra \citep{hor86}
 after converting the datacube intensity units from Jy/beam to Jy/pixel using the beam size in each frequency channel.
All but one source (ID~3324) have spatial profiles that are consistent with being unresolved.
We extract
 these sources using a 2-dimensional Gaussian profile set by the beam shape.
ID~3324 presents an unsual spatial profile with two peaks separated by $\approx$1 beam width (1.5\arcsec).
Spectra extracted separately at each peak location yield 1D lines with roughly comparable fluxes and velocity centroids
 that are consistent with one another and the rest-optical redshift at less than the $1\sigma$ level.
Given the relatively small area probed by our observations, the probability of a S/N$>$5 noise fluctuation located
 spatially and spectrally close to our targets is small.
However, we cannot rule out the possibility that the second peak of ID~3324 is in fact spurious.
We proceed by extracting the total line flux for ID~3324 using a spatial profile of two point sources aligned with the
 two peak locations, but note that our results in Section~\ref{sec:results}
 do not significantly change if we exclude ID~3324 entirely or only include the \cott\ flux from the peak that is
 more spatially coincident with the \textit{HST} imaging centroid.
Spectra of non-detections are extracted using a point source profile at the target centroid location
 from \textit{HST}/WFC3 F160W imaging (rest-optical).
Employing boxcar extractions yields consistent results with slightly larger uncertainties.
The extracted 1D \cott\ spectra are displayed in the bottom panels of each subplot in Figure~\ref{fig:obs}.

Integrated \cott\ line fluxes are measured by fitting single Gaussian profiles to the extracted 1D spectra, with
 the uncertainty on the flux estimated from the covariance matrix.
For non-detections, we fit a single Gaussian where the velocity width
 is fixed to be the same as that of the strong rest-optical lines
 (H$\alpha$ and [O\iii]$\lambda$5007) and the centroid is allowed to vary $\pm$50~km~s$^{-1}$ from the systemic redshift.
The resulting uncertainties are used to infer 3$\sigma$ upper limits on the \cott\ flux. 
Employing bandpass integrations or double Gaussian profiles for sources that may be spectrally double-peaked (IDs 19753 and 2672)
 yield line fluxes that are consistent within the uncertainties with the single Gaussian values.
The 6 robust detections have integrated \cott\ S/N=$3.8-5.4$, while the two
 tentatively detected lines have significances of 2.8$\sigma$ and 2.3$\sigma$.

One source, ID~19753, is robustly detected in \cott\ but
 is not included in the analysis because of ambiguity in whether the rest-optical
 line emission is associated with the region dominating the \cott\ emission.
ID~19753 displays a clumpy morphology with two dominant components displaying a large color difference, with a red clump to the east
 and a blue western clump that dominates the UV.
It is probable that the rest-optical line emission originates from the UV-bright blue component since the H$\alpha$/H$\beta$ ratio
 implies relatively low nebular reddening (\ebvgas=0.16).
Dust continuum emission from ALMA Band~6 observations are offset from the blue component and are more spatially coincident
 with the red component \citep{shi22}.
Due to the large beam size of the Band~3 observations, we cannot determine whether the \cott\ emission is co-spatial
 with the blue or red component, though it is likely that the CO emission aligns with the dust emission.
Given the likelihood of the \cott\ and rest-optical lines of ID~19753 arising from distinct spatial components,
 we remove this source and proceed with a final sample of 13 galaxies.

\subsection{Redshifts, Line Ratios, and SFRs}

Rest-optical emission line fluxes are measured by fitting Gaussian line profiles to the slit-loss corrected 1D science spectra
 from the MOSDEF survey \citep{kri15}.
[Ne\iii]$\lambda$3869, H$\beta$, [O\iii]$\lambda$4959, and [O\iii]$\lambda$5007 are fit independently with single Gaussian profiles.
The [O\ii]$\lambda\lambda$3726,3729 doublet is fit with a double Gaussian with the same line width for each component and
 a rest-frame offset between the two lines fixed to the expected value of 2.78~\AA.
The [S\ii]$\lambda\lambda$6716,6731 doublet is fit simultaneously with a double Gaussian, but with independent line
 parameters for each component.
H$\alpha$ and [N\ii]$\lambda\lambda$6548,6584 are fit simultaneously with a triple Gaussian where the width of the
 three lines are tied but the [N\ii] centroids can vary slightly from their expected position relative to H$\alpha$.
In all cases, the continuum is set to the best-fit model from SED fitting such that stellar
 Balmer absorption is accounted for in the best-fit H$\alpha$ and H$\beta$ fluxes.
Systemic redshifts are inferred from the best-fit centroids of the highest-S/N line
 (H$\alpha$ or [O\iii]$\lambda$5007).

Emission-line fluxes are corrected for the effects of dust by using the H$\alpha$/H$\beta$ ratio to calculate
 \ebvgas, assuming an intrinsic ratio of 2.86 \citep{ost06} and the Milky Way extinction curve of \citet{car89}.
\citet{red20} derived a nebular attenuation curve directly from $z\sim2$ MOSDEF measurements that is consistent with
 the Milky Way curve.
All emission-line ratios utilized for metallicity estimates are calculated using dust-corrected line fluxes.
Star-formation rates (SFRs) are derived from dust-corrected H$\alpha$ luminosities using the H$\alpha$ conversion factor
 of \citet{hao11} adjusted to a \citet{cha03} IMF.
The star-formation rate surface density is calculated as $\Sigma_{\text{SFR}}=\text{SFR}/2\pi R_{\text{eff}}^2$, where
 $R_{\text{eff}}$ is the rest-optical half-light elliptical semimajor axis derived from \textit{HST}/WFC3 F160W imaging
 as cataloged by \citet{van14}.
We define the offset from the star-forming main sequence at fixed \mstar\ as a function of redshift
as \delsfms$\equiv$log(SFR/SFR$_{\text{MS}}(M_*, z)$), where we adopt the SFR$_{\text{MS}}(M_*, z)$ parameterization
 of \citet{spe14}, which matches the mean SFR-\mstar\ relation of MOSDEF star-forming galaxies at $z\sim2.3$ and $z\sim3.3$
 \citep{san21}.

\subsection{Stellar Masses}

Stellar masses are derived from spectral energy distribution (SED) fitting of the extensive broadband photometry in the
 COSMOS field from the 3D-HST survey catalogs \citep{ske14,mom16}.
Photometry was fit using the SED fitting code FAST \citep{kri09} in combination with the flexible stellar population
 synthesis models of \citet{con09}, assuming constant star-formation histories, solar stellar metallicity,
 a \citet{cha03} IMF, and a \citet{cal00} attenuation curve.
Because these models do not include a nebular emission component, photometric measurements were corrected for
 the contribution of strong rest-optical emission lines prior to fitting as described in \citet{san21},
 a necessary step due to the large emission-line equivalent widths common at $z>2$ \citep{red18}.
Correcting the broadband photometry results in lower stellar masses by 0.12~dex on average in our sample,
The two galaxies with the highest rest-optical equivalent widths have mass estimates reduced by 0.35~dex.
The resulting best-fit stellar continuum model is used in the emission-line fitting, as described above.

\subsection{Gas-Phase Metallicities}\label{sec:metallicity}

Gas-phase metallicities, given as 12+log(O/H), are derived from reddening-corrected rest-frame
 optical line ratios.
We employ metallicity calibrations derived from
 composite spectra of extreme local galaxies that are analogs of $z\sim2$ galaxies from \citet{bia18}.
The functional form of the calibrations we use are given in Appendix~\ref{app:metallicity} and Table~\ref{tab:metallicity}.
These calibrations reflect the significant evolution of \hii\ region ionization conditions between
 $z\sim0$ and $z>1$ \citep[e.g.,][]{ste14,ste16,sha15,sha19,san16a,san20b,str17,str18,run20,top20a,top20b}
 and provide a good match to the existing sample of $z>1$ galaxies with direct-method metallicities \citep{san20}.
Since we are interested in measuring O/H, we consider metallicities based on line ratios involving $\alpha$ elements
 (i.e., O, Ne) to be more robust than those involving N since N/O has a large scatter at fixed O/H
 in \hii\ regions and galaxies \citep[$\sim$0.2~dex; e.g.,][]{pil11,str17}
 due to the differing nucleosynthetic production channels of N and $\alpha$ elements.
Accordingly, we establish a hierarchy of preferred metallicity indicators based on the emission lines available for a given source.

The most-preferred metallicity estimate is derived by simultaneously fitting [O\iii]$\lambda$5007/H$\beta$,
 [O\iii]$\lambda$5007/[O\ii], and (when available) [Ne\iii]/[O\ii], following the approach of \citet{san21} and referred
 to here as O2O3Ne3.
If [O\ii] measurements are not available, O3N2$\equiv$([O\iii]$\lambda$5007/H$\beta$)/([N\ii]$\lambda$6584/H$\alpha$) is
 used to derive metallicities.
If [O\iii] and/or H$\beta$ are lacking, then N2=[N\ii]$\lambda$6584/H$\alpha$ is used.
O3N2 is preferred over N2 because metallicities estimated from O3N2 are less affected by N/O variations
 than those based on N2 alone.
As described in Appendix~\ref{app:metallicity}, the \citet{bia18} O3N2 and N2 calibrations have been renormalized
 such that they yield consistent metallicities on average with those derived from O2O3Ne3 for galaxies at $z\sim2$.

In our sample, 10 galaxies have detections of [O\ii], [O\iii], and H$\beta$ (2 with [Ne\iii] as well) for which O2O3Ne3
 metallicities can be derived.
All 13 targets have detections of H$\beta$, [O\iii], H$\alpha$, and [NII] such that O3N2 and N2 metallicities are
 available.
In the following analysis, the oxygen abundance of each source is taken to be the best available estimate
 based on the preference of O2O3Ne3, then O3N2, then N2 unless otherwise specified.
We obtain consistent results if we instead use O3N2 metallicities uniformly for the full sample, with the main
 quantitative difference being that the slopes of the anti-correlations presented in Sec.~\ref{sec:residuals}
 are slightly different.
We define the offset from the mass-metallicity relation at fixed \mstar\ as a function of redshift as
 \delmzr$\equiv12+\log($O/H$)-\text{MZR}(M_*, z)$.
Here, MZR$(M_*, z)$ is the parameterization of the evolving mass-metallicity relation
 derived in Appendix~\ref{app:mzrz}:
\begin{equation}\label{eq:fitmzrz}
\text{MZR}(M_*,z) = 8.80 - 0.28\times\log(1+[M_*/M_0(z)]^{-1.08})
\end{equation}
 where $\log(M_0(z)/\text{M}_{\odot})=9.90 + 2.06\times\log(1+z)$.

\subsection{Composite \cott\ Spectra}

Given that \cott\ emission is not robustly detected for approximately half of our sample,
 we employ spectral stacking techniques to include information from all of the targets in our analysis.
We create composite spectra by converting the extracted 1D spectra from flux density to luminosity density and
 shifting the frequency axis to velocity offset based on the systemic redshift measured from the rest-optical lines.
We then combine the individual spectra by taking the unweighted mean luminosity density at intervals
 of 50~km~s$^{-1}$ velocity offset.
We do not use any weighting because the on-source integration times and sensitivities
 varied widely from source-to-source, such that the few sources with long integration times dominate the
 stacks if inverse-variance weighting is used.
The composite \cott\ spectrum is then converted back to flux density using the mean redshift of the included individual galaxies.
The composite \cott\ line flux is measured by fitting a Gaussian profile to the composite 1D spectrum,
 and is presented in Table~\ref{tab:mosdefalma}.

We create composite \cott\ spectra of all of the galaxies in our sample (13 sources; labeled ``stack-all'') and
 the subset that are not robustly detected (8 sources; ``stack-nondet'').
We also stack in two bins of various galaxy properties divided such that there are roughly equal numbers of galaxies per bin.
The galaxy properties according to which we bin include 
 offset from the MS (\delsfms; ``stack-delsfms''),
 and  metallicity (12+log(O/H); ``stack-oh'').
Splitting the sample in \sigsfr\ results in identical bins as stack-delsfms,
 while binning according to offset from the mass-metallicity relation (\delmzr)
 yields the same results as stack-oh.
The mass, SFR, effective radius, and metallicity associated with each composite \cott\ spectrum and
 given in Table~\ref{tab:mosdefalma} are
 the mean log(\mstar), log(SFR), rest-optical $R_{\text{eff}}$, and 12+log(O/H)
 of the individual galaxies included in each stack.

\begin{figure}
 \includegraphics[width=\columnwidth]{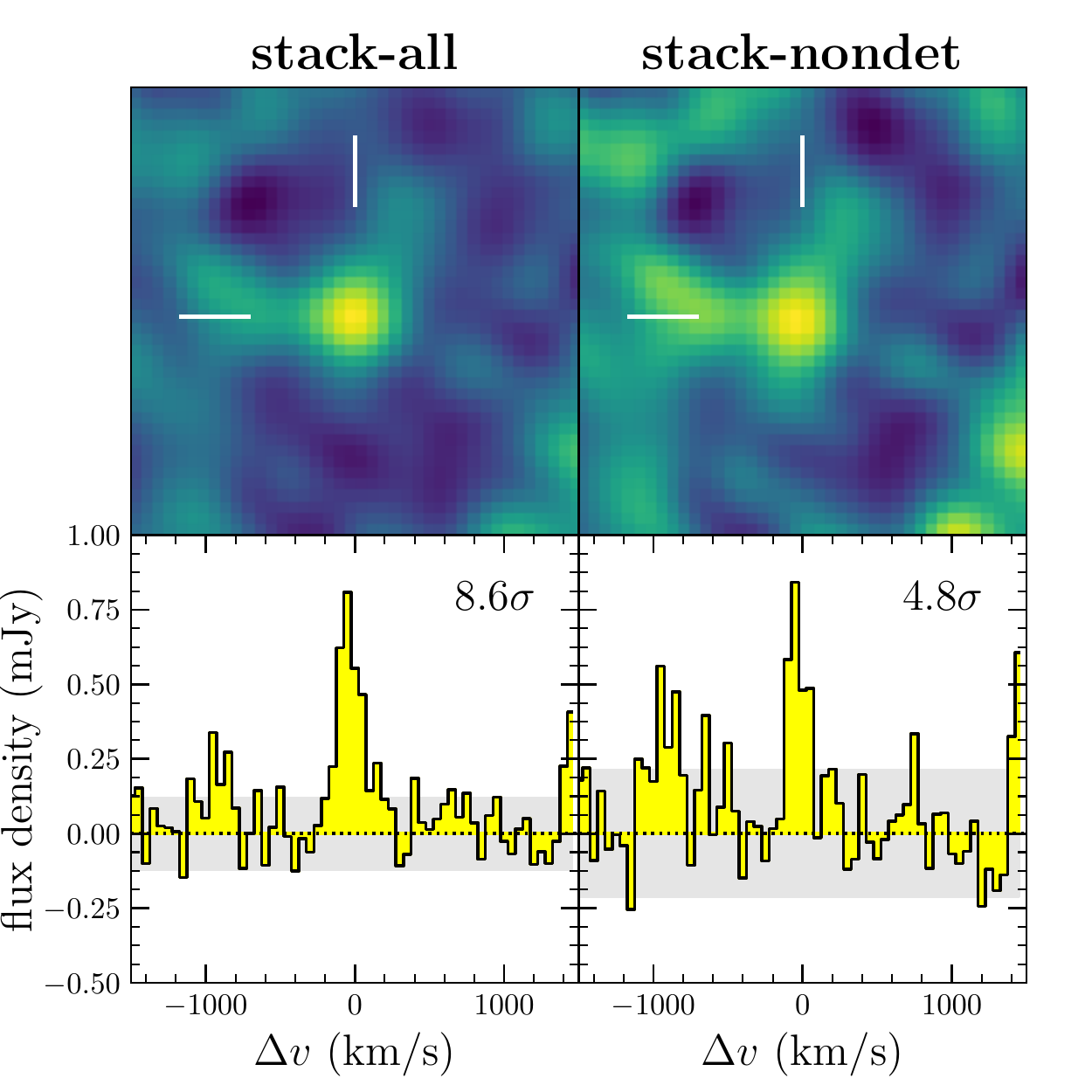}
 \centering
 \caption{
Example 2D (top) and 1D (bottom) composite \cott\ spectra for the stack-all (left; 13 galaxies)
 and stack-nondet (right; 8 galaxies) samples.
The white crosshair is aligned with the expected spatial location of emission based on \textit{HST} imaging centroids.
The integrated \cott\ line S/N is given in the upper right corner of the bottom panels.
The similarity of the noise patterns in the left and right columns arises because there are 8 galaxies in common between
 the two stacks.
}\label{fig:stacks}
\end{figure}

Example composite \cott\ 1D spectra are shown in the bottom row of Figure~\ref{fig:stacks} for the
 stack-all and stack-nondet samples.
The integrated \cott\ significance is $\ge4.8\sigma$ for all stacked spectra used in this paper
 (stack-nondet has the lowest significance).
The top row displays 2D stacks constructed by producing datacubes using \textsc{tclean}
 with a $uv$ taper of 3\arcsec\ and a spatial sampling of 0.5\arcsec/pixel to homogenize the beam sizes,
 integrating the channels within $\pm$300~km~s$^{-1}$ of the systemic redshift,
 and taking the mean flux value at each spatial position relative to the \textit{HST} imaging centroid.
While we utilize composite \cott\ fluxes from the stacked 1D spectra in the following analysis,
 line fluxes derived by spatially integrating the 2D composites agree with the 1D fluxes within 20\%.

\begin{table*}
\movetabledown=2.25in
\begin{rotatetable*}
 \centering
 \caption{Galaxy properties, \cott\ measurements, and derived molecular gas properties of the MOSDEF-ALMA sample and composite spectra.
 }\label{tab:mosdefalma}
 \setlength{\tabcolsep}{4pt}
 \renewcommand{\arraystretch}{1.2}
 \begin{tabular}{ l l l l l l l l l l l l l l }
   \hline\hline
ID  &  R.A.  &  Dec.  &  $t_{\text{int}}$  &  RMS  &  $z_{\text{spec}}$  &
$\log{\left(\frac{M_*}{\text{M}_\odot}\right)}$  &
log$\left(\frac{\text{SFR}}{\text{M}_\odot/\text{yr}}\right)$  &
12+log(O/H)  &
$R_{\text{eff}}$  &
\scott  &
log$\left(\text{L}^{\prime}_{\text{CO}(3-2)}\right)$  &
log$\left(\frac{M_{\text{mol}}}{\text{M}_{\odot}}\right)$  &
log$\left(\mu_{\text{mol}}\right)$  \\
 & \scriptsize{J2000} & \scriptsize{J2000} & \scriptsize{hours} & \scriptsize{$\frac{\text{Jy}}{\text{beam}}$} & & & & & \scriptsize{kpc} & \scriptsize{Jy km~s$^{-1}$}
 & 
 & & \\
   \hline
13296  &  10:00:27.623  &  +02:18:55.05  &  2.82  &  88  &  2.1672  &  10.40$\pm$0.04  &  1.46$\pm$0.16  &  8.81$\pm$0.03  &  3.4  &  0.139$\pm$0.026  &  9.54$\pm$0.08  &  10.44$\pm$0.08  &  0.04$\pm$0.09  \\
19753\tablenotemark{a}  &  10:00:18.182  &  +02:22:50.32  &  0.83  &  198  &  2.4694  &  10.33$\pm$0.08  &  1.83$\pm$0.04  &  ---  &  4.1  &  0.250$\pm$0.051  &  9.90$\pm$0.09  &  ---  &  ---  \\
13701\tablenotemark{b}  &  10:00:27.052  &  +02:19:09.98  &  2.82  &  86  &  2.1659  &  10.57$\pm$0.04  &  2.19$\pm$0.10  &  8.81$\pm$0.02  &  4.1  &  0.147$\pm$0.032  &  9.57$\pm$0.10  &  10.46$\pm$0.10  &  -0.11$\pm$0.11  \\
2672  &  10:00:31.073  &  +02:12:25.91  &  1.36  &  143  &  2.3074  &  10.50$\pm$0.05  &  1.99$\pm$0.16  &  8.61$\pm$0.03  &  4.1  &  0.207$\pm$0.047  &  9.77$\pm$0.10  &  10.80$\pm$0.10  &  0.30$\pm$0.11  \\
5094  &  10:00:33.688  &  +02:13:48.67  &  0.64  &  174  &  2.1715  &  10.34$\pm$0.01  &  1.88$\pm$0.27  &  8.71$\pm$0.06  &  5.0  &  0.298$\pm$0.076  &  9.88$\pm$0.11  &  10.77$\pm$0.11  &  0.43$\pm$0.11  \\
3324  &  10:00:35.618  &  +02:12:47.280  &  0.87  &  192  &  2.3072  &  10.55$\pm$0.01  &  1.80$\pm$0.15  &  8.84$\pm$0.03  &  4.8  &  0.302$\pm$0.079  &  9.93$\pm$0.12  &  10.82$\pm$0.12  &  0.27$\pm$0.12  \\
19985  &  10:00:14.484  &  +02:22:57.98  &  0.23  &  322  &  2.1882  &  10.20$\pm$0.07  &  2.29$\pm$0.04  &  8.36$\pm$0.01  &  1.3  &  0.219$\pm$0.079  &  9.75$\pm$0.16  &  11.19$\pm$0.16  &  0.99$\pm$0.18  \\
24763  &  10:00:13.608  &  +02:26:04.786  &  1.50  &  154  &  2.4649  &  10.46$\pm$0.05  &  1.32$\pm$0.13  &  8.66$\pm$0.06  &  5.1  &  0.081$\pm$0.035  &  9.41$\pm$0.20  &  10.36$\pm$0.20  &  -0.10$\pm$0.21  \\
20062  &  10:00:16.436  &  +02:23:00.79  &  0.45  &  254  &  2.1857  &  10.24$\pm$0.04  &  2.36$\pm$0.03  &  8.49$\pm$0.01  &  1.5  &  $<$0.159  &  $<$9.61  &  $<$10.84  &  $<$0.60  \\
3666  &  10:00:18.607  &  +02:12:57.72  &  0.39  &  265  &  2.0859  &  10.25$\pm$0.04  &  1.94$\pm$0.05  &  8.45$\pm$0.01  &  3.6  &  $<$0.268  &  $<$9.80  &  $<$11.09  &  $<$0.84  \\
4497  &  10:00:17.153  &  +02:13:25.94  &  0.72  &  193  &  2.4413  &  10.35$\pm$0.00  &  1.98$\pm$0.14  &  8.65$\pm$0.03  &  3.1  &  $<$0.154  &  $<$9.68  &  $<$10.65  &  $<$0.30  \\
4930  &  10:00:29.037  &  +02:13:43.66  &  1.18  &  154  &  2.2265  &  10.44$\pm$0.05  &  1.74$\pm$0.20  &  8.63$\pm$0.06  &  5.7  &  $<$0.120  &  $<$9.50  &  $<$10.50  &  $<$0.06  \\
5814  &  10:00:40.591  &  +02:14:18.22  &  0.16  &  456  &  2.1266  &  10.48$\pm$0.04  &  2.14$\pm$0.07  &  8.72$\pm$0.02  &  4.7  &  $<$0.298  &  $<$9.86  &  $<$10.76  &  $<$0.28  \\
9971  &  10:00:34.449  &  +02:16:54.47  &  0.57  &  218  &  2.4108  &  10.26$\pm$0.05  &  2.01$\pm$0.06  &  8.55$\pm$0.01  &  1.7  &  $<$0.163  &  $<$9.70  &  $<$10.83  &  $<$0.57  \\
   \hline\hline
\multicolumn{14}{c}{Composite Spectra}  \\
   \hline
\multicolumn{3}{c}{name}  &  bin  &  $N_{\text{gal}}$  &  
$z_{\text{spec}}$  &
$\log{\left(\frac{M_*}{\text{M}_\odot}\right)}$  &
log$\left(\frac{\text{SFR}}{\text{M}_\odot/\text{yr}}\right)$  &
12+log(O/H)  &
$R_{\text{eff}}$  &
\scott  &
log$\left(\text{L}^{\prime}_{\text{CO}(3-2)}\right)$  &
log$\left(\frac{M_{\text{mol}}}{\text{M}_{\odot}}\right)$  &
log$\left(\mu_{\text{mol}}\right)$  \\
   \hline
\multicolumn{3}{c}{stack-all}  &  ---   &  13  &  2.250  &  10.39$\pm$0.03  &  1.93$\pm$0.09  &  8.64$\pm$0.04  &  3.7  &  0.158$\pm$0.018  &  9.63$\pm$0.05  &  10.61$\pm$0.05  &  0.22$\pm$0.06   \\
\multicolumn{3}{c}{stack-nondet}  &  ---  &  8  &  2.266  &  10.33$\pm$0.03  &  1.97$\pm$0.09  &  8.56$\pm$0.04  &  3.3  &  0.144$\pm$0.030  &  9.59$\pm$0.09  &  10.70$\pm$0.10  &  0.37$\pm$0.10   \\
\multicolumn{3}{c}{stack-oh}  &  low  &  6  &  2.234  &  10.31$\pm$0.05  &  2.05$\pm$0.10  &  8.51$\pm$0.04  &  3.0  &  0.146$\pm$0.030  &  9.59$\pm$0.09  &  10.79$\pm$0.09  &  0.48$\pm$0.10  \\
\multicolumn{3}{c}{stack-oh}  &  high  &  7  &  2.264  &  10.45$\pm$0.04  &  1.83$\pm$0.13  &  8.74$\pm$0.03  &  4.3  &  0.173$\pm$0.024  &  9.67$\pm$0.06  &  10.56$\pm$0.06  &  0.11$\pm$0.07   \\
\multicolumn{3}{c}{stack-delsfms}  &  low  &  6  &  2.274  &  10.45$\pm$0.03  &  1.70$\pm$0.13  &  8.71$\pm$0.04  &  4.7  &  0.165$\pm$0.022  &  9.66$\pm$0.06  &  10.55$\pm$0.06  &  0.10$\pm$0.07   \\
\multicolumn{3}{c}{stack-delsfms}  &  high  &  7  &  2.229  &  10.34$\pm$0.05  &  2.13$\pm$0.07  &  8.58$\pm$0.06  &  2.9  &  0.154$\pm$0.031  &  9.61$\pm$0.09  &  10.69$\pm$0.09  &  0.35$\pm$0.11   \\
   \hline\hline
 \end{tabular}
\begin{flushleft}
\tablecomments{ID numbers are from the 3D-HST v4.1 photometric catalog \citep{mom16}.
RMS gives the noise per beam over a bandwidth of 50~km~s$^{-1}$ at the frequency of the \cott\ line.
All metallicities are based on O3O2Ne3 except for those of IDs~5094, 4930, and 24763 that use O3N2.
\lcott\ is in units of M$_{\odot}$~(K km~s$^{-1}$ pc$^2$)$^{-1}$.
\mmol\ and \mumol\ are calculated assuming the \aco(O/H) relation of \citet[see Sec.~\ref{sec:aco} and equation~\ref{eq:acc17}]{acc17}.
\cott\ fluxes, luminosities, and derived gas properties are given as 3$\sigma$ upper limits for undetected sources.
The $z_{\text{spec}}$, log(\mstar), log(SFR), 12+log(O/H), and $R_{\text{eff}}$ assigned to each stack are the mean values of the individual galaxies.
}
\tablenotetext{a}{The rest-optical lines and \cott\ emission likely originate from different components of ID~19753. As such, we do not report the metallicity or calculate \mmol\ or \mumol\ for this object (see Sec.~\ref{sec:almaobs}).}
\tablenotetext{b}{The \cott\ spectrum of ID~13701 was extracted from the datacube targeting ID~13296.}
\end{flushleft}
\end{rotatetable*}
\end{table*}

\subsection{Literature CO Sample}\label{sec:lit}

We draw from the literature a supplementary sample of star-forming galaxies at $z=2-3$ with both CO and
 metallicity constraints available.
We selected galaxies at $z=2-3$ with \cott\ measurements from the PHIBSS \citep{tac13} and ASPECS \citep{boo19} surveys.
This literature CO sample was then cross-matched with rest-optical spectroscopic surveys to identify objects for which
 metallicity constraints could be derived.
Objects with [N\ii]/H$\alpha>0.5$ were excluded as probable AGN.
One additional object at $z=1.99$ with rest-optical spectroscopy from \citet{sha20} was included with a \cooz\ measurement
 from \citet{rie20}.
This literature sample comprises 23 galaxies with CO measurements (19 CO detections and 4 upper limits),
 16 (12 CO detections and 4 upper limits) of which have gas-phase metallicity constraints from rest-optical emission lines
 (1 from O3O2Ne3, 6 from O3N2, and 9 from N2).
The properties of the literature sample are presented in Table~\ref{tab:litco}.
Stellar masses and SFRs are taken from the literature sources and converted to a \citet{cha03} IMF when necessary.
Line ratios are calculated using the rest-optical line fluxes from the given spectroscopic O/H references,
 which we then use to derive metallicities as described in Sec.~\ref{sec:metallicity}.
This sample of $z=2-3$ literature sources complements the MOSDEF-ALMA sample in our analysis.

\section{Results}\label{sec:results}

\subsection{Sample Properties}\label{sec:sample}

\begin{figure}
 \includegraphics[width=1.00\columnwidth]{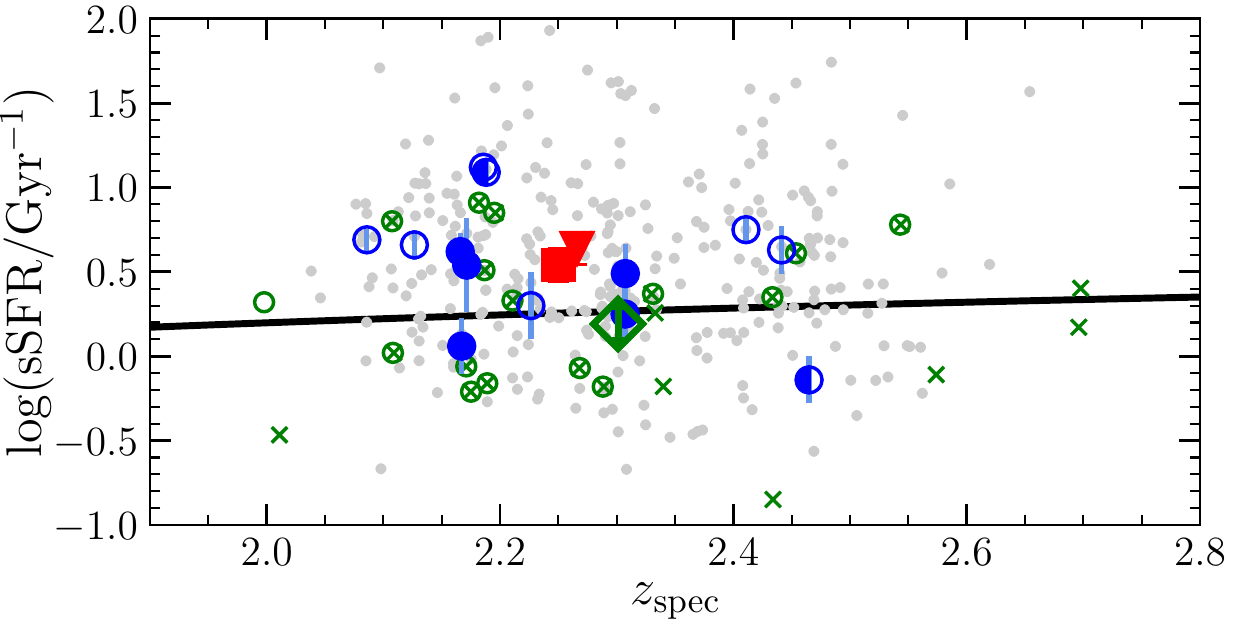}
 \includegraphics[width=1.00\columnwidth]{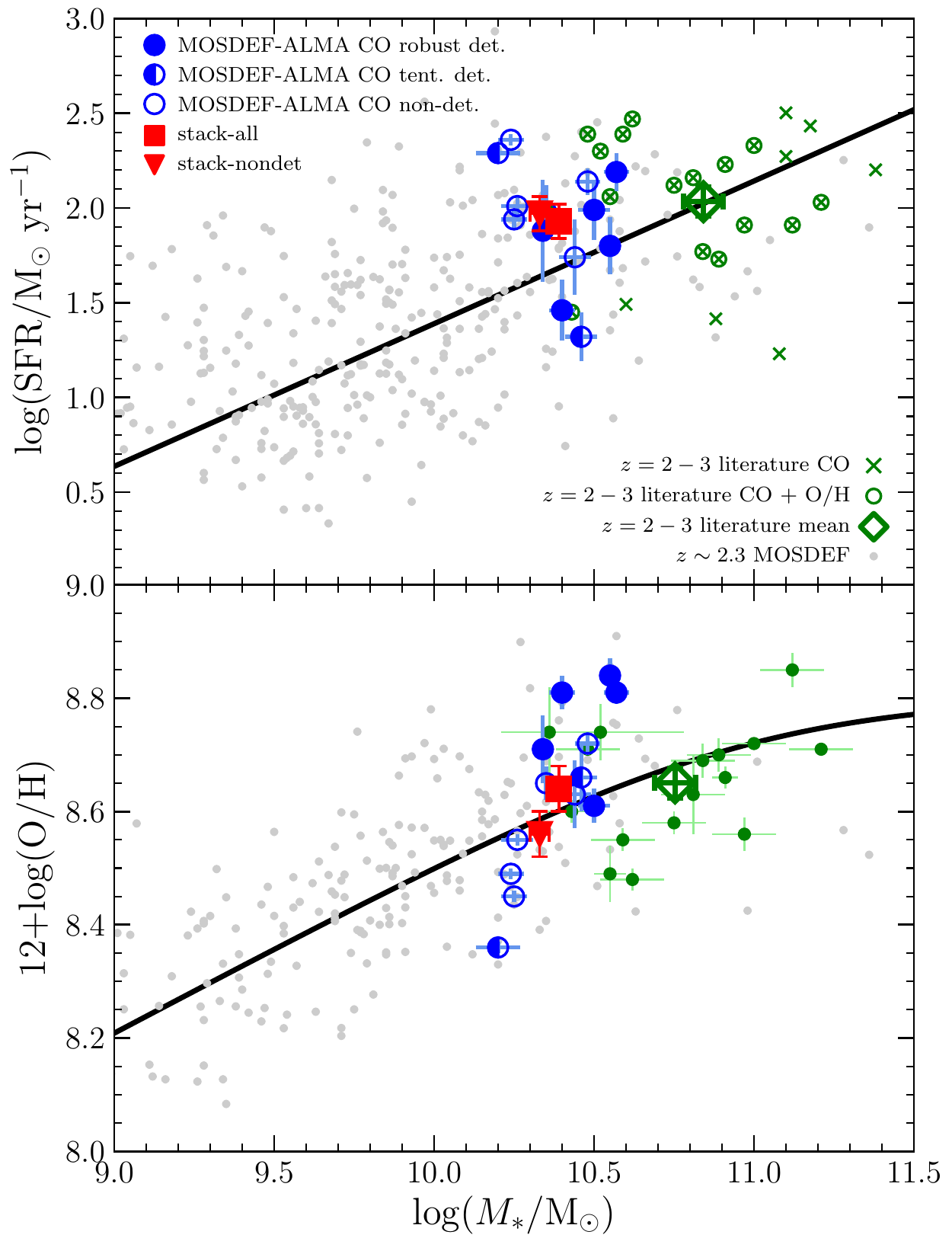}
 \centering
 \caption{
Properties of the $z\sim2.3$ MOSDEF-ALMA sample (blue)
 and the $z=2-3$ literature sample (green).
\textit{Top:} Specific SFR (sSFR=SFR/\mstar) vs.\ redshift.
Robust CO detections from MOSDEF-ALMA are displayed as filled blue circles, while
 tentative detections are half-filled and non-detections are unfilled.
Red points show values for composite spectra.
Green ``x''es denote literature objects with CO measurements, where a green circle indicates those with gas-phase
 metallicity constraints.
The green diamond shows the mean values of the literature sample.
The full sample of $z\sim2.3$ star-forming galaxies from the MOSDEF survey are displayed as small gray points.
The black solid line presents the star-forming main sequence of \citet{spe14} evaluated at $10^{10.5}~\msun$.
\textit{Middle:} SFR vs.\ \mstar, with points as above.
The black line shows the \citet{spe14} parameterization evaluated at $z=2.3$.
\textit{Bottom:} O/H vs.\ \mstar.
This panel only includes literature sources with metallicity constraints (green circles).
The black line displays the parameterization of the mass-metallicity relation given in
 equation~\ref{eq:fitmzrz} evaluated at $z=2.3$.
}\label{fig:sample}
\end{figure}

The redshift, \mstar, and SFR properties of the MOSDEF-ALMA and $z=2-3$ literature CO samples
 are shown in Figure~\ref{fig:sample}.
The two samples are well matched in redshift (top panel)
 with mean redshifts of 2.25 and 2.30, respectively.
The combined sample spans $z=2.0-2.7$ with a mean redshift of 2.28$\pm$0.17, where the uncertainty represents
 the sample standard deviation.

The middle panel of Fig.~\ref{fig:sample} shows SFR vs.\ \mstar.
The MOSDEF-ALMA sample spans a range of stellar masses $\log(M_*/\msun)=10.20-10.57$ with a mean mass of $10^{10.39} \msun$,
 and a range of SFRs $\log(\mbox{SFR}/\msun\mbox{ yr}^{-1})=1.32-2.36$ with a mean SFR of $85~\msun$~yr$^{-1}$.
The MOSDEF-ALMA sample falls 0.25~dex above the MS on average (red square),
 while probing a large range of offsets below and above the MS.
Our CO sample is complementary to existing ones at these redshifts, lying at lower stellar masses than samples from
 the large CO surveys PHIBSS and ASPECS (green points).
The $z=2-3$ literature sample has a mean stellar mass of $10^{10.84}~\msun$, a factor of 2.8 times larger than that of MOSDEF-ALMA,
 with a mean SFR that falls directly on the MS.
In the combined sample, 29/36 sources fall within a factor of 3 of the \citet{spe14} MS.
Offsets from the MS are nearly identical if we instead adopt the MS parameterization of \citet{whi14},
 which matches that of \citet{spe14} at $<0.1$~dex in SFR at fixed \mstar\ across the mass range of the combined sample.

The bottom panel of Fig.~\ref{fig:sample} presents the mass-metallicity relation (MZR) at $z\sim2.3$.
On average, both the MOSDEF-ALMA and literature samples fall on the mean MZR at this redshift (black line,
 equation~\ref{eq:fitmzrz}).
The MOSDEF-ALMA sample in particular displays a wide range of metallicities spanning 12+log(O/H$)=8.36-8.84$
 ($0.5-1.4$~Z$_{\odot}$), with a mean metallicity of 8.64 (0.9~Z$_{\odot}$).

Collectively, the combined sample is representative in both SFR and metallicity of typical $z\sim2.3$ star-forming
 galaxies in the mass range $\log(M_*/\msun)=10.3-11.0$, falling on the mean MS and MZR at this redshift.
The addition of the MOSDEF-ALMA observations expands the sample of $z\sim2$ galaxies with CO measurements to lower
 masses and metallicities than were available to date.
The wide range of offsets both above and below the mean MS and MZ relations makes MOSDEF-ALMA an ideal sample
 with which to search for correlated residuals around mass-scaling relations that are signposts of self-regulated
 baryon cycling.

\subsection{\cott\ Luminosity and Galaxy Properties}\label{sec:lco32}

Before deriving molecular gas masses, we first explore 
 empirical relations between observed \cott\ luminosity and global galaxy properties with the goal
 of assessing whether the CO production efficiency, and thus the CO-to-\htwo\ conversion factor,
 varies systematically.
Integrated \cott\ line fluxes, $S_{\text{CO}}$, are converted to
 total line luminosities using the relation of \citet{sol97}:
\begin{equation}\label{eq:lcosco}
L^{\prime}_{\text{CO}}=3.25\times10^7 S_{\text{CO}} \nu_{\text{obs}}^{-2} D_L^2 (1+z)^{-3}
\end{equation}
where $\nu_{\text{obs}}$ is the observed line frequency in GHz and $D_L$ is the luminosity distance in Mpc.
The MOSDEF-ALMA \cott\ luminosities are given in Table~\ref{tab:mosdefalma}.

\begin{figure}
 \includegraphics[width=\columnwidth]{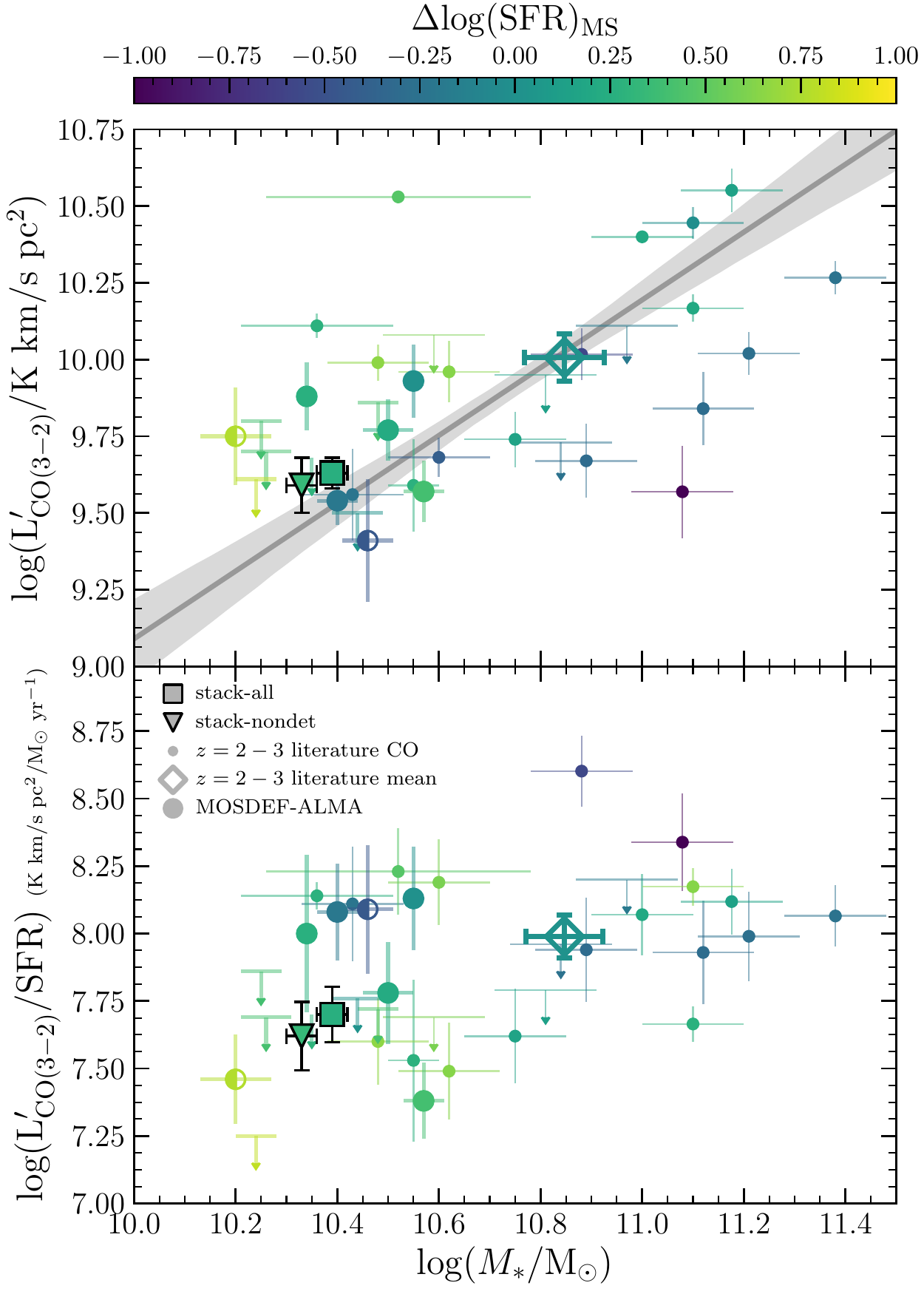}
 \centering
 \caption{
\cott\ luminosity (\lcott, top) and \lcott/SFR (bottom) vs.\ stellar mass.
Gray lines indicate best-fit linear relations from a bivariate weighted orthogonal distance regression.
The light gray shaded region displays the 1$\sigma$ uncertainty bounds on the best-fit line.
In the top panel, the gray line corresponds to equation~\ref{eq:lco32mstar} evaluated at \delsfms=0.
}\label{fig:lco32}
\end{figure}

The relation between \lcott\ and stellar mass is shown in the top panel of Figure~\ref{fig:lco32}.
The points are color-coded by offset from the star-forming main sequence, \delsfms.
We find a significant positive correlation between \lcott\ and \mstar, with an evident secondary dependence on
 \delsfms\ such that galaxies at fixed \mstar\ with higher \lcott\ have higher SFR.
Such a secondary dependence is expected if galaxies above (below) the MS have larger (smaller) molecular gas reservoirs
 than those on the MS, as has been observed in the local universe \citep[e.g.,][]{sai16,sai17,sai22}
 and at high redshifts with dust- and CO-based measurements \citep[e.g.,][]{tac13,tac18,tac20,gen15,sco17,ara20}.
Using an orthogonal distance regression,
 we fit a function that is linear in $\log(M_*)$ and \delsfms\ to the combined sample of individual galaxies (excluding limits),
 and find a best-fit relation of:
\begin{equation}\label{eq:lco32mstar}
\begin{multlined}
\log\left(\frac{\mbox{L}^{\prime}_{\text{CO}(3-2)}}{\footnotesize{\mbox{K km~s$^{-1}$ pc}^2}}\right) = 1.10(\pm0.17)\times\log\left(\frac{M_*}{10^{10.7} \msun}\right) \\
+ 0.71(\pm0.14)\times\log\left(\frac{\mbox{SFR}}{\mbox{SFR}_{\text{MS}}}\right) + 9.86(\pm0.04)
\end{multlined}
\end{equation}
Since this relation is a function of \delsfms, the fact that the MOSDEF-ALMA sample lies above the MS on average
 does not introduce bias.
The stack of the MOSDEF-ALMA non-detections is $1\sigma$ consistent with this best-fit relation, suggesting that
 the exclusion of limits when fitting has not significantly biased this result.
After accounting for measurement uncertainties, the remaining intrinsic scatter in \lcott\ around the best-fit function
 is $\sigma_{\mathrm{int}}=0.17$~dex.
This equation can thus be used to predict the \cott\ luminosity of $z\sim2$ galaxies to within a factor of $\approx$1.5 using
 \mstar\ and SFR alone.
If the SFR term is neglected such that \lcott\ is a function of \mstar\ alone (gray line, top panel of Fig.~\ref{fig:lco32}),
 the intrinsic scatter is 0.31~dex.

The ratio of \lcott\ to SFR is a proxy of the CO production per unit \mmol\ since SFR is tightly coupled to
 \mmol\ via the molecular KS law \citep[e.g.,][]{ken98,ken21,tac13,rey19}.
The bottom panel of Figure~\ref{fig:lco32} displays \lcott/SFR as a function of \mstar.
A positive correlation remains after normalizing \lcott\ by SFR, made clear by the sample means,
 though with very large scatter at fixed \mstar.
No secondary trend with \delsfms\ is evident.
The correlation in the bottom panel of Fig.~\ref{fig:lco32} implies that \cott\ emission is more efficiently
 produced per unit molecular gas mass in high-mass galaxies.

\begin{figure}
 \includegraphics[width=\columnwidth]{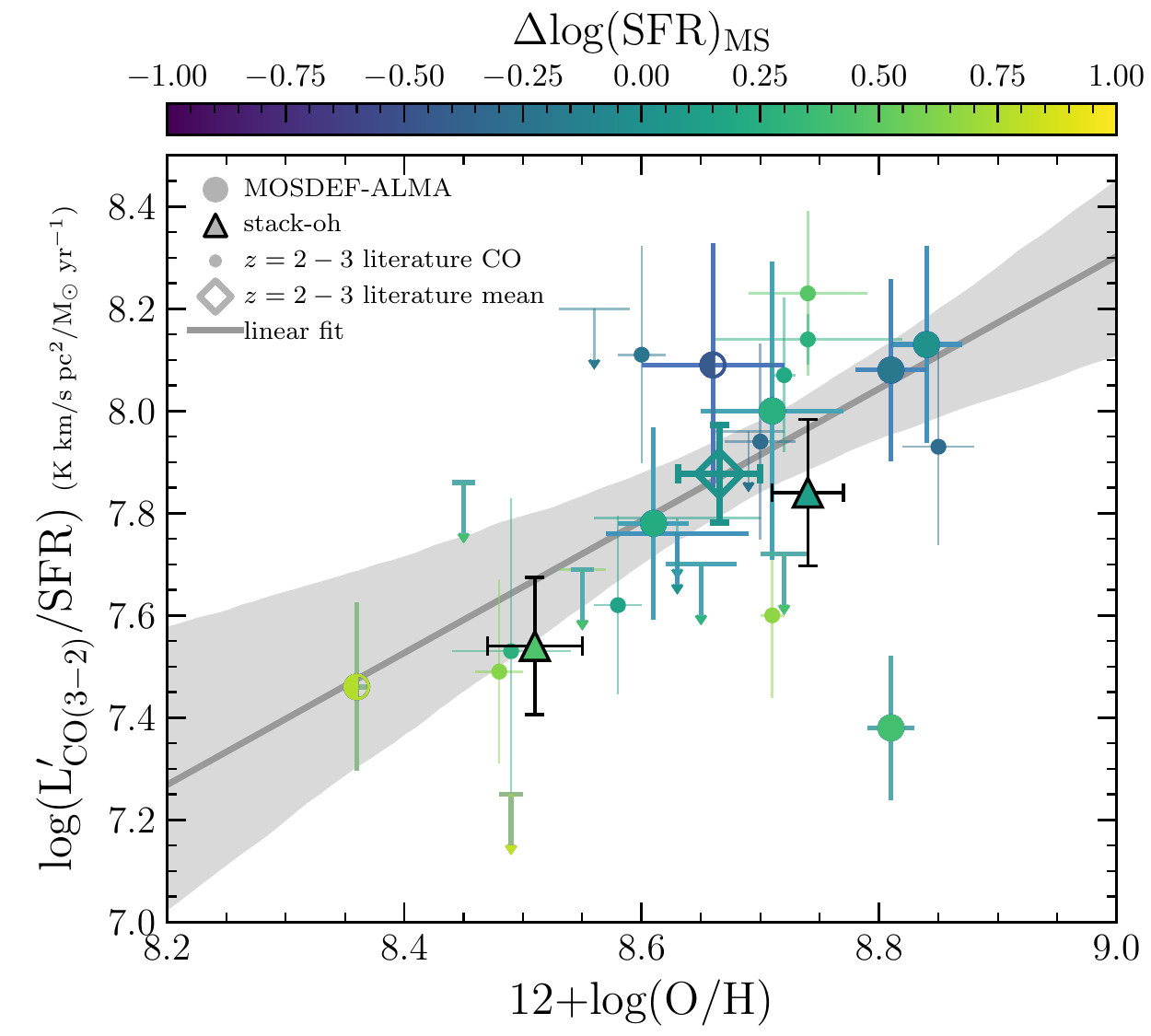}
 \centering
 \caption{
\lcott/SFR vs.\ gas-phase oxygen abundance, with points and lines as in Figure~\ref{fig:lco32}.
Colored triangles denote measurements from composite spectra in two bins of O/H.
The best-fit linear relation is given in equation~\ref{eq:lco32sfroh}.
}\label{fig:lco32sfroh}
\end{figure}

Figure~\ref{fig:lco32sfroh} presents \lcott/SFR as a function of metallicity.
There is a positive correlation between these two quantities, with significantly smaller scatter
 in \lcott/SFR at fixed O/H compared to that at fixed \mstar, suggesting that the relation between \lcott/SFR
 and O/H is more fundamental such that the correlation with \mstar\ emerges because of the mass-metallicity relation.
Applying a Spearman correlation test to the individually detected sources yields
 $\rho_S=0.45$ and a $p\text{-value}=0.05$, indicating a 2$\sigma$ correlation.
If the one significant outlier is excluded (discussed below), the correlation is stronger
 with $\rho_S=0.65$ and $p\text{-value}=0.003$, indicating 3$\sigma$ significance.
Fitting a linear function to the individual galaxies, excluding limits, yields:
\begin{equation}\label{eq:lco32sfroh}
\begin{multlined}
\log\left(\frac{\mbox{L}^{\prime}_{\text{CO}(3-2)}}{\mbox{SFR}}\right) = 1.29(\pm0.44)\times x + 7.91(\pm0.07)
\end{multlined}
\end{equation}
where $x=12+\log(\mbox{O/H})-8.7$, \lcott\ is in units of K~km~s$^{-1}$~pc$^2$, and SFR is in units of $\msun$~yr$^{-1}$.
There is no significant secondary dependence on \delsfms,
 suggesting that the best-fit relation is robust even though the
 MOSDEF-ALMA sample is offset above the MS on average.
The composite spectra agree with the best-fit relation at the $1\sigma$ level, implying that our exclusion of limits
 when fitting has not significantly biased the result.
The intrinsic scatter about this best-fit line is 0.12~dex after accounting for measurement uncertainties.

The single significant outlier in Fig.~\ref{fig:lco32sfroh}, lying $\sim4\sigma$ below the best-fit line,
is ID~13701 in the MOSDEF-ALMA sample, a potential merger with tidal tail-like features and
 two distinct brightness peaks in \textit{HST} imaging (Fig.~\ref{fig:obs}).
It is possible that the low \lcott/SFR of ID~13701 is a result of an enhanced star-formation efficiency (SFR/\mmol)
during a major merger \citep{dim07,sar14,ken21} since \lcott\ traces \mmol, though it is unclear how significant
 such enhancement is expected to be in high-redshift mergers \citep{fre17}.

The positive correlation between \lcott/SFR and O/H demonstrates that the production efficiency of \cott\ 
 is a function of metallicity, with low-metallicity galaxies outputting less \cott\ per unit SFR than metal-rich galaxies.
If the star-formation efficiency (or, equivalently, depletion timescale) is similar across our sample, then
 the correlation in Fig.~\ref{fig:lco32sfroh} implies a decreasing CO luminosity per unit \mmol\ with decreasing
 metallicity in $z\sim2$ star-forming galaxies.
This same trend has been observed in local galaxy samples, manifesting as a metallicity-dependent CO-to-\htwo\ conversion
 factor (\aco) with \aco\ increasing with decreasing metallicity \citep{wil95,ari96,wol10,sch12,bol13,acc17}.
Higher \aco\ at lower O/H, corresponding to lower CO luminosity per unit gas mass, arises because CO is dissociated at
 greater depths into \htwo\ clouds in low-metallicity environments.
CO molecules are photodissociated by far-UV photons and require dust-shielding to prevent dissociation \citep[e.g.,][]{wol10,glo11}.
Metal-poor galaxies have lower dust-to-gas ratios \citep[e.g.,][]{san13,dev19}
 and low-metallicity massive stars produce harder and more intense UV radiation fields,
 such that there is an increasing fraction of CO-dark \htwo\ gas with decreasing metallicity.
The trend in Fig.~\ref{fig:lco32sfroh} thus suggests that a metallicity-dependent \aco\ is required to accurately translate
 observed CO luminosities of $z\sim2$ galaxies into molecular gas masses.

\subsection{The CO-to-\htwo\ Conversion Factor at $z\sim2$}\label{sec:aco}

Given the evidence for a metallicity-dependent \aco\ at $z\sim2$ presented above,
 we utilize a combined sample of MOSDEF-ALMA and literature targets at $z>1$ with CO and rest-optical spectra
 to constrain the form of the \aco$-$O/H relation at high redshifts.
\citet{gen12} presented evidence for a \aco-O/H relation at high redshift using a sample of $\sim40$
 $z>1$ main-sequence galaxies, but the majority of their sample lacked spectroscopic metallicity
 constraints and instead had metallicities inferred from the MZR,
 with only a subset of 9 CO-detected galaxies having metallicities based on [N\ii]/H$\alpha$ ratios.
The analysis below represents a significant improvement on the foundational results of \citet{gen12}
 by using a sample composed entirely of galaxies with spectroscopic metallicities.

To maximize the sample size for the \aco\ analysis, we supplement the combined $z=2-3$ MOSDEF-ALMA
 and literature CO+O/H sample with 14 additional galaxies at $z>1$ with the necessary observations
 that were not included in our primary
 sample because they either fall outside of the target redshift range (i.e., at $z=1-2$ or $z>3$), are extreme starbursts
 with \delsfms$\ge$10, or are strongly gravitationally lensed.
Sources in this supplementary $z>1$ sample are given in Table~\ref{tab:litco}.
Thus, this expanded sample is only used for the \aco\ analysis, while the remainder of the paper employs only the
 more homogeneous $z=2-3$ MOSDEF-ALMA and literature samples described in Sec.~\ref{sec:lit}.
The \aco\ sample includes 43 galaxies spanning $z=1.08-3.22$ with a mean redshift of 2.12.
Of these 43 sources, all have metallicity constraints from rest-optical line ratios, 33 have CO detections,
 and 10 have CO upper limits.

The CO-to-\htwo\ conversion factor \aco\ is defined as
\begin{equation}\label{eq:aco}
\alpha_{\mathrm{CO}}=\frac{M_{\mathrm{mol}}}{\mathrm{L}^{\prime}_{\text{CO}(1-0)}}=\frac{r_{J1}M_{\mathrm{mol}}}{\mathrm{L}^{\prime}_{\text{CO}(J-J\text{-}1)}}
\end{equation}
where \mmol\ is the total molecular gas mass including a 36\% contribution from Helium,
 and $r_{J1}$ is the excitation correction factor to convert higher-$J$ CO luminosities to that of the ground transition.
We adopt $r_{31}=0.55$ \citep{tac18} and $r_{21}=0.76$ \citep{dad15}.
Our \aco\ sample contains 33 galaxies with \cott\ measurements, 7 with \coto\, and 3 with \cooz.
Our results are thus most sensitive to the assumed $r_{31}$ value, whereas the adopted $r_{21}$ will have only a minor effect.

Constraining \aco\ requires an estimate of \mmol\ that is independent of CO emission.
We employ two methods of estimating \mmol: from dynamical masses and the molecular KS law.
Both techniques require the galaxy size in the calculations, so we required a half-light radius
 measurement in our selection criteria.
We operate under the assumption that the effective radii of rest-optical continuum emission,
 H$\alpha$ emission, and molecular gas are similar for $z\sim2$ galaxies,
 as has been found at $z\sim1-2$ by comparing resolved H$\alpha$ and CO sizes to $HST$ imaging \citep{tac13,for19}.
We describe the calculations of \mmol\ using each method below.

\subsubsection{Dynamical mass method}

The dynamical mass \mdyn\ provides a measure of the total mass in a galaxy, including stars, gas, and dark matter.
We assume that $z\sim2$ galaxies are dominated by baryons within the effective radius such that dark matter
 is negligible \citep{gen17,gen20,pri21},
 and that the total gas mass in $z>1$ galaxy disks is dominated by the molecular component \citep{tac18}.
Under these assumptions, \mmol\ can be inferred from \mdyn\ using
\begin{equation}\label{eq:mdynmmol}
M_{\text{mol}} \approx M_{\text{dyn}} - M_*
\end{equation}
In the \aco\ sample, 13 objects have \mdyn\ constraints based on measured rotation, either from
 H$\alpha$ velocity mapping \citep[7 galaxies;][]{for06,for18}, CO velocity mapping \citep[1 galaxy;][]{swi11},
 or forward-modeling of tilted emission lines in slit spectra \citep[5 galaxies;][]{pri16,pri20}.

For targets without measured rotation, \mdyn\ can be estimated via the virial mass equation
\begin{equation}\label{eq:mdyn}
M_{\text{dyn}} = \frac{k R_{\text{eff}}\sigma_v^2}{G}
\end{equation}
where $G$ is the gravitational constant, $\sigma_v$ is the velocity dispersion measured from emission line widths
 (corrected for instrumental resolution), $k$ is the virial coefficient, and $R_{\text{eff}}$ is the half-light
 elliptical semimajor axis.
The value of the virial coefficient depends on the mass and velocity distribution of stars and gas in the source.
Rather than assuming a value for $k$ from the literature, we calibrate $k$ such that 
 dispersion-based \mdyn\ estimated using equation~\ref{eq:mdyn} matches rotation-based \mdyn\ on average for the
 13 targets with measured rotation.
This process yields a best-fit value of $k=8.6$, very close to the virial coefficient for an exponential profile
 (S{\'e}rsic index $n=1$) from \citet{cap06}.
Our best-fit virial coefficient is somewhat higher than has been used for disks and spheroids in past work
 \citep[$k\approx3-6$; e.g.,][]{pet01,cap06,erb06c,pri16},
 though this difference is at least partially explained by the fact that we do not perform inclination corrections
 such that our best-fit $k$ encompasses both the virial coefficient and an average inclincation correction factor.
Excluding the 13 objects with measured rotation, we find that 26 of the remaining galaxies have published line widths,
 for which we calculate \mdyn\ using equation~\ref{eq:mdyn}.
We thus have \mdyn\ constraints for a sample of 39 galaxies (13 from rotation, 26 from line widths),
 for which we estimate \mmol\ with equation~\ref{eq:mdynmmol} and \acodyn\ using equation~\ref{eq:aco}.
Seven objects have \mstar$>$\mdyn\ such that the inferred \mmol\ and \acodyn\ are unphysically negative,
 which we attribute to the large systematic scatter in deriving \mdyn\ from spatially unresolved measurements.
For these sources, we instead calculate 3$\sigma$ upper limits on \mmol\ and \acodyn.

The molecular gas mass can also be estimated by applying the molecular KS relation between
 the surface densities of SFR (\sigsfr) and molecular gas mass (\sigmol) \citep[e.g.,][]{ken98,ken21,tac13,rey19}.
We calculated \sigsfr\ according to
\begin{equation}
\Sigma_{\text{SFR}}=\frac{\text{SFR}}{2\pi r_{\text{eff}}^2}
\end{equation}
converted \sigsfr\ to \sigmol\ based on the adopted KS law
\begin{equation}
\Sigma_{\text{SFR}}=c \Sigma_{\text{mol}}^n
\end{equation}
 with \sigsfr\ in units of $\msun$~yr$^{-1}$~kpc$^{-2}$ and \sigmol\ in units of $\msun$~pc$^{-2}$,
 and finally derived \mmol\ using
\begin{equation}
M_{\text{mol}}=2\pi r_{\text{eff}}^2 \Sigma_{\text{mol}}
\end{equation}
where the same value of $r_{\text{eff}}$ is used in the first and last steps.
We adopt the molecular KS law of \citet{tac13} adjusted for the different $r_{31}$ assumed in that work,
 with $\log(c)=-2.96$ and $n=1.05$.
This \mmol\ estimate is then applied in equation~\ref{eq:aco} to infer \acoks.

\subsubsection{The \aco$-$O/H relation at $z\sim2$}

\begin{figure*}
 \includegraphics[width=\columnwidth]{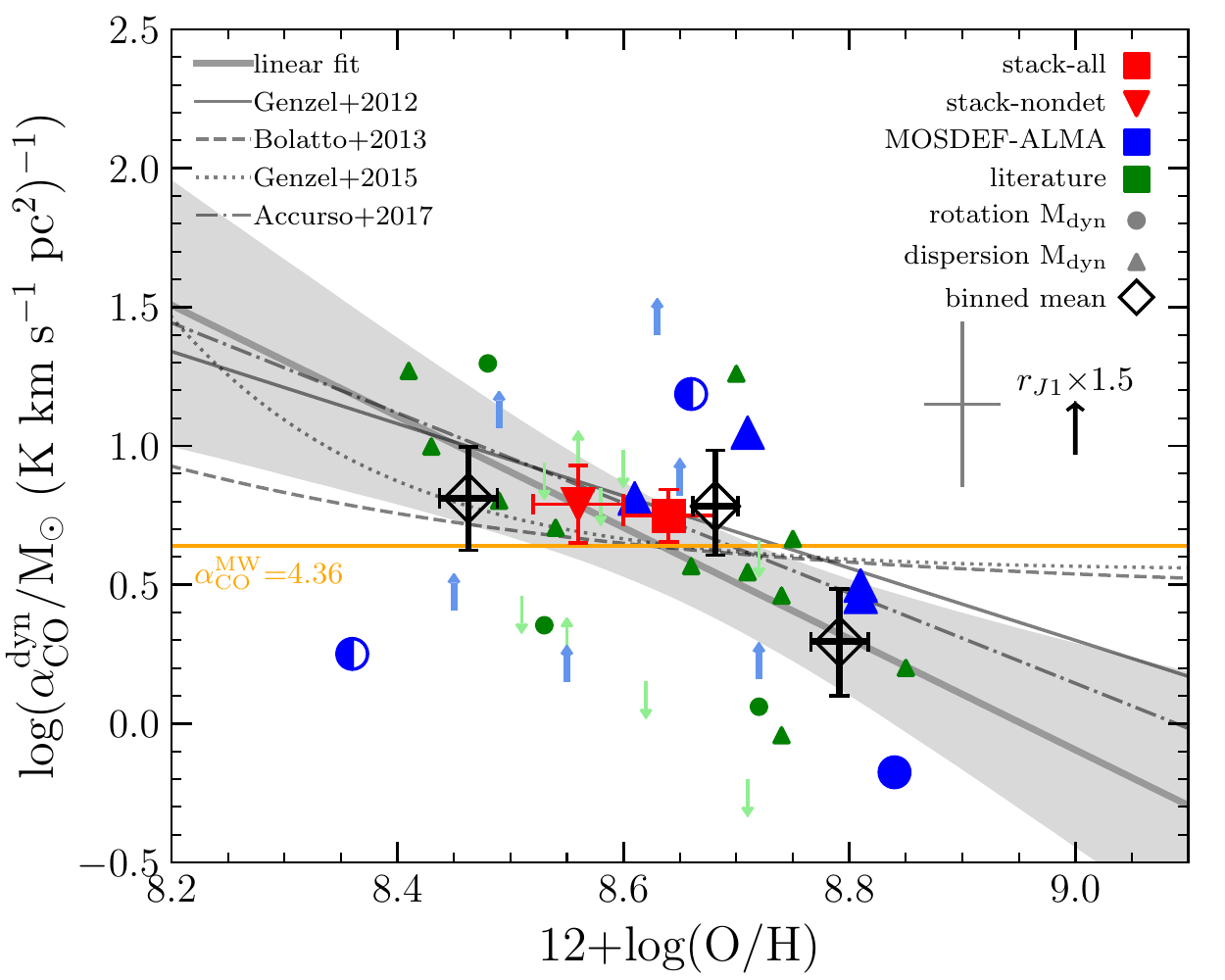}
 \includegraphics[width=\columnwidth]{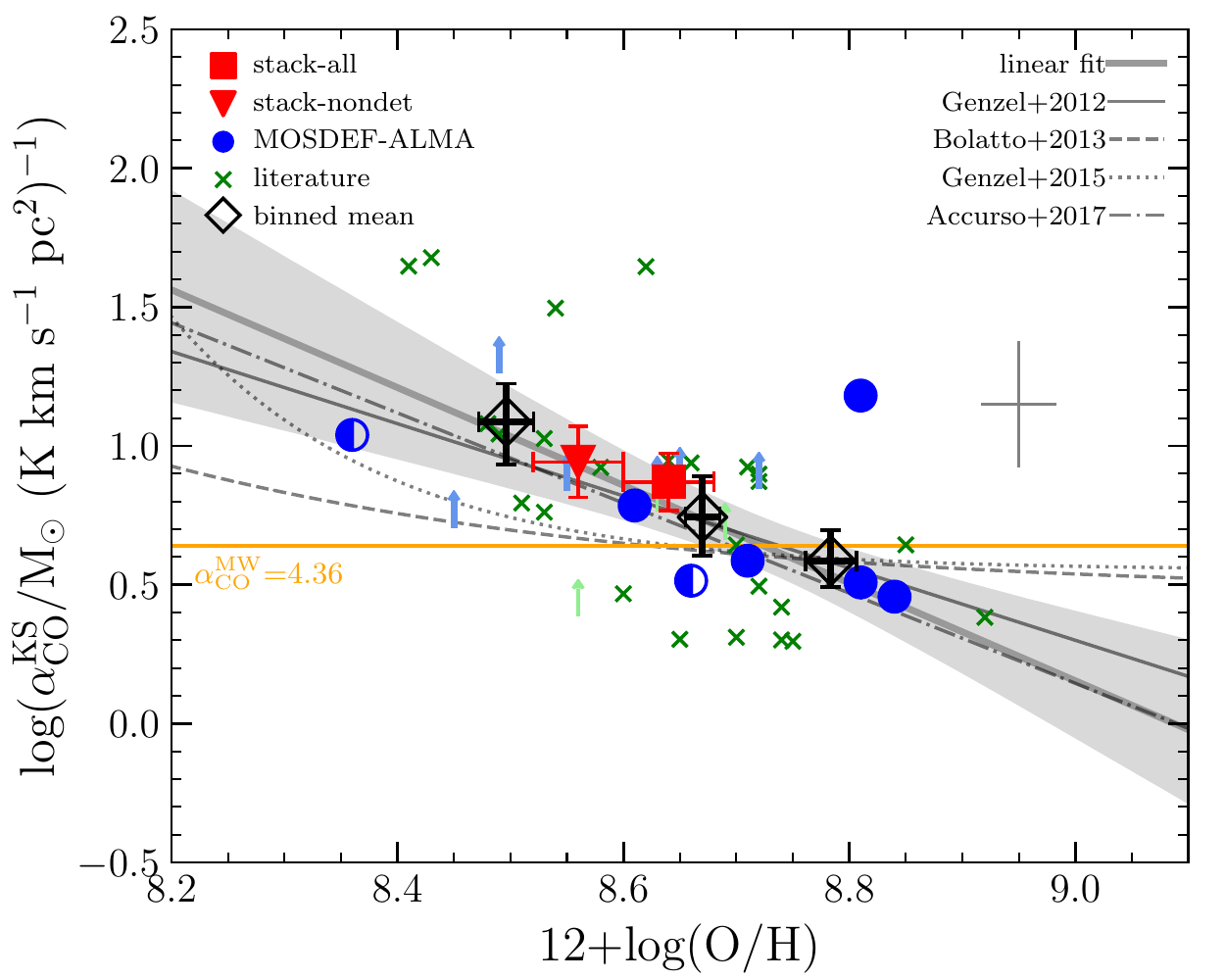}
 \centering
 \caption{
The CO-to-\htwo\ conversion factor, \aco, vs.\ gas-phase oxygen abundance.
In the calculation of \aco, \mmol\ has been inferred from the difference between dynamical and stellar masses
 (left), or the molecular KS relation of \citet{tac13} (right).
In the left panel, galaxies with \mdyn\ derived from modeling observed rotation are displayed as circles,
 while triangles denote those for which \mdyn\ is estimated from rest-optical line widths and radii.
In both panels, black open diamonds present mean \aco\ values in three bins of O/H, with an equal number of
 sources in each bin.
The gray error bars on the right side of each plot display the mean statistical uncertainty of the individual galaxies.
The black arrow in the left panel demonstrates how the inferred \aco\ values would change if the excitation correction
 factor $r_{J1}$ was larger by 50\%.
Thin gray lines display \aco-O/H relations from \citet[solid]{gen12}, \citet[dashed]{bol13}, \citet[dotted]{gen15},
 and \citet[dash-dotted]{acc17}.
The orange line indicates the Milky Way \aco\ value.
The thick gray line shows the best-fit relation (equations~\ref{eq:acodynoh} and~\ref{eq:acoksoh}), with the
 1$\sigma$ uncertainty bound shown in gray shading.
}\label{fig:aco}
\end{figure*}

Figure~\ref{fig:aco} shows \aco\ as a function of metallicity using the dynamical mass method (left panel)
 and the KS law method (right panel).
The black diamonds show mean values of the individual objects (excluding limits) in 3 bins of O/H.
With both methods, we find a trend of decreasing \aco\ with increasing O/H.
In each panel, only one source at 12+log(O/H$)>8.7$ has inferred $\alpha_{\text{CO}}>10$, while several galaxies
 exceed this value at lower metallicities.
Likewise, \aco\ lower than the Milky Way value of 4.36 is common at 12+log(O/H$)>8.6$, while at lower metallicities
 there are only two galaxies that have \acodyn$<$4.36 and none with \acoks\ below the Galactic value.
A Spearman correlation test indicates $\rho_S=-0.55$ and $p\text{-value}=0.0098$ for \acodyn,
 and $\rho_S=-0.62$ and $p\text{-value}=1.1\times10^{-4}$ for \acoks, rejecting the null hypothesis of no correlation
 at 2.6$\sigma$ and $3.3\sigma$, respectively.

The best-fit relations to the individual galaxies (excluding limits; thick gray lines in Fig.~\ref{fig:aco}) are:
\begin{equation}\label{eq:acodynoh}
\log(\alpha_{\text{CO}}^{\text{dyn}}) = -2.00(\pm0.97)\times x + 0.51(\pm0.16)
\end{equation}
\begin{equation}\label{eq:acoksoh}
\log(\alpha_{\text{CO}}^{\text{KS}}) = -1.76(\pm0.70)\times x + 0.68(\pm0.09)
\end{equation}
where $x=12+\log(\mbox{O/H})-8.7$ and \aco\ is in units of \acounits.
The MOSDEF-ALMA stacks are consistent with the best-fit relations, suggesting that the exclusion of limits
 has not introduced a significant bias.
These best-fit relations are fully consistent with one another despite the different methods employed to
 estimate \aco.
They are also generally consistent with existing \aco$-$O/H calibrations \citep{gen12,gen15,bol13,acc17},
 though the $z\sim2$ data favor a steeper slope than that of \citet{bol13}
 and are in particularly good agreement with the \citet{acc17} relation, which has a slope of $-1.623$ and
 a normalization at 12+log(O/H$)=8.7$ of $0.63\pm0.165$.
Our fits also broadly agree with \aco(O/H) at $z=2$ derived from the SIMBA cosmological simulations \citep{dav20}.

A number of systematic uncertainties may affect the position of the $z\sim2$ sample in the \aco$-$O/H plane.
First, the metallicity calibration can shift galaxies horizontally.
We have employed the high-redshift analog calibrations of
 \citet[see Appendix~\ref{app:metallicity}, Table~\ref{tab:metallicity}]{bia18}.
If we instead use the $z\sim0$ calibrations of \citet{pet04} or \citet{cur17,cur20b}, the metallicities
 decrease by $\approx0.15$~dex on average.
Given that the majority of the sample has \cott\ measurements, the value of the $r_{31}$ excitation correction factor,
 which is not well constrained at high redshift, can shift galaxies vertically.
If we instead assume $r_{31}$ that is 1.5 times larger (i.e., $r_{31}=0.84$ as
 found by \citealt{rie20} for five $\sim10^{11}~\msun$ galaxies at $z\approx2.6$), the inferred \aco\ would decrease by $0.18$~dex.
Finally, the assumed molecular KS law and virial coefficient affect \acoks\ and \acodyn, respectively.

While these systematic uncertainties are a significant concern for the absolute normalization of the \aco$-$O/H relation,
 they each shift the majority of galaxies in our sample in the same direction
 unless there is a strong systematic dependence of $r_{31}$ or the virial coefficient on metallicity.
It is plausible that $r_{31}$ increases with decreasing metallicity since the combination of reduced dust shielding and
 harder and more intense UV radiation fields at low metallicity is expected to more highly excite CO.
In this scenario, the true \aco(O/H) slope would be steeper than our derivation that assumes a constant $r_{31}$,
 though more high-redshift galaxies with detections of both \cott\ and \cooz\ are needed to robustly assess
 the impact of excitation \citep{boo20,rie20}.
With the currently available constraints, 
 we conclude that \aco\ depends on O/H at $z\sim2$ with similar dependence to that observed at $z\sim0$.
It is crucial to take this metallicity dependence into account to derive accurate gas masses for subsolar metallicity
 galaxies that are common in the high-redshift universe.

\subsection{The Molecular Gas Content of $z\sim2$ Star-Forming Galaxies}\label{sec:mgas}

In this section, we derive molecular gas masses for the MOSDEF-ALMA and $z=2-3$ literature samples,
 and explore relationships between molecular gas content, stellar mass, SFR, and metallicity.
Based on the good agreement with the $z\sim2$ galaxies in Fig.~\ref{fig:aco},
 we adopt the \citet{acc17} relation between \aco\ and metallicity\footnote{The \citet{acc17} \aco\ parameterization
 is a function of offset in SFR from the MS in addition to metallicity.
 However, the SFR dependence is very weak
 such that even a large offset of 1~dex from the MS changes \aco\ by only 0.06~dex(15\%).
 We ignore the SFR term and evaluate the \citet{acc17} relation on the MS such that \aco\ is only a function of O/H.}
 with a floor at the Milky Way value of $\alpha_{\text{CO}}^{\text{MW}}=4.36$~\acounits\ such that super-solar metallicity
 galaxies do not have significantly smaller values of \aco:
\begin{equation}\label{eq:acc17}
log(\alpha_{\text{CO}}) = 
\begin{cases}
14.752 - 1.623x\,, & \text{if } x \le 8.7\\
\log(4.36)\,, & \text{if } x > 8.7
\end{cases}
\end{equation}
where $x=12+\log(\text{O/H})$.
Molecular gas masses are calculated assuming this \aco(O/H) relation and $r_{31}=0.55$,
 where the latter is adopted for consistency with \citet{tac18}.
For the 7 literature sources without spectroscopic metallicity constraints,
 O/H from the MZR($M_*,z$) of equation~\ref{eq:fitmzrz} is used for \mmol\ calculations,
 and these sources are excluded from any plots and analyses that directly include metallicity.
The derived molecular gas properties of the MOSDEF-ALMA sample and stacks are presented in Table~\ref{tab:mosdefalma},
 while those of the $z=2-3$ literature sample are given in Table~\ref{tab:litco}.

\subsubsection{Molecular gas content and stellar mass}

\begin{figure}
 \includegraphics[width=\columnwidth]{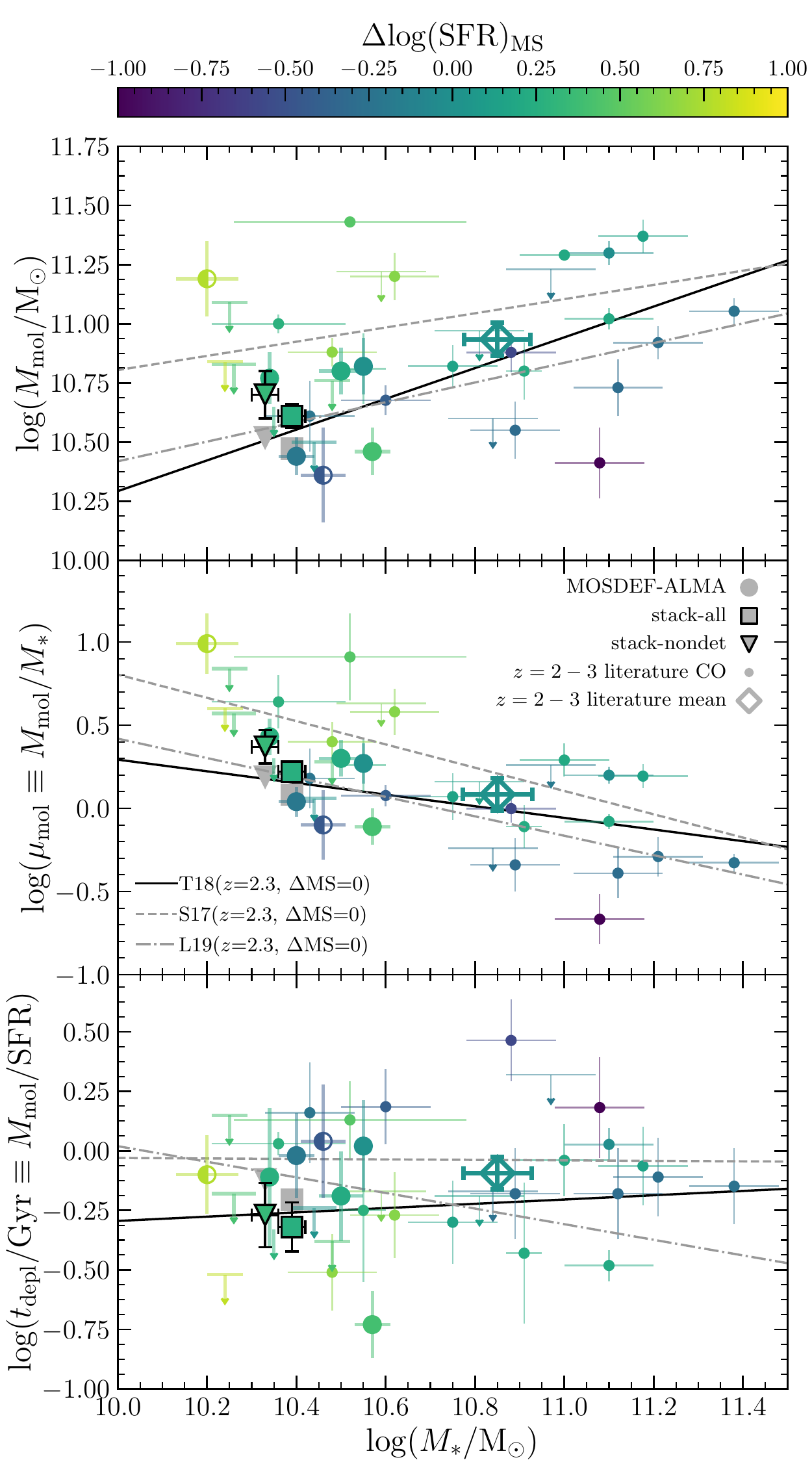}
 \centering
 \caption{
Molecular gas mass (top), gas fraction (middle), and depletion time (bottom) as a function of stellar mass.
Points are color-coded according to offset from the star-forming main sequence of \citet{spe14}.
The black solid line in each panel shows the scaling relations of \citet{tac18} evaluated at $z=2.3$
 and on the main-sequence.
Gray points show the position of the composite spectra (stack-all and stack-nondet) after correcting for
 their offset from the MS according to the $\Delta$SFR$_{\mathrm{MS}}$ dependence of
 the \citet{tac18} relations.
The gray dashed and dot-dashed lines display the scaling relations of \citet{sco17} and \citet{liu19},
 respectively, evaluated at $z=2.3$ for targets on the main sequence.
}\label{fig:mgas}
\end{figure}

The molecular gas masses, \mmol, and molecular gas fractions, \mumol=\mmol/\mstar,
 are shown as a function of \mstar\ in the
 top and middle panels of Figure~\ref{fig:mgas}, with points color-coded by offset from the star-forming main sequence.
On average, the MOSDEF-ALMA sample has lower \mmol\ and higher \mumol\ than the more-massive literature sample,
 indicating a positive correlation between \mmol\ and \mstar\ and an anti-correlation between \mumol\ and \mstar.
\mmol\ is typically larger than \mstar\ below $10^{10.8}~\msun$, with an average \mumol$\approx$2 at $10^{10.4}~\msun$.
Consequently, gas mass is expected to dominate the baryonic mass of $z\sim2$ galaxies at $\lesssim10^{10.4}~\msun$.

There is a significant additional dependence on SFR such that galaxies with higher SFR at fixed
 \mstar\ have larger \mmol, as has been observed in previous works \citep[e.g.,][]{tac13,tac18,tac20,sco17,liu19}.
The black line shows the \citet{tac18} scaling relation\footnote{While the scaling relation reported in \citet{tac18} is for \mumol, this function can be converted to \mmol\ by muliplying by \mstar.} evaluated at $z=2.3$ and \delsfms=0.
While the sample average of the literature sources falls on this relation, the MOSDEF-ALMA stacks fall
 slightly above it because the sample lies 0.25~dex above the MS on average.
If we correct \mmol\ of the MOSDEF-ALMA stacks assuming the SFR dependence from \citet{tac18}
 ($M_{\text{mol}}\propto \Delta\text{SFR}_{\text{MS}}^{0.53}$; see also Fig.~\ref{fig:delsfms}, left),
 then the composites fall on the \citet{tac18} main-sequence relation.
We thus find that the \citet{tac18} scaling relation reliably reproduces the average \mmol\ and \mumol\ of
 $z\sim2$ star-forming galaxies down to at least $\sim10^{10.2}~\msun$.

The bottom panel of Figure~\ref{fig:mgas} presents the molecular gas depletion timescale, \tdepl=\mmol/SFR.
We do not find evidence for any significant dependence of \tdepl\ on \mstar\ across log($M_*/\msun)=10.2-11.2$.
In the sample averages, we find slightly lower \tdepl\ at lower \mstar, though the MOSDEF-ALMA and literature
 mean values are consistent with one another at the $\approx$1.5$\sigma$ level.
A dependence on offset from the MS is apparent, with higher-SFR galaxies at fixed \mstar\ having shorter \tdepl.
The gray points once again show the MOSDEF-ALMA stacks corrected for their offset from the MS
 according to the dependence of the \citet{tac18} scaling relation ($t_{\text{depl}}\propto\Delta\text{SFR}_{\text{MS}}^{-0.44}$).
The SFR-corrected stack of all MOSDEF-ALMA targets has log(\tdepl/Gyr$)=-0.20\pm0.10$ at $10^{10.4}~\msun$, while
 the $z=2-3$ literature mean is log(\tdepl/Gyr$)=-0.09\pm0.07$ at $10^{10.85}~\msun$.
We thus find that the molecular gas depletion timescale of $z\sim2$ main-sequence star-forming galaxies
 at log($M_*/\msun)\sim10-11.5$ is $t_{\text{depl}}=600-800$~Myr with no significant \mstar\ dependence,
 shorter by a factor of $\sim2$ than that of main-sequence $z\sim0$ galaxies at similar \mstar\ with
 $t_{\text{depl}}\approx1.4$~Gyr \citep{sai16,sai17}.

\subsubsection{The $z\sim2$ molecular KS law}

\begin{figure}
 \includegraphics[width=\columnwidth]{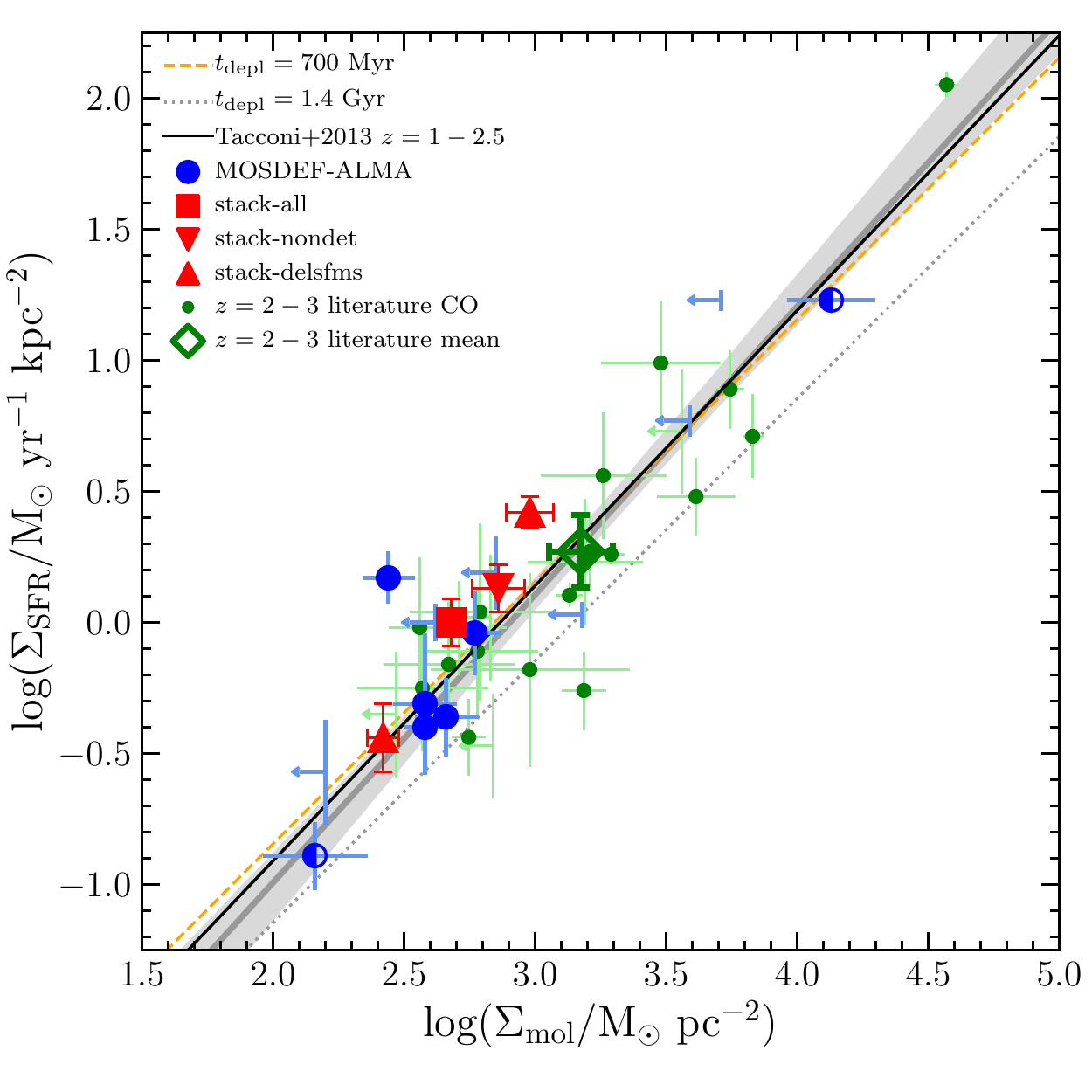}
 \centering
 \caption{
The molecular Kennicutt-Schmidt law relating \sigsfr\ and \sigmol.
Upward red triangles denote composite spectra in two bins of \sigsfr.
The best linear fit (gray line) is very similar to the $z>1$ KS law from \citet{tac13}
 and to a line of constant depletion time \tdepl=700~Myr (orange dashed line), but is
 offset from the \citet{tac13} fit to $z=0$ data (dotted line).
}\label{fig:ks}
\end{figure}

Figure~\ref{fig:ks} shows \sigsfr\ vs.\ \sigmol\ for $z=2-3$ galaxies.\footnote{Our adopted \aco(O/H) relation from \citet{acc17} was not derived using the KS relation.  These authors employed radiative transfer modeling of [C\ii] and \cooz\ to infer \aco.  As such, this \aco(O/H) relation does not imprint a certain KS relation by construction.}
We find a tight correlation between these quantities that is well fit by a linear relation
 with constant $t_{\text{depl}}=700$~Myr (orange dashed line), in agreement with the
 findings of \citet{tac13} based on the PHIBSS $z=1.0-1.5$ and $z=2.0-2.5$ samples, the latter of
 which is included in our literature sample.
The best-fit relation from fitting the individual galaxies (excluding limits) is:
\begin{equation}\label{eq:ks}
\begin{multlined}
\log\left(\frac{\Sigma_{\text{SFR}}}{\msun \text{yr}^{-1} \text{kpc}^{-2}}\right) = \\
1.09(\pm0.09)\left[\log\left(\frac{\Sigma_{\text{mol}}}{\msun \text{pc}^{-2}}\right) - 3\right] + 0.10(\pm0.05)
\end{multlined}
\end{equation}
Our best-fit relation is fully consistent with the $z=1-2.5$ relation from PHIBSS \citep{tac13}.
The best-fit slope at $z=2-3$ is consistent with the molecular KS law in the local universe that is generally found to
 be near-linear with either integrated or spatially-resolved observations \citep[e.g.,][]{ken98,big08,ler08,rey19,ken21},
 although the normalization is offset from what is appropriate for typical integrated $z\sim0$ main-sequence galaxies with
 $t_{\text{depl}}\approx1-1.5$~Gyr (gray dotted line).

We find that the $z\sim2$ molecular KS law is very tight over nearly 3 orders of magnitude in \sigmol\ and \sigsfr,
 with an intrisic scatter
 of 0.07~dex (18\%) in \sigsfr\ around the best-fit line after accounting for measurement uncertainties.
The MOSDEF-ALMA results demonstrate that $z\sim2$ galaxies down to $10^{10.2}~\msun$ fall on a well-defined KS relation,
 implying that equation~\ref{eq:ks} can be used to accurately infer \mmol\ (and \mdyn, assuming negligible dark matter) for existing large samples of $z\sim2$ galaxies
 with SFR and size measurements in deep extragalactic legacy fields.
This implication is especially beneficial for existing large rest-optical spectroscopic surveys,
 which have mean sample stellar masses of $\approx10^{10}~\msun$ (e.g., MOSDEF, \citealt{kri15}; KBSS-MOSFIRE, \citealt{ste14}).

\subsubsection{Correlated residuals around mean scaling relations}\label{sec:residuals}

We now search for correlations among the residuals around three mean scaling relations:
 the SFR-\mstar\ relation (i.e., star-forming main sequence; MS), the mass-metallicity relation (MZR), and the
 \mmol-\mstar\ relation.
Models of galaxy growth governed by a self-regulating baryon cycle predict that the scatter around these
 scaling relations will be (anti-)correlated \citep[e.g.,][]{lil13,dav17,der17,tor19}.
We adopt parameterizations of each of these mean relations as a function of \mstar\ and redshift:
 SFR-\mstar\ from \citet{spe14}, MZR from equation~\ref{eq:fitmzrz},
 and \mmol-\mstar\ of \citet{tac18} evaluated on the MS ($\Delta\text{SFR}_{\text{MS}}=0$).
The residuals around each of these relations are derived by substracting the logarithmic scaling relation value
 (at matched \mstar\ and $z$) from the measured value for each object.
We adopt the following terms for these residuals: \delsfms, \delmzr, and \delmmol.

\begin{figure*}
 \includegraphics[width=\columnwidth]{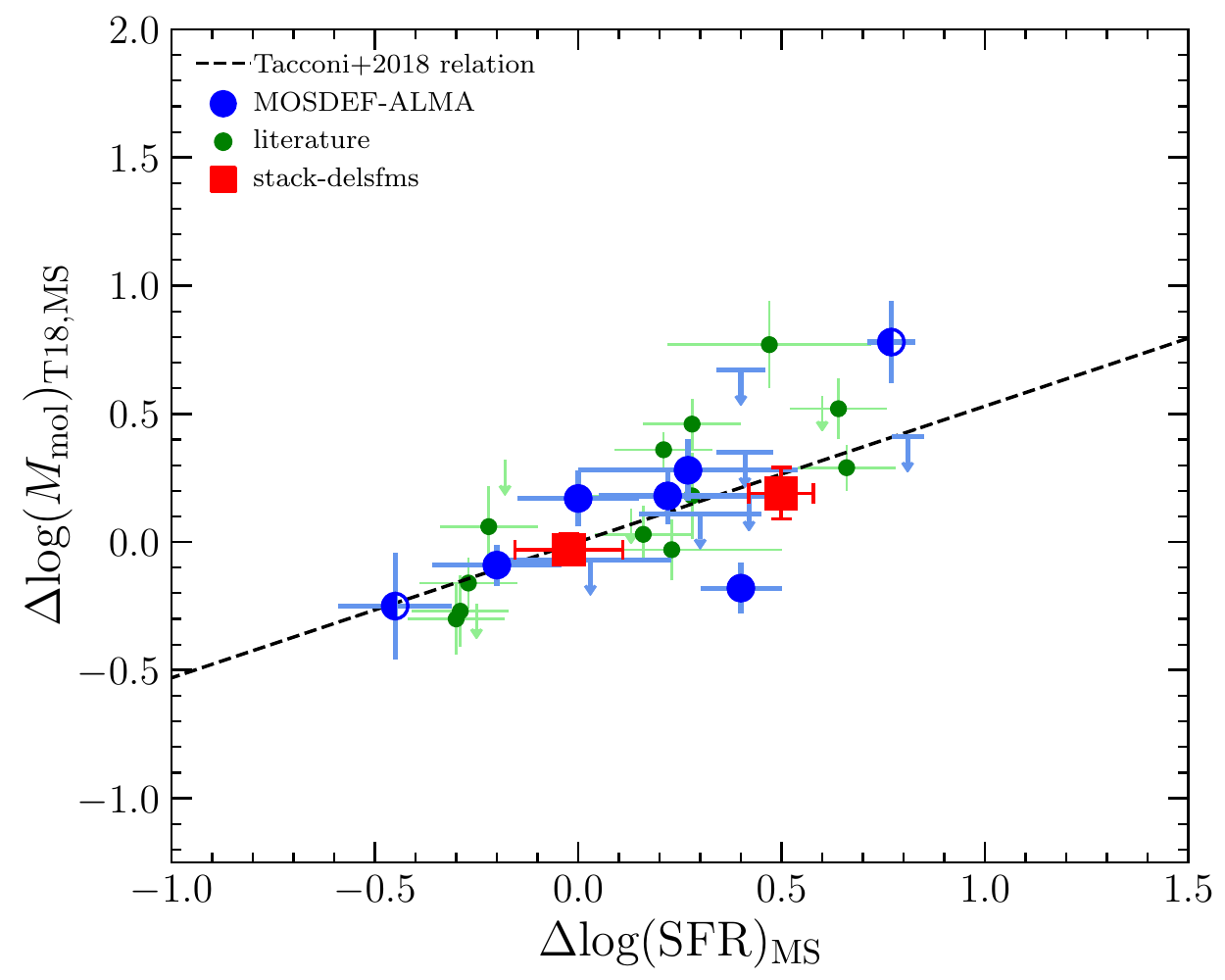}
 \includegraphics[width=\columnwidth]{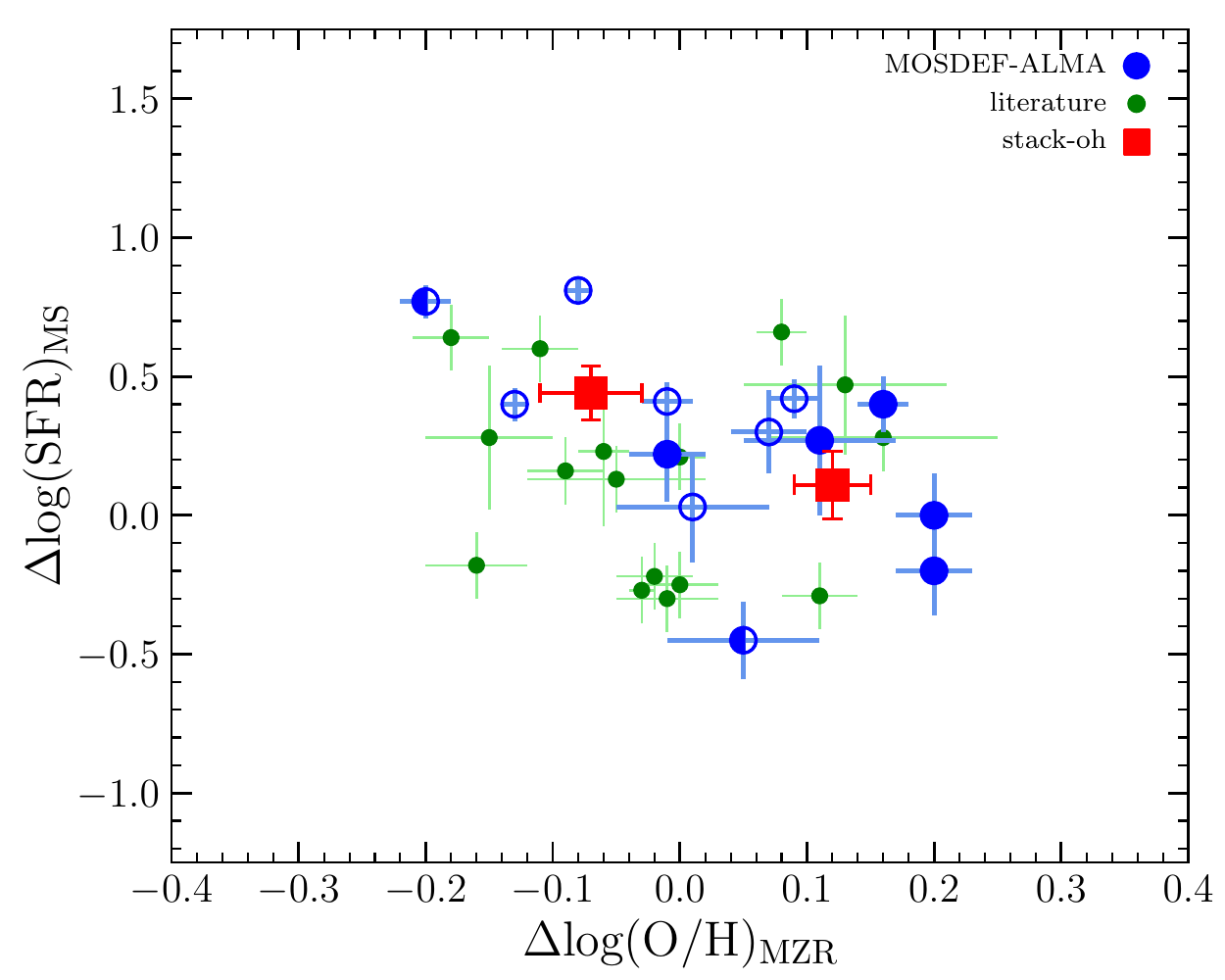}
 \centering
 \caption{
Residual plots comparing offsets from mean scaling relations at fixed \mstar.
\textit{Left:} Offset from the \citet{tac18} \mmol\ scaling relation evaluted at $\Delta$SFR$_{\mathrm{MS}}=0$
 vs.\ offset from the star-forming main sequence.
The dashed black line shows the best-fit form of this correlation from \citet{tac18}:
 $\Delta\mbox{M}_{\mathrm{mol}}\propto\Delta\mbox{SFR}^{0.53}$.
\textit{Right:} Offset from the star-forming main sequence vs.\ offset from the mass-metallicity relation.
In each panel, red squares denote MOSDEF-ALMA composites in two bins of the $x$-axis variable.
}\label{fig:delsfms}
\end{figure*}

The left panel of Figure~\ref{fig:delsfms} presents \delmmol\ vs.\ \delsfms\ for the MOSDEF-ALMA and $z=2-3$
 literature samples.
This comparison is a different way of viewing the \delsfms\ dependence of \mmol\ at fixed \mstar\
 shown in the top panel of Figure~\ref{fig:mgas}.
We find a positive correlation between these quantities, indicating that galaxies with higher SFR have
 larger \mgas\ at fixed \mstar.
The correlation is highly significant: a Spearman correlation test yields $\rho_S=0.79$ with a $p$-value=$5.6\times10^{-5}$.
The trends of stacked MOSDEF-ALMA data in two bins of \delsfms\ (red squares) and the individual objects are
 consistent with the SFR dependence of the \citet{tac18} scaling relation ($M_{\text{mol}}\propto \text{SFR}^{0.53}$ at fixed \mstar)
 and the existence of a tight KS relation at $z\sim2$.

The relation between \delsfms\ and \delmzr\ is shown in the right panel of Figure~\ref{fig:delsfms}.
Focusing on the MOSDEF-ALMA sample, we find evidence of an anti-correlation between these quantities
 for the individual galaxies, where galaxies above the MS fall below the MZR.
This trend is more clearly evident in the MOSDEF-ALMA composites in two bins of \delmzr\ (red squares).
The observed anti-correlation is a manifestation of the SFR-FMR
 in which galaxies at fixed \mstar\ with higher SFR have lower O/H.
The inverse correlation of \delmzr\ on \delsfms\ has been observed with much higher significance using both
 individual galaxies and stacked spectra of a considerably larger ($N\sim250$) MOSDEF $z\sim2.3$ star-forming sample
 \citep{san18,san21}.
Here, we confirm the existence of a SFR-FMR among the 13 MOSDEF-ALMA targets.

There is not a significant correlation present for the literature CO sample in the right panel of Figure~\ref{fig:delsfms}.
This apparent lack of a SFR-FMR may be due to the small sample size or the impact of systematic uncertainties on O/H,
 or could be the true physical scenario for this sample.
The small sample size of only 16 galaxies can make it difficult to recover the secondary dependence of O/H on SFR
 in addition to the primary dependence on \mstar, considering that there is both intrinsic scatter in the SFR-FMR
 and several literature sources have fairly large statistical uncertainties on O/H.
The metallicities of the literature sample are derived from a heterogeneous set of strong-line ratios:
 9 based on N2, 6 from O3N2, and 1 from O3O2Ne3.
While we have made efforts to control systematics between O/H derived from each of these indicators
 (see Appendix~\ref{app:metallicity}), metallicites based on N2 (and O3N2 to a lesser extent)
 are strongly affected by object-to-object variations in N/O
 that are commonly a factor of 2 at fixed O/H \citep[][Fig.~\ref{fig:b18oh}]{pil11,str17}.
In contrast, the MOSDEF-ALMA sample is more uniform in metallicity indicator, with 11 based on O3O2Ne3
 (unaffected by N/O) and 2 derived using O3N2 (less strongly affected by N/O than N2).

It is also possible that there truly is no dependence of O/H on SFR at fixed \mstar\ in the literature sample.
In the local universe, the SFR-dependence in the SFR-FMR significantly weakens and may disappear entirely or even
 invert at high masses ($>10^{10.5}~\msun$; e.g., \citealt{man10,yat12}).
That the massive literature sample (mean $M_*=10^{10.85}~\msun$) displays no SFR dependence while the
 less-massive MOSDEF-ALMA sample (mean $M_*=10^{10.4}~\msun$) does could thus indicate a behavior similar
 to that of the $z\sim0$ SFR-FMR at high masses.
Uniform full-coverage of the rest-optical lines for a larger sample of such massive $z\sim2$
 star-forming galaxies is required to discern whether the lack of a correlation is the result of systematic effects
 or is real.

\begin{figure}
 \includegraphics[width=\columnwidth]{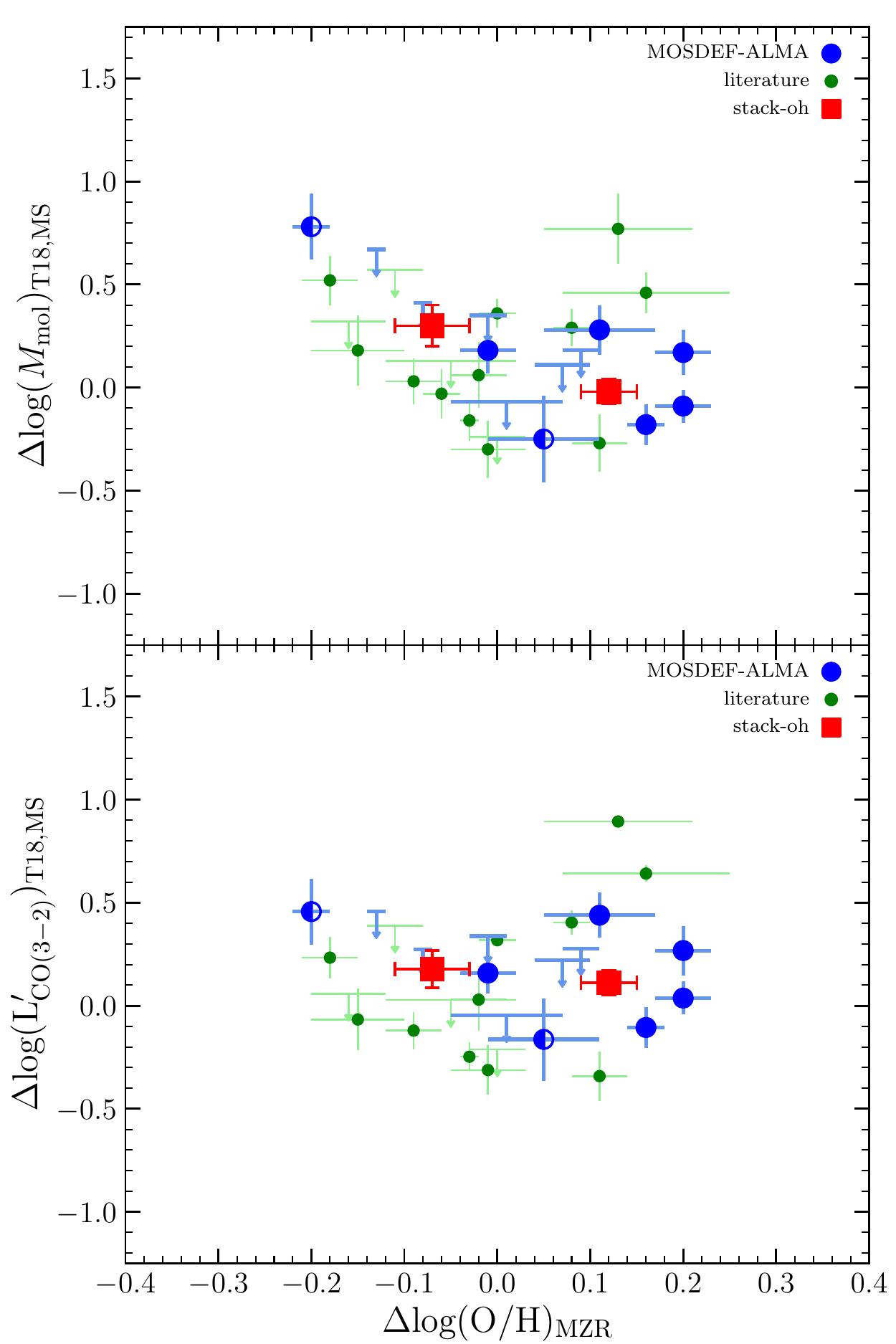}
 \centering
 \caption{
\textit{Top:} Offset from the \citet{tac18} \mmol\ scaling relation evaluted at $\Delta$SFR$_{\mathrm{MS}}=0$
 vs.\ offset from the mass-metallicity relation.
\textit{Bottom:} Offset from the mean \lcott-\mstar\ relation vs.\ offset from the mass-metallicity relation.
The mean \lcott\ as a function of \mstar\ and redshift is inferred as follows.
Beginning with the \citet{tac18} \mmol\ relation evaluated at $\Delta$SFR$_{\mathrm{MS}}=0$, \lcooz($M_*,z$) is
 derived by dividing \mmol($M_*,z$) by \aco(O/H),
 where O/H is taken from the MZR($M_*,z$) parameterization of equation~\ref{eq:fitmzrz}.
The resulting function is then multiplied by $r_{31}$ to arrive at \lcott($M_*,z$).
}\label{fig:delmgasdelmzr}
\end{figure}

The top panel of Figure~\ref{fig:delmgasdelmzr} displays \delmmol\ vs.\ \delmzr.
While the trend among individual MOSDEF-ALMA galaxies is unclear due to the large number of limits,
 there is a clear anti-correlation present for the two stacks in bins of \delmzr (red squares).
This anti-correlation indicates the existence of a \mstar-\mmol-O/H relation in which
 galaxies at fixed \mstar\ with higher \mmol\ have lower O/H.
This relation has a clear connection to the SFR-FMR (i.e., \mstar-SFR-O/H relation) since the SFR depends
 on the amount of molecular gas present via the molecular KS relation (Fig.~\ref{fig:ks}), such that
 the SFR-FMR likely emerges because this more fundamental Gas-FMR exists among \mstar, \mmol, and O/H.
We thus expect that if a SFR-FMR is present, a Gas-FMR will be present as well, as found in the MOSDEF-ALMA sample.
We do not find evidence for a Gas-FMR among the more-massive $z=2-3$ literature sample, consistent with
 the lack of a SFR-FMR in the right panel of Fig.~\ref{fig:delsfms}.

Interpreting the top panel of Figure~\ref{fig:delmgasdelmzr} is complicated by the fact that the derived
 \mmol\ depends on O/H through \aco(O/H), such that galaxies with decreasing O/H increases the inferred \mmol.
To understand the effect \aco(O/H) has on the \delmmol$-$\delmzr\ anti-correlation, we replace \delmmol\ with
 $\Delta$log(L$^{\prime}_{\text{CO}(3-2)}$), the difference in observed \lcott\ and the mean \lcott\ at fixed \mstar.
The mean \lcott\ as a function of \mstar\ and $z$ is inferred by dividing the \citet{tac18} \mmol($M_*,z,\Delta\text{SFR}_{\text{MS}}=0$)
 relation by \aco(O/H), where O/H is inferred from the MZR($M_*,z$) given in equation~\ref{eq:fitmzrz},
 and finally multiplying by $r_{31}$.
When evaluated at $z=2.3$, the resulting \lcott($M_*,z$) relation is similar to the best-fit
 \lcott\ vs.\ \mstar\ relation in the top panel of Figure~\ref{fig:lco32},
 such that we obtain consistent results if that relation is instead assumed.

The bottom panel of Figure~\ref{fig:delmgasdelmzr} displays
 $\Delta$log(L$^{\prime}_{\text{CO}(3-2)})_{\text{T18,MS}}$\ vs.\ \delmzr.
The distribution of MOSDEF-ALMA targets and stacks is flat, such that \lcott\ has no dependence on O/H at fixed \mstar.
Consequently, the anti-correlation between \delmmol\ and \delmzr\ is significantly affected by how \aco\ depends on
 metallicity.
If \aco\ has no dependence on O/H, then no Gas-FMR is found in the $z\sim2$ MOSDEF-ALMA sample.
However, we have presented strong evidence for an O/H-dependent \aco\ in Fig.~\ref{fig:aco}.
Furthermore, the inverse relation between \aco\ and O/H is well-established in the local universe
 \citep[e.g.,][]{bol13,acc17} and is thought to be
 driven by physics that appear to behave similarly at higher redshifts, namely a correlation between metallicity and dust
 content \citep[Popping et al., in prep.]{red10,sha20,sha21,shi22}, and an anti-correlation between metallicity and UV radiation field intensity.
We thus conclude that the evidence for an anti-correlation between \mmol\ and O/H at fixed \mstar\ in the
 MOSDEF-ALMA sample is robust, though the exact slope of this inverse relation will depend on the \aco(O/H) relation assumed.

\section{Discussion}\label{sec:discussion}

\subsection{Signatures of baryon cycling at $z\sim2$}

Our analysis of the relations among \mstar, SFR, metallicity, and gas content
 provides strong evidence for the existence of baryon cycling processes that govern the growth
 of $z\sim2$ massive galaxies ($\log(M_*/\msun)\sim10.0-11.5$).
The typical depletion times of $700$~Myr are only $\approx$25\% of the age of the universe at $z=2.3$.
Since a star-forming main sequence exists at this redshift, accretion of gas from the IGM and/or CGM
 is required to sustain star formation in these galaxies, as has been pointed out by previous studies
 \citep[e.g.,][]{tac13}.
However, the gas fractions are not large enough to explain the ISM metallicities at $z\sim2$.
For the typical $\mu_{\text{mol}}\approx1.5$ and assuming neutral atomic gas is negligible,
 a simple accreting box model with an accretion rate equal to the SFR predicts 12+log(O/H)=9.1,
 assuming the stellar O yield of core-collapse SNe from a \citet{cha03} IMF.
This value is considerably higher than the average observed 12+log(O/H$)=8.6-8.7$.
Considering the large amount of evidence for ubiquitous high-velocity outflowing gas around high-redshift
 star-forming galaxies \citep[e.g.,][]{ste10,for19,wel22},
 strong gas outflows present a probable candidate mechanism to further decrease the effective yield and drive
 model ISM metallicities lower toward observed values.

\subsubsection{Implications for outflow mass loading}\label{sec:outflows}

The mass outflow rate (\mdotout) is often parameterized in the outflow mass loading factor (\etaout=\mdotout/SFR)
 that encapsulates the efficiency with which feedback-driven galactic winds remove mass (and metals) from galaxies.
We place quantitative constraints on \etaout\ by employing the ideal gas-regulator model of \citet{lil13}.
In this formalism, the ISM metallicity \zism\ is a function of \etaout\ and the gas fraction \mugas=\mgas/\mstar:
\begin{equation}\label{eq:l13}
Z_{\text{ISM}} = Z_0 + \frac{y}{1 + \eta_{\text{out}}(1-R)^{-1} + \mu_{\text{gas}}}
\end{equation}
where $y$ is the stellar metal yield, $R$ is the fraction of stellar mass returned to the ISM through
 stellar evolutionary processes, and $Z_0$ is the metallicity of accreting gas.
We assume $Z_0\ll Z_{\text{ISM}}$ such that accreting metals are negligible.
We adopt a stellar oxygen yield of $y_O=0.033$ as a mass fraction (equivalent to 12+log(O/H)$_y=9.45$
 by number density) and $R=0.45$, both values appropriate for core-collapse SNe enrichment with a \citet{cha03} IMF \citep{vin16}.
Finally, we assume that the total gas mass is dominated by the molecular component
 in high-redshift star-forming galaxies such that \mgas$\approx$\mmol \citep{tac18},
 as found in studies of the atomic-to-molecular hydrogen fraction using
 semi-analytic models \citep{lag11,pop14,pop15}.

\begin{figure}
 \includegraphics[width=\columnwidth]{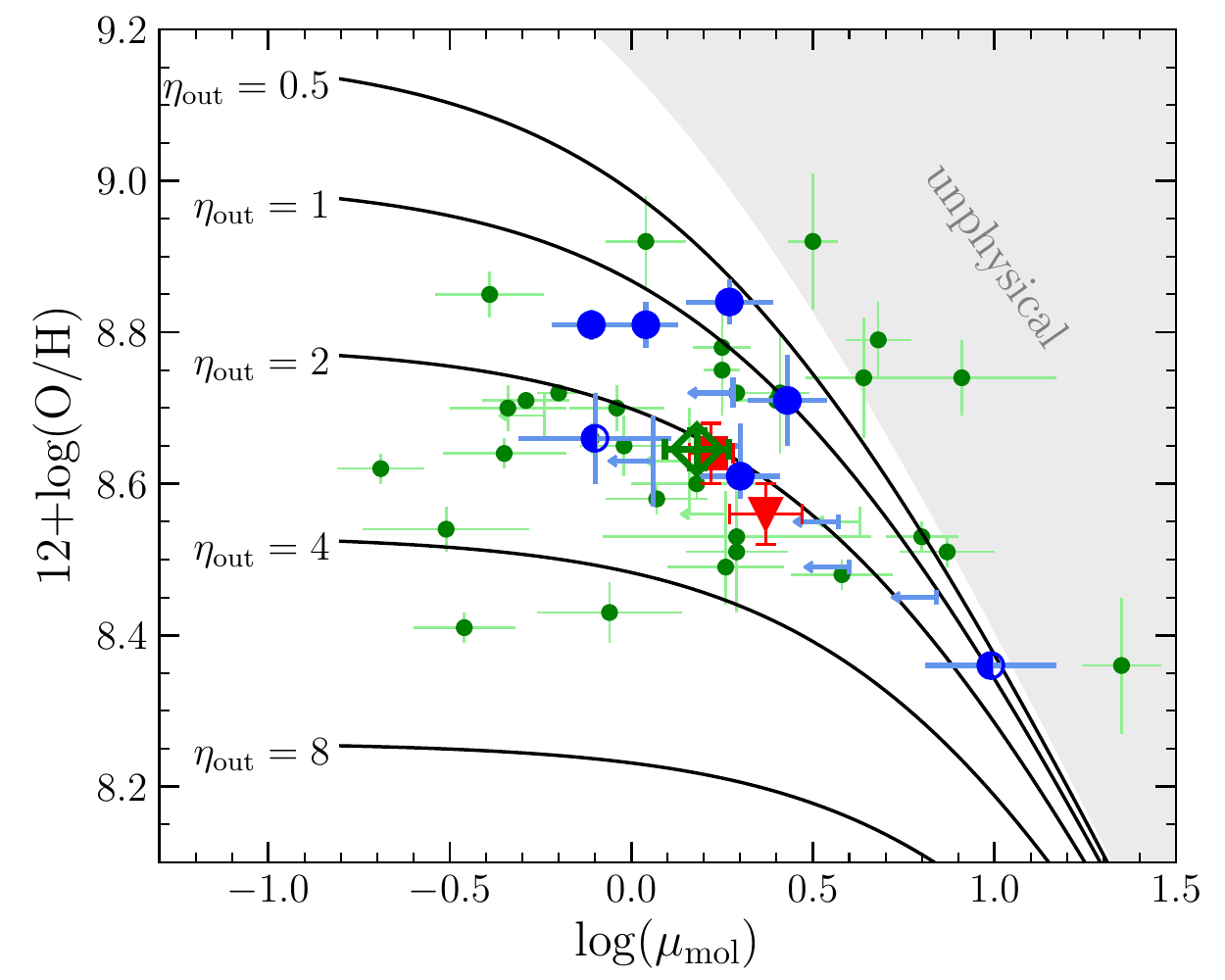}
 \includegraphics[width=\columnwidth]{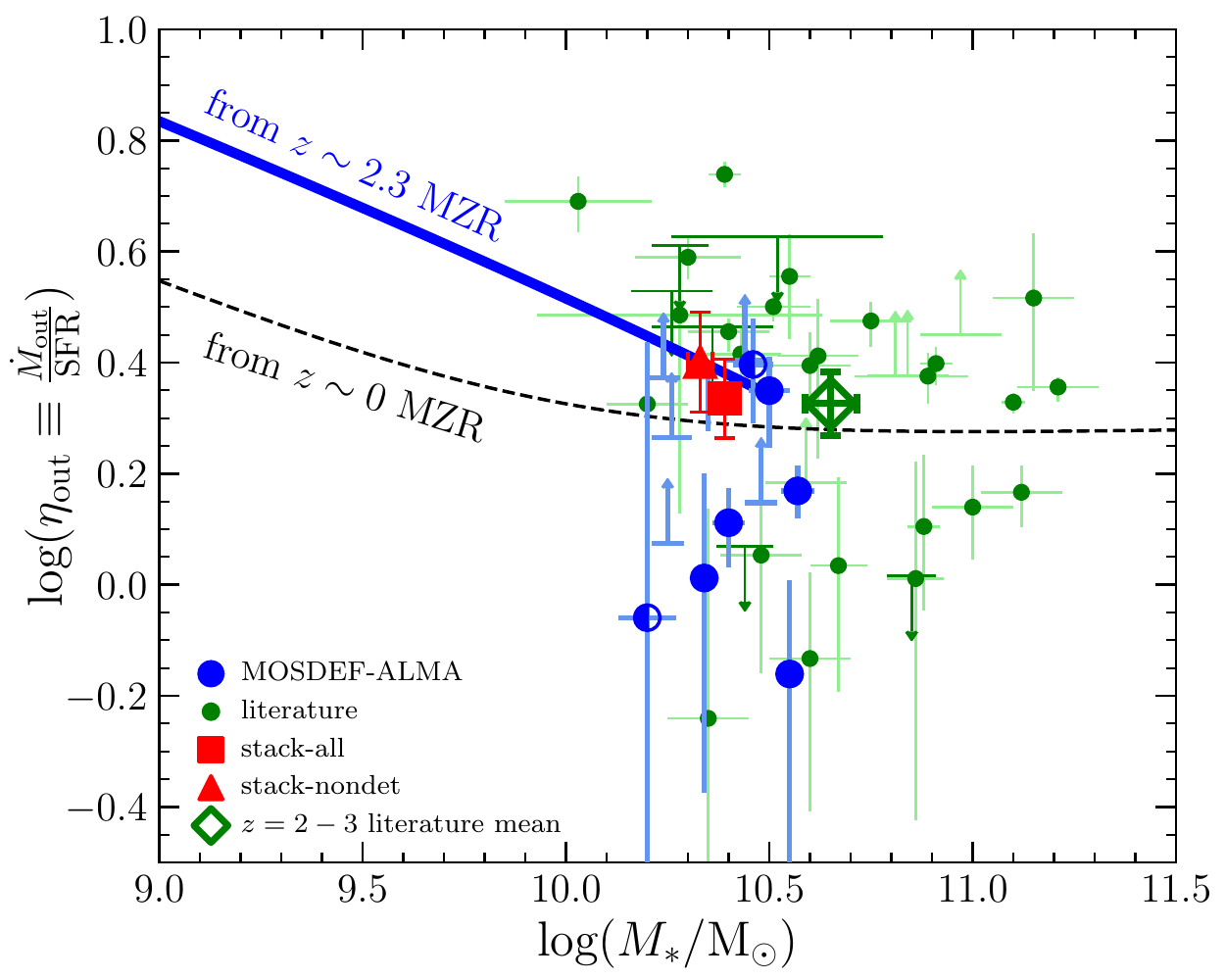}
 \centering
 \caption{
\textit{Top:} Gas-phase metallicity vs.\ molecular gas fraction.
Black lines show the gas-regulator model of \citet{lil13} evaluated at outflow mass-loading factors of
 $\eta_{\mathrm{out}}=0.5$ to~8.
The gray shaded region requires $\eta_{\mathrm{out}}<0$ and is thus unphysical.
\textit{Bottom:} \etaout\ vs.\ \mstar, where \etaout\ is inferred from the \citet{lil13} gas-regulator model.
The thick blue line shows an inference of \etaout($M_*$) at $z\sim2.3$ obtained by combining the
 best-fit $z\sim2.3$ MZR from \citet{san21} and the \citet{tac18} \mumol\ scaling relation evaluated at $z=2.3$ and
$\Delta$SFR$_{\mathrm{MS}}=0$.
The dashed black line shows \etaout($M_*$) at $z\sim0$ inferred from the combination of the local MZR of \citet{cur20b}
 and the $z\sim0$ \mugas\ scaling relation of \citet{sai16} evaluated on the MS.
}\label{fig:etaout}
\end{figure}

The top panel of Figure~\ref{fig:etaout} displays metallicity as a function of \mumol.
The black lines show gas-regulator models of equation~\ref{eq:l13} with \etaout\ varying between 0.5 and 8.
We find that the $z\sim2$ CO sample is generally described by models with \etaout\ ranging from $0.5-4$,
 with an average value of $\eta_{\text{out}}\approx2$.
Only 5 galaxies lie in the unphysical region of parameter space that would mathematically require $\eta_{\text{out}}<0$,
 though 4 of these are $\approx1\sigma$ consistent with the physical region.
The data are incompatible with significantly lower stellar yields (i.e., $y_O=0.015$ for a \citet{sal55} IMF),
 which would place a large fraction of the $z\sim2$ sample in the unphysical regime.
Such low yields could only be accommodated if $Z_0$ is comparable to \zism.

\citet{suz21} found that $z\sim3.3$ galaxies at $\sim10^{10.5}~\msun$ have metallicities significantly
 lower than $z\sim0$ and $z\sim1.6$ galaxies at fixed \mugas\ (where \mugas\ includes both molecular and atomic
 components for local galaxies).
These authors concluded that there is a sharp increase in \etaout\ at fixed \mugas\ between $z\sim1.6$ and $z\sim3$.
We find that our $z\sim2$ sample, with a mean value of 12+log(O/H$)=8.6-8.7$ at \mugas=1.6, falls near the
 distribution of the local and $z\sim1.6$ galaxies at fixed \mugas, implying no strong evolution in \etaout\ at
 fixed \mugas\ over $z=0-2.5$.
The apparent evolution in \etaout\ at fixed \mugas\ at $z>3$ found by \citet{suz21} may be due to systematics
 associated with converting dust mass to \mmol\ and the metallicity calibrations these authors adopted.
Obtaining joint CO and metallicity constraints for $z>3$ main-sequence galaxies would aid in understanding
 whether \etaout\ is different at $z>3$.

Using our measured values of O/H and \mumol, we solve equation~\ref{eq:l13} for \etaout\ and present the
 resulting values as a function of \mstar\ in the bottom panel of Figure~\ref{fig:etaout}.
The majority of the combined $z\sim2$ sample has $\eta_{\text{out}}=1-4$, with an average value of $\approx$2
 and no apparent dependence on \mstar\ over log($M_*/\msun)\sim10.2-11.2$.
The blue line displays an inference of the average \etaout\ as a function of \mstar\ for lower-mass main-sequence
 galaxies, obtained by inputting the best-fit $z\sim2.3$ MZR of \citet{san21} and the \citet{tac18} \mumol\ scaling
 relation into equation~\ref{eq:l13}, assuming the same $y_O$ and $R$ as above.
As found by \citet{san21}, \etaout\ decreases with increasing \mstar\ up to $10^{10.5}~\msun$ as
 $\eta_{out}\propto M_*^{-1/3}$, at which point the blue line aligns with the average value of the higher-mass
 CO sample.
The CO-based results suggest that \etaout\ does not continue falling with increasing \mstar\ above $10^{10.5}~\msun$, but
 instead flattens out.
A similar behavior is inferred on average for $z\sim0$ galaxies (dashed black line).
Since $\eta_{\text{out}}>1$ even at the highest masses, the flattening of \etaout\ with respect to \mstar\ is responsible
 for the flattening of the MZR at high masses (Fig.~\ref{fig:sample}).
As \citet{san21} noted, the lower-mass inferences based on the MZR suggest that, at fixed \mstar\, \etaout\ is larger
 at $z\sim2$ than at $z\sim0$.
This does not appear to be the case at $>10^{10.5}~\msun$, where both low- and high-redshift galaxies are inferred to
 have similar \etaout\ on average.

In the EAGLE simulations, \citet{mit20} found that \etaout\ falls roughly as a power law with \mstar\ even out to
 very high masses when AGN feedback is not included, while including AGN feedback causes \etaout\ to flatten out
 above halo masses of $\sim10^{12}~\msun$ ($M_*\sim10^{10}~\msun$).
\citet{nel19} see a flattening of \etaout\ toward high masses and eventual steep increase of \etaout\ with \mstar\ at
 very high masses that are attributed to AGN feedback in IllustrisTNG.
Other analyses of cosmological simulations have found that the flattening of the MZR at high masses requires
 the inclusion of AGN feedback \citep[e.g.,][]{der15,der17,ma16}.

While we have excluded AGN from our sample, the tracers used to identify AGN (X-ray luminosity, near-IR colors tracing
 hot dust, and rest-optical line ratios) can only identify galaxies with contemporaneous high rates of black hole accretion.
Galaxies with AGN activity in the recent past but low levels of black hole accretion
 in the present would correctly be identified as star-forming, though AGN feedback may have a lasting effect on
 ISM gas via material removal or gas heating that makes it easier for SNe feedback to drive outflows.
This scenario is consistent with the idea that super-massive black hole accretion rates trace SFRs
 (in turn tracing gas inflow rates) on average \citep{mad14}, while maintaining significant variability
 such that galaxies alternate between an ``on'' AGN phase and ``off''
 star-formation dominated phase \citep{hic14,hec14}.
We speculate that the transition from a power-law \etaout($M_*$) to a flat \etaout\ occurs at the mass above which
 the AGN duty cycle becomes high enough that the imprint of AGN feedback on ISM gas does not have time to be
 erased between consecutive AGN ``on'' phases, providing a physical driver for observed MZR flattening at high masses.
The occurrence rate of strong AGN rises steeply above the mass at which the MZR begins
 to flatten at $z\sim0$ \citep[e.g.,][]{kau04,air19}.
Using the KMOS$^{\text{3D}}$ sample at $z=0.6-2.7$,
 \citet{for19} found that the occurrence rates of both AGN and AGN-driven outflows grow steeply with increasing
 \mstar\ and exceed 10\% at $M_*>10^{10.5}~\msun$, similar to the mass where we find the onset of
 MZR and \etaout\ flattening at $z\sim2$.
While these results are generally consistent with intermittent AGN feedback as a driver of high-mass MZR flattening at $z\sim2$,
 robustly evaluting this scenario will require better quantitative constraints on the turnover mass of the
 $z\sim2$ MZR to directly compare to measures of AGN frequency as a function of \mstar.

\subsubsection{The role of accretion rate in scaling relations at $z\sim2$}

The (anti-)correlated residuals around the MS, MZR, and \mmol-\mstar\ relation
 displayed in Figures~\ref{fig:delsfms} and~\ref{fig:delmgasdelmzr} demonstrate the existence of
 a multi-dimensional relation among \mstar, SFR, \mmol, and metallicity in the $z\sim2$ MOSDEF-ALMA sample.
This relation is such that, at fixed \mstar, galaxies with higher gas masses have higher SFR and lower metallicity.
Such trends are found in a range of cosmological simulations \citep[e.g.,][]{ma16,dav17,dav19,der17,tor18,tor19},
 and are a staple of models of galaxy evolution based on a self-regulating baryon cycle \citep{dav12,lil13,pen14}.
These trends have also been observed in $z\sim0$ star-forming galaxy populations as the SFR-FMR and Gas-FMR
 \citep{man10,lar10,bot13,bot16a,bot16b,hug13,lar13,bro18}.
Evidence for a SFR-FMR at $z>1$ has been found previously \citep{zah14b,san18,san21,hen21},
 while hints of a Gas-FMR have been found at $z\sim1.4$ \citep{sek16a}.
Here, we confirm the interdependent nature of these four properties and the simultaneous existence of
 the SFR-FMR and Gas-FMR at high redshift for the first time.

The correlated residuals in \mmol, SFR, and metallicity can be explained if deviations of \mmol\ from the mean
 reflect variations in the gas inflow rate, \mdotin.
The observed trends then indicate that higher \mdotin\ drives larger SFR due to the availability of more cold gas
 while simultaneously diluting metals in the ISM as metal-poor gas is mixed in, lowering O/H.
This scenario implies that the scatter of the MS, MZR, and \mmol-\mstar\ relations are all
 primarily driven by variations in accretion rates,
 a conclusion reached by previous works modeling the SFR-FMR and Gas-FMR
 \citep{day13,for14,bro18,tor19}.
The simultaneous existence of a correlation between \delmmol\ and \delsfms\ and an anti-correlation between
 \delmmol\ and \delmzr\ requires that SFR and metallicity respond to the accretion of fresh gas quickly,
 faster than the timescale of significant variations in the inflow rate,
 as is seen in cosmological simulations \citep{mur15,tor18,tor19}.
Indeed, the dynamical timescales of $z\sim2$ galaxy disks (on which galaxy-averaged SFR and metallicity can vary)
 are a few tens of Myr \cite[e.g.,][]{for06,red12},
 while halo dynamical times (on which \mdotin\ from smooth gas accretion is expected to vary)
 are roughly an order of magnitude larger \citep[e.g.,][]{tor18}.
The overall tightness of these three scaling elations (1$\sigma$ scatter=$0.1-0.3$~dex at fixed \mstar)
 then suggests that large variations in \mdotin\ averaged over the SFR and O/H response timescale
 (i.e., disk dynamical time, $\sim30-50$~Myr) are rare.

Comparing the observed quantitative slopes of these correlated residual relations to those produced in numerical
 simulations and semi-analytic models of galaxy formation presents an avenue to make
 high-order tests of accretion and feedback models \citep[e.g.,][]{dav17,tor19,pan20}.
The slopes are not well-constrained with the current $z\sim2$ MOSDEF-ALMA sample due to the small sample size
 and data quality.
This work motivates a larger and deeper survey of CO in $z\sim2$ galaxies with accompanying metallicity
 measurements to quantitatively constrain the secondary dependence on gas mass.

\subsection{ISM metal mass and retention}\label{sec:zmass}

\begin{figure}
 \includegraphics[width=\columnwidth]{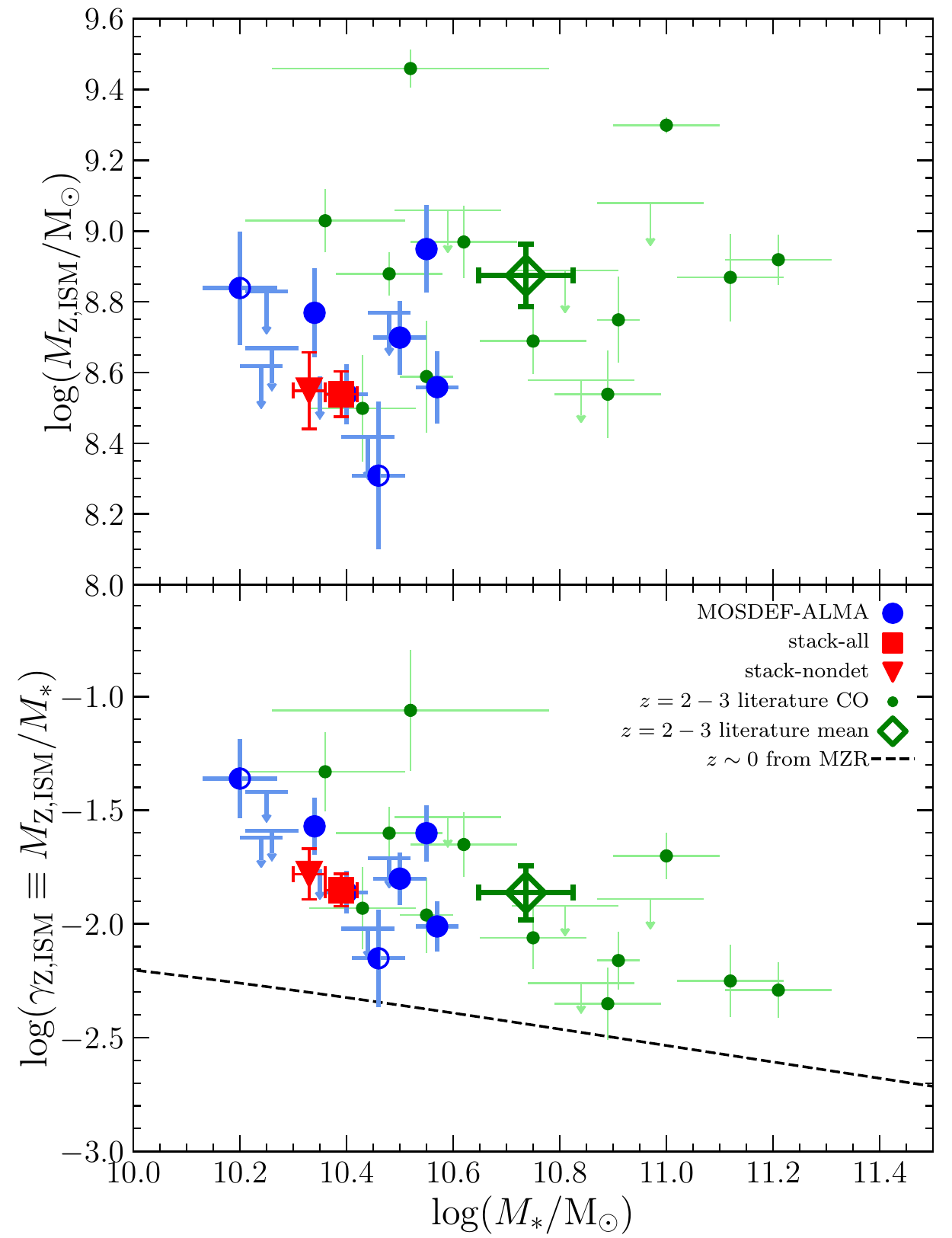}
 \centering
 \caption{
\textit{Top:} Total metal mass in the gas-phase ISM vs.\ stellar mass.
\textit{Bottom:} The retained metal yield of the ISM vs.\ stellar mass.
The black dashed line represents the mean relation at $z\sim0$ derived by combining the MZR of \citet{cur20b} with
 \mgas\ (\mhi+\mmol) inferred from the scaling relation of \citet{sai16} evaluated on the main sequence.
}\label{fig:mzyz}
\end{figure}

The total metal mass in the ISM can be estimated from the combination of metallicity and gas mass.
We convert from the number density 12+log(O/H) to a mass fraction $Z_O=M_O/M_{\text{gas}}$
 assuming He makes up 36\% of the gas mass, and then use the ratio of the oxygen and total metal
 stellar yields $y_O/y_Z=0.6$\footnote{We used the IMF-integrated core-collapse SNe yields of \citet{vin16} for a \citet{cha03} IMF and the \citet{nom13} yield tables.
The O-to-total metals mass fraction is lower for systems significantly enriched by Type~Ia SNe
 (e.g., 0.43 for a solar abundance pattern).
There is evidence that the enrichment of $z\sim2$ star-forming galaxies is
 dominated by core-collapse SNe with little contribution from Type~Ia \citep{ste16,str18,san20,top20a,top20b}.}
 to infer the total gas-phase ISM metal mass fraction $Z_{\text{ISM}}=M_Z/M_{\text{gas}}$.
We then multiply \zism\ by \mmol, assuming \mmol$\approx$\mgas\ for the $z\sim2$ sources, to obtain the mass of metals
 in the ISM, \mzism.
The top panel of Figure~\ref{fig:mzyz} shows that \mzism\ increases with increasing \mstar\ on average (i.e., for the composites),
 as expected since both metallicity and \mmol\ increase with increasing \mstar.
The typical ISM metal mass of $z\sim2$ star-forming galaxies is $M_{\text{Z,ISM}}\approx10^{8.7}~\msun$ at $M_*=10^{10.5}~\msun$.

The ratio of gas-phase ISM metal mass to galaxy stellar mass has been used as a measure of a galaxy's ability to
 retain metals since \mstar\ is approximately proportional to the total metals produced over a galaxy lifetime
 \citep{ma16,tor19}.
\citet{tor19} defined this quantity as the ISM metal retention efficiency, $\gamma_{\text{Z,ISM}}=M_{\text{Z,ISM}}/M_*$,
 and argued that evolving gas fractions, not outflow efficiencies, control the evolution of the MZR because
 higher redshift galaxies in IllustrisTNG have higher $\gamma_{\text{Z,ISM}}$ at fixed \mstar\
 and are thus inferred to be more efficient at retaining gas-phase metals than their low-redshift counterparts.
We plot $\gamma_{\text{Z,ISM}}$ vs.\ \mstar\ in the bottom panel of Figure~\ref{fig:mzyz}, and find that
 the combined $z\sim2$ sample indeed displays higher ISM metal retention efficiencies at fixed \mstar\ by a
 factor of $\sim3$ than the mean value at $z\sim0$ (black dashed line).

This result implies that $z\sim2$ galaxies at $\log(M_*/\msun)\sim10.5-11$ are less efficient at ejecting metals,
 seemingly in conflict with the result based on gas-regulator models that \etaout\ is the same on average
 at $z\sim2$ and $z\sim0$ over this mass range (Fig.~\ref{fig:etaout}, bottom panel).
In their analysis of the MZR in the IllustrisTNG simulations,
 \citet{tor19} pointed out the tension between $\gamma_{\text{Z,ISM}}$
 that implies that gas fraction primarily governs the MZR shape and its evolution (see also \citealt{ma16}),
 and analyses based on gas-regulator or equilibrium models
 in which outflow efficiency is inferred to play the larger role \citep[e.g.,][]{pee11,dav12,lil13,san21}.
\citet{tor19} argued that the importance of outflows is overestimated when using current gas-regulator models
 because they are not properly treating potentially important physical processes known to
 occur in numerical simulations, notably the recycling of enriched previously-outflowing material \citep{fin08,mur15,mur17,ang17,pan20}.
Theoretical work has shown close connections between \etaout\ and \mugas\ or \siggas\ \citep{hay17,kim20},
 such that the action of outflows vs.\ gas fraction in setting the ISM metallicity may not be cleanly separable.
We will revisit the apparent tension between Figures~\ref{fig:etaout} and~\ref{fig:mzyz} at the end of Sec.~\ref{sec:metalbudget}.

\subsection{A budget of metals in massive $z\sim2$ star-forming galaxies}\label{sec:metalbudget}

We now combine the ISM metal mass calculated above in Sec.~\ref{sec:zmass} with estimates of the metal mass
 in stars and dust to perform a simple metal budget analysis.
We carry out this analysis on the sample averages of the MOSDEF-ALMA and $z=2-3$ literature samples.
We estimate the mass of metals locked in stars, \mzstars, by multiplying the stellar metallicity by \mstar,
 assuming the stellar metallicity is equal to the current gas-phase metallicity.
This approach likely overestimates the true \mzstars\ since ISM metallicity increases with time on average
 such that only recently formed stars will have metallicity equal to that of the current ISM, while previous
 generations of stars should have lower metallicity.
The mass of metals in dust, \mzdust, is taken to be the total dust mass since dust grains are essentially entirely
 composed of metals.
To estimate \mzdust, we combine the gas mass with the dust-to-gas ratio as a function of O/H from
 \citet{dev19}\footnote{We use their parameterization based on the \citet{pet04} O3N2 metallicity calibration.},
 which is consistent with current dust-to-gas constraints at $z\sim2$ \citep[Popping et al., in prep.]{sha20}.
As before, we assume that \mgas$\approx$\mmol\ in high-redshift star-forming galaxies \citep{lag11,pop14,pop15,tac18}.
The lifetime metal production is calculated by first dividing \mstar\ by (1-R) to infer the lifetime total stellar mass produced,
 and multiplying the resulting value by the stellar metal yield.
We adopt $y_Z=0.051$ and $R=0.45$ for a \citet{cha03} IMF \citep{vin16}.

\begin{figure}
 \includegraphics[width=\columnwidth]{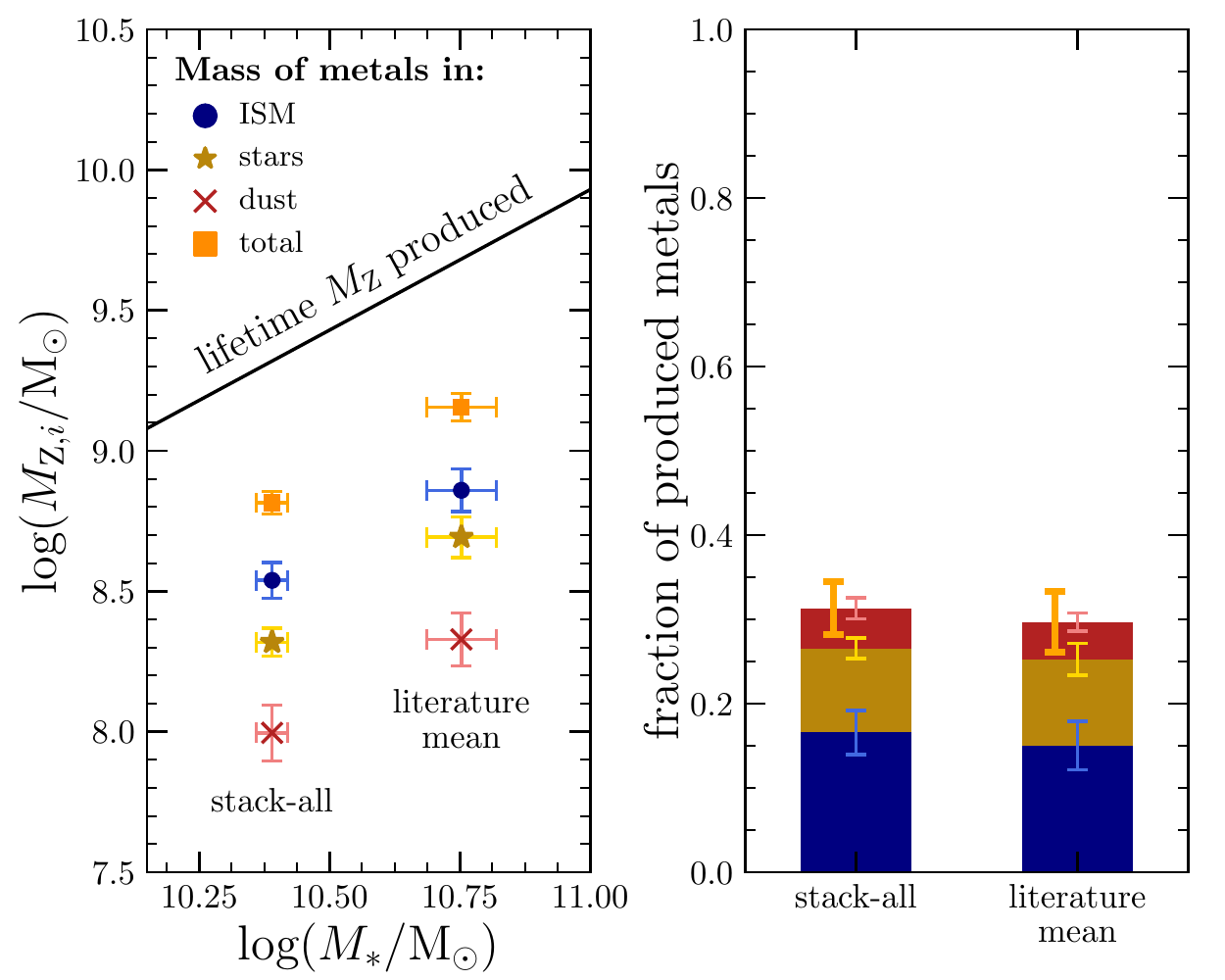}
 \centering
 \caption{
A budget of metals in star-forming galaxies at $z=2-3$.
Data are shown for the composite spectrum of all MOSDEF-ALMA targets (stack-all) at lower mass and the mean of the $z=2-3$
 literature CO+O/H sample at higher mass.
\textit{Left:} The mass of metals in different phases vs. \mstar.
Metal mass has been estimated for the gas-phase ISM (blue circle), stars (yellow star),
 interstellar dust (red ``x''), and the sum of these three components representing the total metal mass accounted
 for within the galaxy (orange square).
The black line shows the estimated total metal mass produced.
\textit{Right:} Fraction of total produced metals accounted for in the ISM, stars, and dust, obtained by dividing the metal mass
 in each phase by the lifetime metal mass produced.
The orange error bar shows the cumulative uncertainty on the sum.
The $approx$70\% of produced metals that are unaccounted for are inferred to be lost to the CGM and IGM.
}\label{fig:metalbudget}
\end{figure}

The left panel of Figure~\ref{fig:metalbudget} displays the mass of metals in the ISM, stars, dust, and the sum total
 of metals accounted for inside galaxies compared to the lifetime metal mass produced for the $z\sim2$ samples.
We find that the majority of metals inside $z\sim2$ galaxies reside in the gas-phase ISM, even at very high \mstar.
This is unlike what is found at $z\sim0$, where only low-mass galaxies at $M_*<10^{10}~\msun$ have most of their metals
 in the ISM, while the majority of metals in more massive local galaxies resides in stars \citep{pee14,opp16,mur17}.
This difference can be attributed to the steep increase of \mugas\ with redshift: $z\sim2$ galaxies have roughly an order
 of magnitude larger \mgas\ than $z\sim0$ galaxies at fixed \mstar.
Dust appears to be a subdominant destination for metals at $z\sim2$, similarly to $z\sim0$.
Our dust mass estimates, based on a dust-to-gas ratio, are $\sim0.2$~dex lower than those derived from
 ALMA dust-continuum measurements of $z\sim2.3$ galaxies at similar \mstar\ by \citet{shi22}.
Even if \mzdust\ was larger by that amount, it would remain smaller than \mzism\ and \mzstars.

The fraction of metals accounted for relative to the lifetime metal mass produced is presented in the right
 panel of Figure~\ref{fig:metalbudget}.
The metal mass accounted for in the ISM, stars, and dust makes up only $30\pm4\%$ of the total metals produced,
 implying that the majority of produced metals have been ejected from the galaxies in our samples
 and reside in either the CGM or IGM.
This value is very similar to the 25\% average metal retention fraction derived for $z\sim0$ galaxies by \citet{pee14}
 using similar metal budgeting techniques,
 and is also close to that of $\sim10^{10}~\msun$ galaxies at $z=2$ in the FIRE simulations \citep{mur17}.
The metal loss fraction of 70\% does not show variation across the mass range spanned by the $z\sim2$ samples,
 again similar to the behavior at $z\sim0$ \citep{pee14,opp16} despite large differences in global properties
 such as \mugas, SFR, and \sigsfr\ that are thought to be related to outflow strength.
This analysis predicts a large mass of metals in the CGM of $z\sim2$ star-forming galaxies,
 with $M_{\text{Z,CGM}}\sim10^9~\msun$ at $M_*=10^{10.5}~\msun$.
Significant metal absorption from low- and high-ionization species has been observed in the CGM of $z\sim2$
 galaxies \citep[e.g.,][]{ade05,ste10,rud19}.
Recently, \citet{rud19} found that the mass of metals in the CGM of $M_*\sim10^{10}~\msun$ $z\sim2$ galaxies is
 $>25\%$ of \mzism\ alongside evidence that metal ejection from the halo into the IGM may be common.
Improved observational constraints on the CGM metal mass of high-redshift galaxies are needed to see if the
 missing metals in Figure~\ref{fig:metalbudget} can be accounted for.
 
The fact that the total metal retention fraction is found to be $\approx$30\% for galaxies at $\log(M_*/\msun)\sim10.5$
 at \textit{both} $z\sim0$ and $z\sim2$ implies that they have a similar efficiency of metal ejection,
 consistent with our inferences of \etaout\ at both redshifts in the bottom panel of Fig.~\ref{fig:etaout}.
We can now understand the apparent tension in the evolution of the ISM metal retention efficiency, $\gamma_{\text{Z,ISM}}$,
 at fixed \mstar\ in Fig.~\ref{fig:mzyz} toward higher metal retention with increasing redshift.
As a consequence of their large gas fractions (i.e., $M_{\text{gas}}>M_*$), $z\sim2$ galaxies store the largest fraction of their
 metals in the gas-phase ISM, with smaller contributions from stars and dust.
In contrast, gas fractions are low at $z\sim0$ ($M_{\text{gas}}\approx0.2M_*$ at $\sim10^{10.5}~\msun$), such that
 the vast majority of metals are stored in stars and only a small part is in the ISM
 for local galaxies in this mass range \citep{pee14}.
Thus, the evolution of $\gamma_{\text{Z,ISM}}$ at fixed \mstar\ in Fig.~\ref{fig:mzyz} reflects an increasing fraction of
 total metals stored in the gas-phase ISM with increasing redshift as a consequence of higher \mugas,
 rather than a higher overall metal retention (and lower metal ejection efficiency) at higher redshift.

We therefore do not find any tension between analyses based on gas-regulator models and metal retention fractions,
 but stress that galaxy metal retention (and its evolution with redshift) cannot be properly evaluated using
 metals in the gas-phase ISM alone because the relative contribution of this phase to the total metal mass
 changes as a function of both \mstar\ and redshift.
This finding suggests that the missing physics (e.g., outflow recycling) in current gas-regulator models may not
 be required to obtain a reasonably accurate understanding of the origin and evolution of the MZR,
 though this problem ultimately requires the development of analytic models including these high-order processes
 to evaluate the impact of their inclusion or exclusion.
In Fig.~\ref{fig:etaout}, inferences at lower masses from the MZR suggest that \etaout\ increases with increasing redshift
 below $10^{10}~\msun$.
Accordingly, we expect the total metal retention fraction to be lower at $z\sim2$ than at $z\sim0$ at fixed \mstar\ for
 low-mass systems.
This motivates an analysis of the metal budget at $z\sim2$ across a wider mass range, which we will address in future work.

\subsection{Cold gas content scaling relations at $z\sim2$}

Recent studies have calibrated scaling relations of \mmol, \mumol, and \tdepl\ as a function of
 redshift, \mstar, and SFR (or, equivalently, offset from the MS)
 by combining samples with Rayleigh-Jeans dust continuum and/or CO line emission observations spanning $z=0-4$
 \citep{gen15,sco17,tac18,tac20,liu19,wan22}.
These scaling relations potentially have great utility by providing estimates of the cold gas content of galaxies
 based on galaxy properties that are easier to directly constrain.
However, at $z\ge2$, the samples used to calibrate such scaling relations are composed of massive galaxies
 ($M_*\gtrsim10^{10.5}$) such that using these relations to estimate gas masses of lower-mass high-redshift galaxies
 requires extrapolation.
Furthermore, the vast majority (and in some works entirety) of galaxies in the $z>1$ calibration samples
 have gas mass measurements based on dust continuum emission, while only a small fraction have CO-based \mmol.
Using the combination of the lower-mass MOSDEF-ALMA and more-massive literature CO samples in this work,
 we can evaluate which cold gas scaling relations are reliable at $z\sim2$.

The scaling relations of \citet{sco17}, \citet{tac18}, and \citet{liu19} are displayed in each
 panel of Figure~\ref{fig:mgas}, all evaluated at $z=2.3$ for galaxies on the MS.
We find that the scaling relation of \citet{tac18} reliably reproduces CO-based \mmol\ and \mumol\ of
 main-sequence galaxies on average (gray square and colored diamond) down to $10^{10.4}~\msun$, in
 particular having a slope that matches the observed trend with \mstar.
In contrast, the \citet{sco17} and \citet{liu19} relations have slopes as a function of \mstar\ that are
 too shallow for \mmol\ and too steep for \mumol, suggesting that these relations will overestimate
 gas mass when extrapolating to $M_*\lesssim10^{10}~\msun$.
The normalization of the \citet{sco17} relation is too high across the full range of masses probed by
 the $z\sim2$ CO sample, while the \citet{liu19} relation underestimates \mmol\ and \mumol\ at $\sim10^{11}~\msun$.
The good match between the relation of \citet{tac18}
 and the CO-based mean values suggests that the \citet{tac18} parameterization is more reliable when
 extrapolating to lower masses.

We found little dependence of \tdepl\ on \mstar\ in Figure~\ref{fig:mgas}, with lower-mass galaxies having
 marginally smaller \tdepl.
Both the \citet{sco17} relation that has no mass dependence and
 the relation of \citet{tac18} with $t_{\text{depl}}\propto M_*^{0.09\pm0.05}$ are consistent with the
 observed \mstar\ scaling within the uncertaines.
The \citet{liu19} dependence ($t_{\text{depl}}\propto M_*^{-0.52}$) is clearly too steep to match
 the CO-based \tdepl\ and would severely overestimate \tdepl\ at lower \mstar.
The \citet{sco17} and \citet{tac18} relations bracket the observations in normalization, with the former
 slightly higher and the latter slightly lower.

The SFR dependence of \mmol\ and \mumol\ at fixed \mstar\ is nearly identical for \citet{tac18} and \citet{liu19}
 ($\mu_{\text{mol}}\propto(\text{SFR/SFR}_{\text{MS}})^{0.53}$), matching the observed trend in Figure~\ref{fig:delsfms}, left panel.
\citet{sco17} find $\mu_{\text{mol}}\propto(\text{SFR/SFR}_{\text{MS}})^{0.32}$, significantly shallower than
 the CO-based constraints at $z\sim2$.
In summary, we find that the \citet{tac18} scaling relations provide the best overall match to the CO-based observations for
 estimates of \mmol, \mumol, and \tdepl\ as a function of \mstar\ and SFR at $z\sim2$,
 and provide a particularly good match in mass-dependence suggesting extrapolation to lower masses is
 reliable.\footnote{While the majority of our $z=2-3$ literature CO sample was included in the \citet{tac18} calibration sample,
 $<$25\% of their $z>1$ sample has CO measurements, with the vast majority based on dust continuum.
The good match of the \citet{tac18} scaling relations to the CO-based measurements is thus not by construction
 since this small subset of their total sample would not dominate during the fitting process.}

\subsection{Prospects for detecting CO of low-mass high-redshift galaxies with ALMA}

With the MOSDEF-ALMA sample, we have pushed the stellar mass of $z\sim2$ main-sequence galaxies with CO-based \mmol\ down
 by a factor of 2, from $10^{10.7}~\msun$ to $10^{10.4}~\msun$.
It is of interest to investigate the prospects of detecting CO emission in even lower-mass $z\sim2$ galaxies
 to expand the dynamic range of parameter space probed by CO in the \mstar-SFR plane.
We estimated the integrated \cott\ line flux, \scott, for $z=2.3$ galaxies on the MS by combining
 equations~\ref{eq:lco32sfroh} and~\ref{eq:fitmzrz} to estimate \lcott/SFR as a function of \mstar,
 using the $z=2.3$ MS of \citet{spe14} to convert to \lcott,
 and finally obtaining \scott\ with equation~\ref{eq:lcosco}.
We then inferred \scott\ for offsets above and below the MS using the best-fit SFR dependence of \lcott\ at fixed \mstar:
 L$^{\prime}_{\text{CO}(3-2)}\propto(\text{SFR/SFR}_{\text{MS}})^{0.71}$ (equation~\ref{eq:lco32mstar}).

\begin{figure}
 \includegraphics[width=\columnwidth]{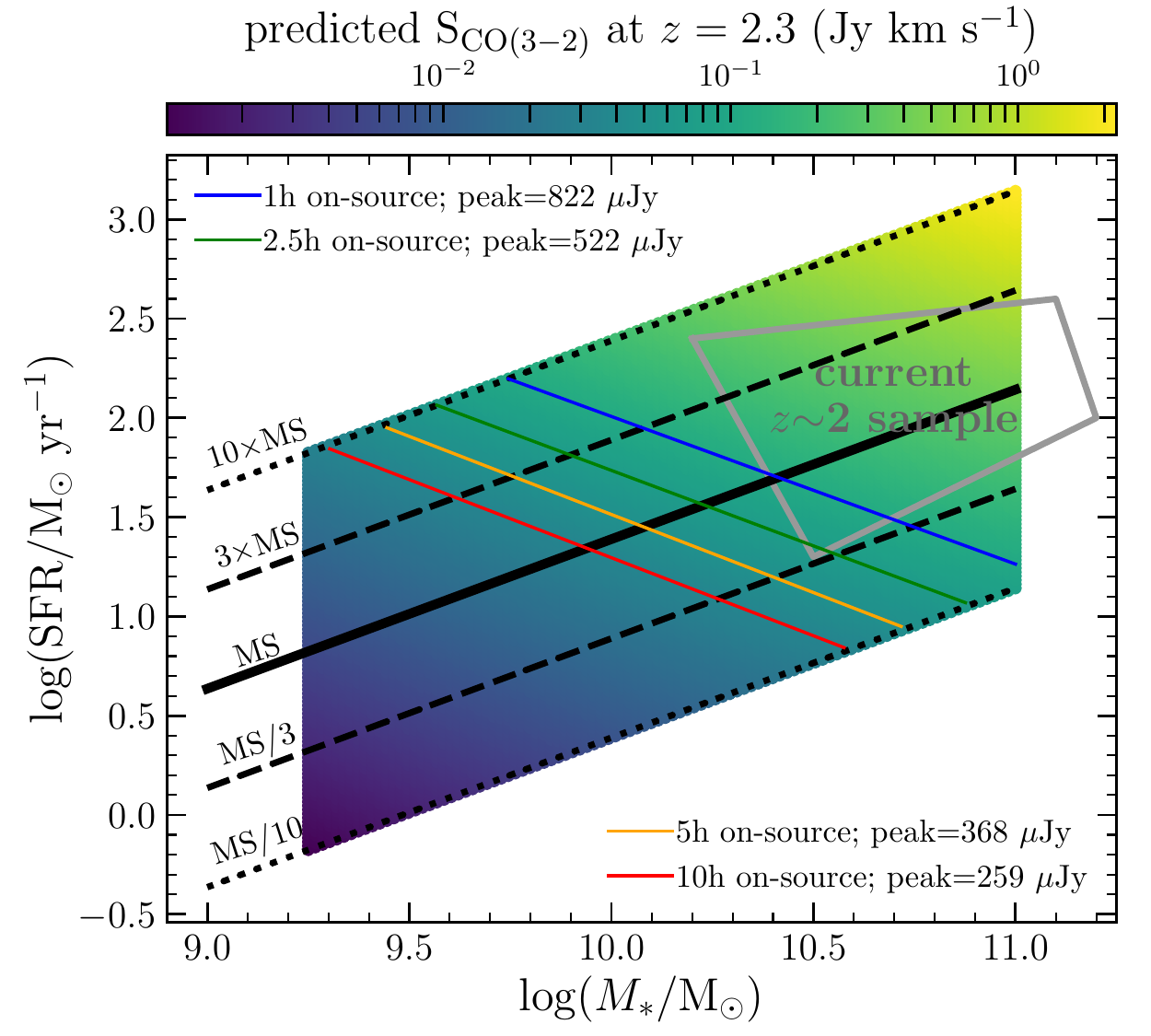}
 \centering
 \caption{
Predicted integrated \cott\ line flux, \scott, as a function of \mstar\ and SFR for galaxies at $z=2.3$.
\scott\ is estimated as described in the text.
Colored lines show contours with integrated \cott\ S/N=4 in a given on-source inegration time with ALMA,
 with the integration time and flux density of the line peak given in the legend.
The RMS sensitivity over a 50~km~s$^{-1}$ bandwidth is equal to the peak flux density divided by 3.5.
The range of \mstar\ and SFR for the current $z\sim2$ CO sample employed in this work is outlined
 by the thick gray line.
The solid black line traces the star-forming main sequence at $z=2.3$ \citep{spe14}, while the dashed
 and dotted lines show locations 3 and 10 times above/below the MS.
}\label{fig:alma}
\end{figure}

The predicted \scott\ at $z=2.3$ across the \mstar-SFR plane is displayed in Figure~\ref{fig:alma}.
We converted \scott\ to the peak line flux density assuming a velocity FWHM of 150 km~s$^{-1}$, the average value
 for $z\sim2.3$ galaxies in the MOSDEF survey spanning $\log(M_*/\msun)=9.0-10.5$ \citep{pri16}.
We then used this peak flux density to estimate the required integration time with ALMA to detect \cott\ across
 the \mstar-SFR plane.
We employed the ALMA Sensitivity Calculator\footnote{\url{https://almascience.eso.org/proposing/sensitivity-calculator}}
 to calculate the RMS sensitivity for different on-source integration times.
We assumed an observed frequency of 104.8~GHz (\cott\ at $z=2.3$) that falls in ALMA Receiver Band~3,
 an equatorial target (e.g., the COSMOS field),
 and a bandwidth of 50~km~s$^{-1}$ equal to 1/3 of the line FWHM for the RMS calculations.
Using Monte Carlo simulations, we found that integrated line S/N=4 is obtained if the RMS value integrated over
 1/3 of the line FWHM is 3.5 times smaller than the line peak.
For a given integration time, we identified the contour in the \mstar-SFR plane for which the line peak is 3.5 times
 the RMS value, above which galaxies will have integrated \cott\ S/N$>$4.

The blue, green, orange, and red lines show contours where the integrated line flux has S/N=4 in on-source integration times
 of 1, 2.5, 5, and 10~hours, respectively.
Integrations of $\sim1$~h on-source reach the current mass limit of $10^{10.4}~\msun$ on the MS.
At $z\sim2.3$, a survey with 5~h on-source will reach $10^{10}~\msun$ on the MS,
 $10^{9.7}~\msun$ for moderate starbursts (3$\times$MS),
 and $10^{9.4}~\msun$ for extreme starbursts (10$\times$MS).
Such a survey could nearly double the parameter space containing galaxies with CO detections at $z\sim2$ relative
 to the current sample, probing a regime that significantly overlaps with samples from large spectroscopic
 surveys for which detailed metallicity, SFR, dust reddening, and ionization constraints are available.
Detecting \cott\ in main-sequence galaxies at masses significantly below $10^{10}~\msun$ would require prohibitively
 long integrations with ALMA and is thus not a feasible route to constrain the gas masses of these targets.
More promising strategies to reach lower masses include measuring CO in gravitationally-lensed galaxies,
 where even modest magnification factors of $3-5$ would lead to significant gains in efficiency,
 or deriving \mmol\ from the Rayleigh-Jeans dust-continuum, which has proven to be detectable with ALMA in shorter
 integration times than are required for CO \citep[e.g.,][]{kaa19,ara20,suz21,shi22}.

\section{Summary and Conclusions}\label{sec:summary}

We have analyzed ALMA observations of \cott\ for a sample of 13 moderate-mass
 main-sequence galaxies at $z\sim2.3$ that uniquely have near-infrared spectra from the MOSDEF survey
 covering the full set of strong rest-optical emission lines, from which we have derived gas-phase metallicities.
We supplemented the MOSDEF-ALMA sample with a sample of more massive $z=2-3$ main-sequence galaxies from the
 literature with existing CO and metallicity constraints.
The combination of cold gas content and metallicity information provides a powerful tool to constrain baryon cycling
 in the high-redshift universe.
Our main conclusions are as follows:
\smallskip

(i) We characterized the dependence of \lcott\ on \mstar, SFR, and O/H at $z\sim2$, finding that
 \lcott/SFR increases with O/H in a tight relation (Fig.~\ref{fig:lco32sfroh}).
This result implies that CO luminosity per unit gas mass is lower in low-metallicity galaxies, carrying
 important implications for the conversion factor between CO luminosity and molecular gas mass.

(ii) We estimated the CO-to-\htwo\ conversion factor \aco\ for a sample of $z=1.1-3.2$ galaxies with
 spectroscopic metallicity constraints using two techniques based on dynamical masses and the molecular KS law.
With both methods, we found a significant dependence of \aco\ on O/H such that \aco\ increases with decreasing O/H
 (Fig.~\ref{fig:aco}), a trend that is physically driven by less dust shielding and more intense UV radiation fields
 in low-metallicity systems.
The $z\sim2$ relation is consistent with $z\sim0$ \aco(O/H) relations within the uncertainties.

(iii) We found that \mmol\ increases and \mumol\ decreases with increasing \mstar, while both properties display
 a strong secondary positive correlation with SFR at fixed \mstar (Fig.~\ref{fig:mgas}).
The scaling relation of \citet{tac18,tac20} reproduces the observed \mstar\ dependence better than other
 published \mumol\ scaling relations \citep{sco17,liu19}, suggesting it provides a more reliable extrapolation
 if applied to lower-mass $z\sim2$ samples.
The molecular depletion timescales are 700~Myr on average and do not display any dependence on \mstar across
 $\log(M_*/\msun)=10.2-11.4$.

(iv) A tight near-linear molecular KS law exists at $z\sim2$ (Fig.~\ref{fig:ks}, equation~\ref{eq:ks}),
 providing a reliable means of estimating \mmol\ indirectly from SFR and size measurements for
 large high-redshift star-forming samples.

(v) In the $z\sim2$ MOSDEF-ALMA sample, we found that residuals around the SFR-\mstar (star-forming main sequence),
 O/H-\mstar\ (mass-metallicity relation), and \mmol-\mstar\ relations are correlated
 (Figs.~\ref{fig:delsfms} and~\ref{fig:delmgasdelmzr}).
These correlations are such that, at fixed \mstar, galaxies with larger \mmol\ have higher SFR and lower O/H.
This result confirms the simultaneous existence of a SFR-FMR and Gas-FMR at $z\sim2$, both of which have been
 observed at $z\sim0$.
These results suggest that the scatter of both the star-forming main sequence and mass-metallicity relation
 are driven by stochastic variations
 in gas inflow rates that are traced by variations in gas fraction at fixed \mstar.
Better gas mass constraints spanning a wider range of \mstar\ and SFR are required to obtain the quantitative form of
 the high-redshift Gas-FMR and its evolution across cosmic history, which could provide high-order tests of gas
 accretion and feedback models in numerical simulations.

(vi) We used gas-regulator models to infer the outflow mass-loading factors of the $z\sim2$ sample,
 finding $\eta_{\text{out}}=1-4$ with a typical value of 2.5 with no significant dependence on \mstar\ above
 $10^{10.2}~\msun$ (Fig.~\ref{fig:etaout}).
A similar flattening of \etaout\ at high \mstar\ is observed at $z\sim0$.
We conclude that the high-mass flattening of the MZR is driven by a flattening in \etaout,
 which in turn may be caused by the onset of cyclical AGN feedback in high-mass galaxies.

(vii) We performed a metal budgeting analysis, estimating the mass of metals found inside massive $z\sim2$ main-sequence
 galaxies in the gas-phase ISM, dust, and stars (Fig.~\ref{fig:metalbudget}).
Comparing the sum of these phases to the total metal mass produced over a galaxy lifetime showed that
 massive $z\sim2$ star-forming galaxies retain only 30\% of produced metals, implying that two thirds of
 produced metals are ejected into the CGM and/or IGM and demonstrating the important role of
 feedback-driven outflows in removing metals from galaxies.
Future constraints on CGM metal masses at $z\sim2$ can determine whether the missing metals can be accounted for.

(viii) We find that the \mmol, \mugas, and \tdepl\ scaling relations of \citet{tac18} provide the best overall
 match to the $z\sim2$ CO samples in terms of \mstar\ and SFR dependence, suggesting that these relations can be
 extrapolated to lower masses without incurring large systematic biases.
The \mugas\ scaling relations of \citet{sco16} and \citet{liu19} are too steep with \mstar\ such
 that they likely overestimate the molecular gas content of low-mass systems at $z\sim2$.

(ix) We explored the possibility of detecting \cott\ in low-mass $z\sim2$ galaxies with ALMA (Fig.~\ref{fig:alma}).
Main-sequence galaxies at $M_*=10^{10}~\msun$ can be detected in 5~hours on source, while starbursts a factor of
 5 times above the main-sequence can be reached with the same depth down to $10^{9.5}~\msun$.
Detecting CO at lower masses is inefficient with ALMA due to the combined effect of decreasing \mmol\ with
 decreasing \mstar\ and large \aco\ at low metallicity.
Targeting lensed galaxies for CO and deriving \mmol\ from dust continuum observations present promising
possibilites to reach low masses at high redshift.
\smallskip

These results demonstrate the power of combining measurements of metallicity and gas mass to constrain baryon
 cycling for high-redshift galaxies.
Existing samples at $z>2$ with these measurements are small and largely have low-S/N gas measurements (CO or dust continuum)
 that often require stacking to produce significant detections.
Combining samples from the literature potentially introduces systematic effects due to the variety of selection criteria and
 rest-optical line ratios available for metallicity determinations.
The quest for a detailed understanding of baryon cycling at high redshifts would greatly benefit from a deeper survey
 of CO emission in a large and systematically selected sample of galaxies with accompanying rest-optical spectroscopy.
ALMA currently has the capabilities to obtain the required measurements for a large sample at $z\sim2$.

\medskip
\noindent This paper makes use of the following ALMA data: ADS/JAO.ALMA\#2018.1.01128.S. ALMA is a partnership of ESO (representing its member states), NSF (USA) and NINS (Japan), together with NRC (Canada), MOST and ASIAA (Taiwan), and KASI (Republic of Korea), in cooperation with the Republic of Chile. The Joint ALMA Observatory is operated by ESO, AUI/NRAO and NAOJ.
The National Radio Astronomy Observatory is a facility of the National Science Foundation operated under cooperative agreement by Associated Universities, Inc.
Support for this work was provided by NASA through the NASA Hubble Fellowship grant \#HST-HF2-51469.001-A awarded
 by the Space Telescope Science Institute, which is operated by the Association of Universities for Research
 in Astronomy, Incorporated, under NASA contract NAS5-26555.
We acknowledge support from NSF AAG grants AST-1312780, 1312547, 1312764, and 1313171,
 archival grant AR-13907 provided by NASA through the Space Telescope Science Institute,
 and grant NNX16AF54G from the NASA ADAP program.
We also acknowledge a NASA
 contract supporting the ``WFIRST Extragalactic Potential
 Observations (EXPO) Science Investigation Team'' (15-WFIRST15-0004), administered by GSFC.
We additionally acknowledge the 3D-HST collaboration
 for providing spectroscopic and photometric catalogs used in the MOSDEF survey.
The authors wish to recognize and acknowledge the very significant cultural role and reverence that
 the summit of Maunakea has always had within the indigenous Hawaiian community.
We are most fortunate to have the opportunity to conduct observations from this mountain.


\appendix
\restartappendixnumbering

\section{Metallicity Calibrations}\label{app:metallicity}

The set of calibrations used to convert strong-line ratios to metallicities deserves careful
 consideration in analyses where metallicity plays a key role, especially when a uniform set
 of emission lines is not available for the entire sample.
We adopt calibrations derived from the local sample of $z\sim2$ analogs from \citet{bia18} that \citet{san20}
 found to be a good match to actual $z\sim2$ galaxies with direct-method abundances for line ratios
 of [O\iii], [O\ii], [Ne\iii], and H$\beta$.
Here, we describe some important differences in our use of the \citet{bia18} calibration set relative to what is
 reported in that work.
Table~\ref{tab:metallicity} provides the metallicity calibrations adopted in this work as polynomial
 coefficients for use in the following equation:
\begin{equation}\label{eq:calibrations}
\log(R) = \sum_{n=0}^{N} c_n x^n ,
\end{equation}
where $R$ is the line ratio under consideration, $x$=12+log(O/H), and $c_n$ are the coefficients.
Differences from the values in \citet{bia18} are described below.

\begin{table}
 \centering
 \caption{Adopted metallicity calibrations for high-redshift samples, with coefficients given for equation~\ref{eq:calibrations}.
 }\label{tab:metallicity}
 \begin{tabular}{ r r r r r }
   \hline\hline
   line ratio ($R$) & $c_0$ & $c_1$ & $c_2$ & $c_3$ \\
   \hline
   $[$O\iii$]$$\lambda$5007/H$\beta$ & 43.8576 & $-$21.6211 & 3.4277 & $-$0.1747 \\
   $[$O\iii$]$$\lambda$5007/$[$O\ii$]$ & 14.35 & $-$1.70 & --- & --- \\
   $[$Ne\iii$]$$\lambda$3869/$[$O\ii$]$ & 12.38 & $-$1.59 & --- & --- \\
   R$_{23}$\tablenotemark{a} & 138.0430 & $-$54.8284 & 7.2954 & $-$0.32293 \\
   $[$N\ii$]$$\lambda$6584/H$\alpha$ & $-$18.24 & 2.04 & --- & --- \\
   O3N2\tablenotemark{b} & 23.12 & $-$2.56 & --- & --- \\
   \hline\hline
 \end{tabular}
 \tablenotetext{a}{R$_{23}$=([O\iii]$\lambda\lambda$4959,5007+[O\ii]$\lambda\lambda$3726,3729)/H$\beta$}
 \tablenotetext{b}{O3N2=([O\iii]$\lambda$5007/H$\beta$)/([N\ii]$\lambda$6584/H$\alpha$)}
\end{table}

As noted in \citet{san21}, the [O\iii]$\lambda$5007/H$\beta$ calibration coefficients given in equation~17
 of \citet{bia18} are actually the coefficients fit to [O\iii]$\lambda\lambda$4959,5007/H$\beta$, i.e.,
 the ratio including both of the [O\iii] components.
This issue has been confirmed by the authors (F. Bian, private communication).
The normalization of the [O\iii]$\lambda$5007/H$\beta$ calibration in Table~\ref{tab:metallicity} is
 0.126~dex lower than the value reported in \citet{bia18} to correct for this issue, under the assumption that
 [O\iii]$\lambda$5007/$\lambda$4959=2.98 \citep{sto00}.

While strong-line ratios involving the [N\ii]$\lambda$6584 line are commonly used to derive
 the oxygen abundance \citep[e.g., N2 and O3N2;][]{pet04,mar13},
 N-based calibrations are sensitive to N/H such that differences in N/O at fixed O/H between the
 calibrating and target samples will systematically bias the O/H estimates.
Unfortunately, \citet{san20} were unable to test the N2 and O3N2 analog calibrations from \citet{bia18}
 due to the small number of galaxies at $z>1$ with both direct-method metallicities and [N\ii] detections.

\begin{figure}
 \includegraphics[width=0.8\columnwidth]{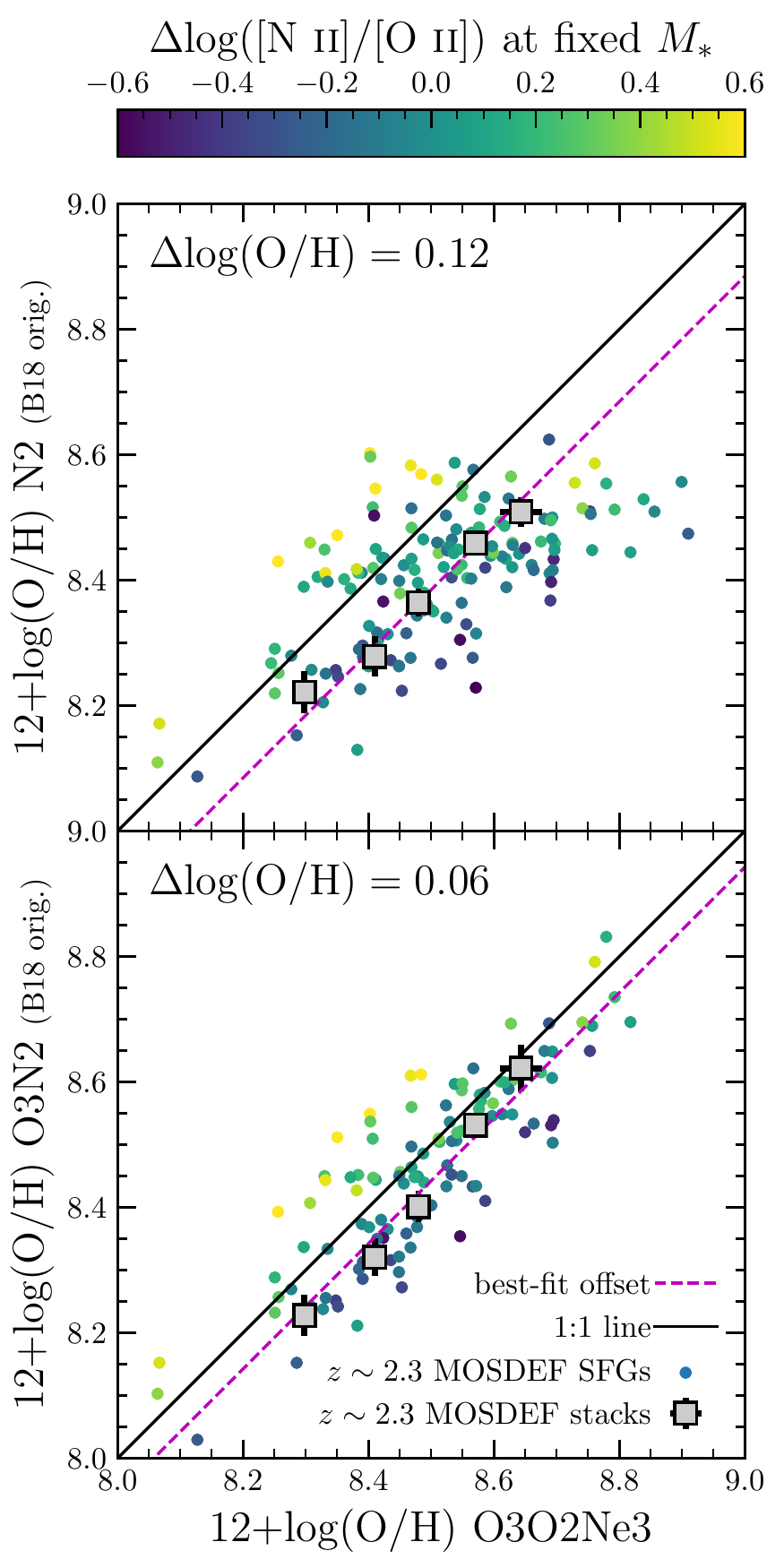}
 \centering
 \caption{
Gas-phase oxygen abundance derived from N2 (top) and O3N2 (bottom) using the original \citet{bia18} calibrations
 vs.\ O/H based on the combination of [O\iii], [O\ii], [Ne\iii], and H$\beta$ (O3O2Ne3).
Gray squares show composite spectra of $z\sim2.3$ star-forming galaxies from \citet{san21}.
Small circles display individual $z\sim2.3$ galaxies, color-coded by the difference in [N\ii]/[O\ii] relative to
 the average value at fixed \mstar.
The solid black line denotes a one-to-one relation.
The dashed pink line presents the best-fit offset between each pair of metallicities, as fit to the composite spectra.
The N2 and O3N2 calibrations given in Table~\ref{tab:metallicity} have been renormalized such that
 N2, O3N2, and O3O2Ne3 yield consistent metallicities on average for the $z\sim2.3$ sample.
}\label{fig:b18oh}
\end{figure}

Figure~\ref{fig:b18oh} presents O/H derived from the original \citet{bia18} analog calibrations
 for the $z\sim2.3$ MOSDEF star-forming galaxy sample of \citet{san21},
 comparing metallicities based on O3N2 and N2 to those
derived from the combination of [O\iii], [O\ii], H$\beta$, and [Ne\iii] (O3O2Ne3).
Gray squares show results for composite spectra in bins of \mstar.
Both the N2 and O3N2 metallicities (tied to the \citet{bia18} scale)
 are lower on average than those based on O3O2Ne3 by 0.12 and 0.06~dex, respectively,
These offsets are nearly constant across the full range of O/H.
To place metallicites derived from N2 and O3N2 on the same scale as those based on O3O2Ne3 (our preferred method;
 see Sec.~\ref{sec:metallicity}), we shift the normalization of the \citet{bia18} analog N2 and O3N2 calibrations
 based on the average offsets reported above, obtaining:
\begin{equation}
12+\log(\mbox{O/H})=8.94+0.49\times \mbox{N2}
\end{equation}
\begin{equation}
12+\log(\mbox{O/H})=9.03-0.39\times \mbox{O3N2}
\end{equation}
For reference, the leading constant factors of the original \citet{bia18} calibrations are 8.82 and 8.97, respectively.
The coefficients in Table~\ref{tab:metallicity} match these renormalized relations and yield consistent metallicities on
 average from N2, O3N2, or O3O2Ne3 for $z\sim2.3$ galaxies.

The individual galaxies are color-coded by the difference in [N\ii]$\lambda$6584/[O\ii] relative to the average
 value at fixed \mstar\ ($\Delta$log([N\ii]/[O\ii])), where the average relation is
log([N\ii]/[O\ii]$)=0.37\times\log(M_*/10^{10}~\mbox{M}_{\odot}) - 0.99$ as fit to the composite spectra of \citet{san21}.
For both N2 and O3N2, there is a clear trend that the offset relative to the O3O2Ne3 metallicities is correlated
 with deviations in [N\ii]/[O\ii].
Furthermore, only galaxies with [N\ii]/[O\ii] $\sim$0.3~dex higher than average at fixed \mstar\ lie on the one-to-one
 line, i.e., have consistent metallicities from O3O2Ne3 and N2 or O3N2 using the original \citet{bia18} analog calibrations.

\begin{figure}
 \includegraphics[width=\columnwidth]{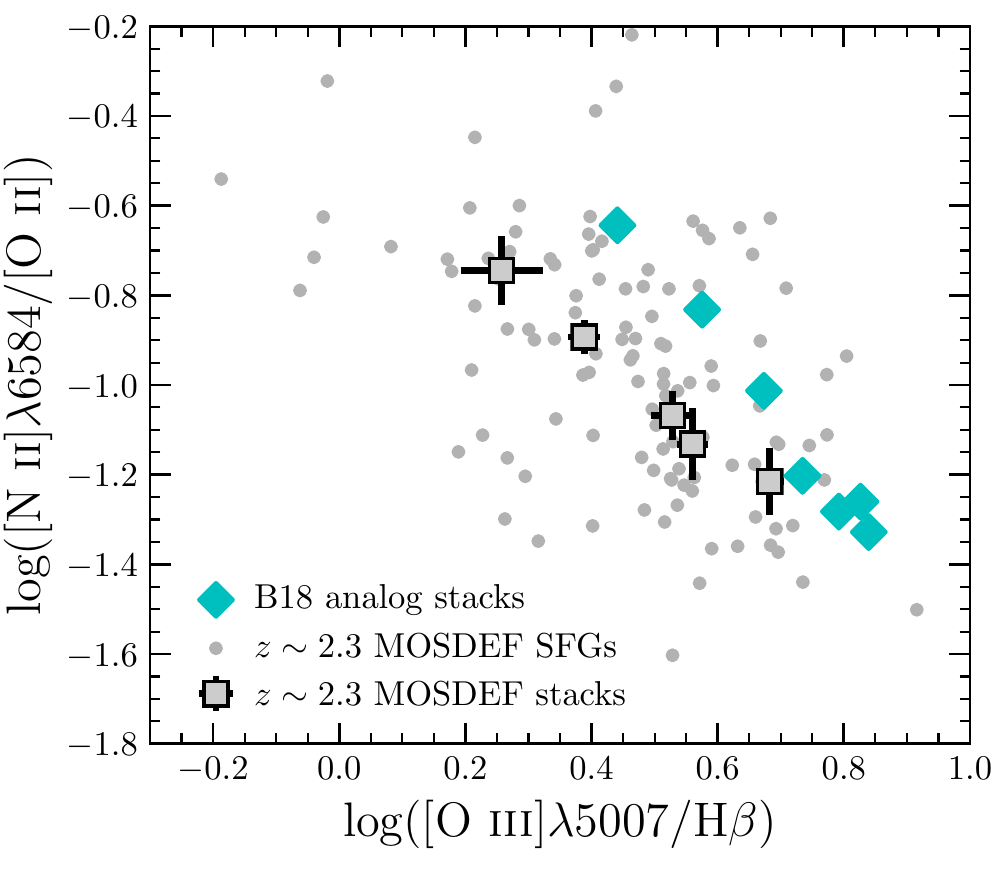}
 \centering
 \caption{
Line ratio diagram of [N\ii]/[O\ii] vs.\ [O\iii]/H$\beta$ for $z\sim2.3$ star-forming galaxies
 (gray circles), composite spectra of $z\sim2.3$ galaxies in bins of \mstar\ (gray squares; \citealt{san21}),
 and composite spectra of local analogs of high-redshift galaxies (cyan diamonds; \citealt{bia18}).
}\label{fig:b18n2o2}
\end{figure}

Metallicities based on N2 and O3N2 are clearly sensitive to variations in N/O (as traced by [N\ii]/[O\ii]) at fixed O/H,
 with N2 more strongly affected than O3N2.
The average offsets observed in Figure~\ref{fig:b18oh} can be explained if the \citet{bia18} analog calibrating sample
 has higher N/O at fixed O/H than the $z\sim2.3$ sample.
Figure~\ref{fig:b18n2o2} presents [N\ii]/[O\ii] vs.\ [O\iii]/H$\beta$ for the $z\sim2.3$ MOSDEF and \citet{bia18}
 analog samples.
The \citet{bia18} stacks have $\sim0.3$~dex larger [N\ii]/[O\ii] than the $z\sim2.3$ sample at fixed [O\iii]/H$\beta$.
Based on the calibration of \citet{str17}, a difference of 0.3~dex in [N\ii]/[O\ii] corresponds to a difference of
 0.2~dex in N/O.
This exercise suggests that the \citet{bia18} analog sample has larger N/O at fixed O/H than actual $z\sim2.3$
 star-forming galaxies and highlights the care that is needed when using local analogs in place of actual
 high-redshift samples.
These analogs were selected to lie in the same region of the [N\ii] BPT diagram as $z\sim2$ samples.
This criterion has clearly failed to select galaxies with identical ISM conditions to those at $z\sim2$,
 consistent with recent work that found local and high-redshift galaxies occupying the same region of the BPT
 diagram maintain some distinct differences in their ISM conditions \citep{run20}.
Improving metallicity calibrations for use at high redshifts will require more careful selection of local analogs
 and, ultimately, an expanded sample of high-redshift sources with direct-method metallicities that will soon be
 enabled by the \textit{James Webb Space Telescope}.

\section{Parameterization of the evovling mass-metallicity relation}\label{app:mzrz}

We utilize recent observational determinations of the mass-metallicity relation
 at different redshifts to obtain a functional form for the evolving mass-metallicity relation
 parameterized by \mstar\ and $z$.
We use the high-redshift measurements of \citet{san21} based on composite spectra of $z\sim2.3$ and $z\sim3.3$
 star-forming galaxies from the MOSDEF survey.
Metallicities of the high-redshift samples are based on O3O2Ne3 and derived using the \citet{bia18} $z\sim2$ analog
 calibrations.
We add the binned data of \citet{cur20b} for a sample of $\sim$150,000 star-forming galaxies
 at $z_{\text{med}}=0.08$ from the Sloan Digital Sky Survey (SDSS) to set the local baseline.
Metallicities of the local sample were determined using calibrations based on stacked spectra of normal $z\sim0$ galaxies
 derived in \citet{cur20b}.
Both calibration sets are on a $T_e$-based metallicity scale.

We fit the $z\sim0$, $z\sim2.3$, and $z\sim3.3$ data with a function of the form
\begin{equation}\label{eq:mzrz}
\text{MZR}(M_*,z) = Z_0 - \gamma/\beta\times\log(1+[M_*/M_0(z)]^{-\beta})
\end{equation}
where $Z_0$ is the high-mass asymptotic metallicity, $\gamma$ is the low-mass power law slope, and
 $\beta$ controls the sharpness of the turnover that occurs at the mass
 $\log(M_0(z)/\text{M}_{\odot})=m_0 + m_1*\log(1+z)$.
This is the same functional form used to fit the $z\sim0$ SFR-FMR by \citet{cur20b}, but with the secondary
 parameter being $(1+z)$ instead of SFR.
We find the following best-fit parameters: $Z_0=8.80$, $\gamma=0.30$, $\beta=1.08$, $m_0=9.90$, and $m_1=2.06$.
This fitting function yields a nearly identical relation to the \citet{cur20b} best-fit $z\sim0$ MZR when
 evaluated at $z=0.08$.

\begin{figure}
 \includegraphics[width=\columnwidth]{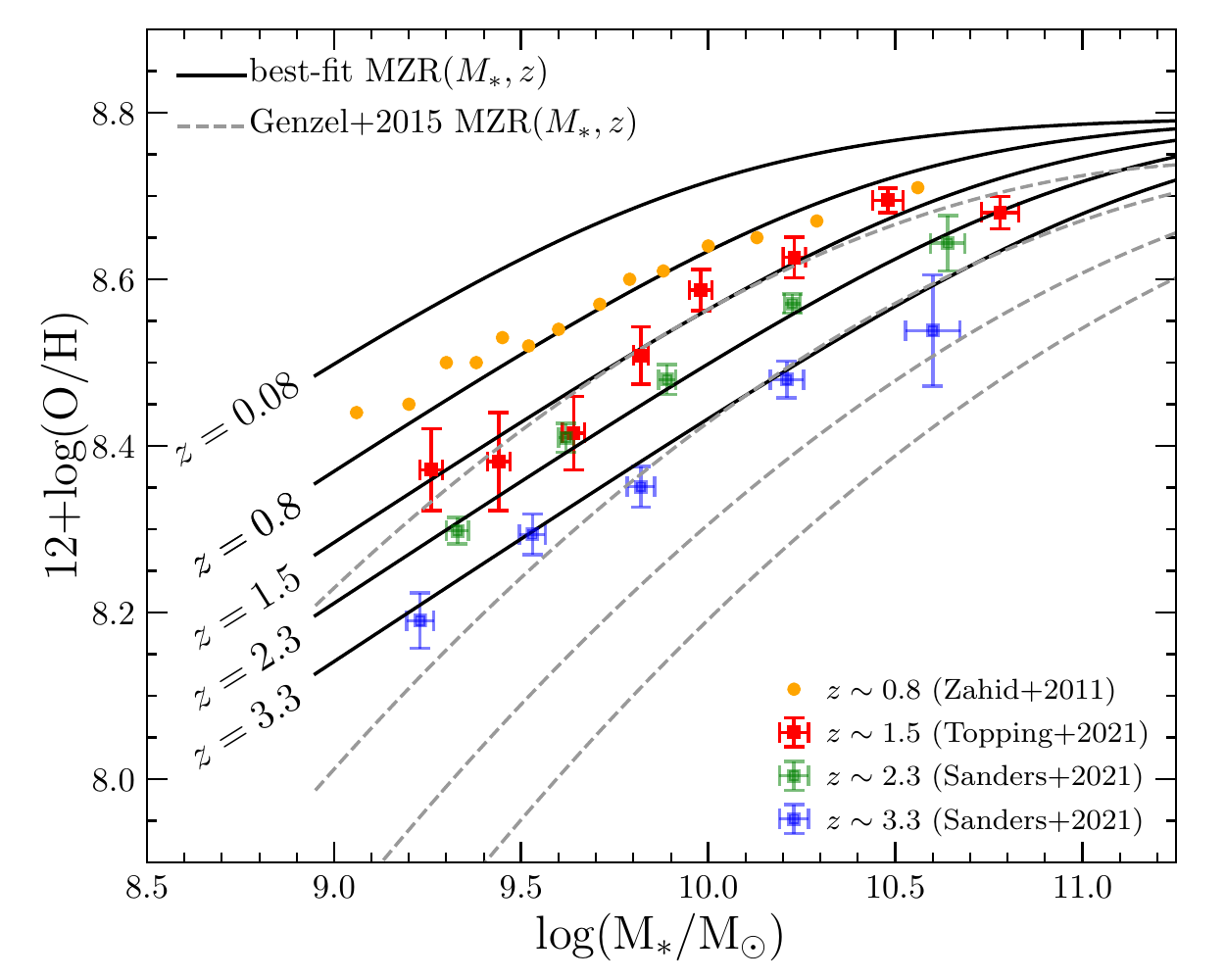}
 \centering
 \caption{
The mass-metallicity relation for samples of star-forming galaxies at a range of redshifts:
 $z\sim0.8$ \citep{zah11}, $z\sim1.5$ \citep{top21}, and $z\sim2.3$ and 3.3 \citep{san21}.
Solid black lines show the best-fit MZR($M_*, z$) to data at $z=0.08$ \citep{cur20b},
 $z\sim2.3$, and $z\sim3.3$, assuming the functional form given in equation~\ref{eq:mzrz}.
Additional black lines display the best-fit MZR($M_*, z$) evaluated at $z=0.8$ and $z=1.5$ in comparison
 to observational MZR data at these redshifts that were not used in the fitting process, demonstrating
 that the best-fit function is robust across the intermediate redshift range as well.
Dashed gray lines display the MZR($M_*, z$) function of \citet{gen15}, evaluated at
 $z=[0.8, 1.5, 2.3, 3.3]$.
}\label{fig:mzrz}
\end{figure}

Figure~\ref{fig:mzrz} shows the best-fit MZR($M_*,z$) alongsize the $z\sim2.3$ and $z\sim3.3$ data
 (green and blue, respectively; $z\sim0$ data is not displayed for clarity),
 displaying a good fit across the entire mass range.
To test the applicability of this fitting function across a wider range of redshifts, we compare to
 samples at $z\sim0.8$ \citep[][DEEP2]{zah11} and $z\sim1.5$ \citep[][MOSDEF]{top21}.
The metallicities of the $z\sim0.8$ sample have been reevaluated using the $R_{23}$ and $O_{32}$ $z\sim0$ calibrations
 of \citet{cur20b}, while those of the $z\sim1.5$ sample have been rederived with the renormalized high-redshift N2 calibration given
 in Table~\ref{tab:metallicity}.
Our best-fit MZR($M_*,z$) function provides a good match at intermediate redshifts as well, demonstrating that it is
 widely applicable over $z=0-3.3$ for galaxies above $10^9$~M$_{\odot}$.

The gray dashed lines show the MZR($M_*,z$) derived by \citet{gen15} and adopted by \citet{tac18,tac20},
 evaluated at $z=[0.8, 1.5, 2.3, 3.3]$.
Our new relation yields higher metallicity at fixed \mstar\ and redshift across the full mass range
 (0.15~dex in O/H at $10^{10.5}$~M$_{\odot}$ and $z=2.3$).
The difference primarily stems from the use of larger and more representative high-redshift datasets in this work
 (see \citealt{san21} for a discussion) and different choices of metallicity calibrations, in particular the use
 of different calibrations at low and high redshifts in this work to reflect evolving ISM ionization conditions.

\section{Literature CO references}

Table~\ref{tab:litco} presents the properties and literature references for the $z=2-3$ literature CO sample and the supplementary $z>1$ \aco\ sample.

\begin{table*}
 \centering
 \caption{Literature CO samples.
 }\label{tab:litco}
 \setlength{\tabcolsep}{4pt}
 \begin{tabular}{ l l l l l l l l l l }
   \hline\hline
\multicolumn{10}{c}{Literature $z=2-3$ CO sample} \\
\hline
ID  &  $z_{\text{spec}}$  &  
$\log{\left(\frac{M_*}{\text{M}_\odot}\right)}$  &
log$\left(\frac{\text{SFR}}{\text{M}_\odot/\text{yr}}\right)$  &
12+log(O/H)  &  Type\tablenotemark{a}  &
log$\left(\frac{M_{\text{mol}}}{\text{M}_{\odot}}\right)$  &
log$\left(\mu_{\text{mol}}\right)$  &
CO Ref.\tablenotemark{b}  &  O/H Ref.\tablenotemark{b}  \\
\hline
zC406690  &  2.1950  &  10.62$\pm$0.10  &  2.47$\pm$0.15  &  8.48$\pm$0.02  &  N2  &  11.20$\pm$0.10  &  0.58$\pm$0.14  &  1  &  2  \\
HDF-BX1439  &  2.1867  &  10.55$\pm$0.05  &  2.06$\pm$0.26  &  8.49$\pm$0.05  &  O2O3Ne3  &  10.81$\pm$0.15  &  0.26$\pm$0.16  &  1  &  3  \\
Q1623-MD66  &  2.1075  &  10.59$\pm$0.10  &  2.39$\pm$0.15  &  8.55$\pm$0.02  &  N2  &  $<$11.22  &  $<$0.63  &  1  &  4  \\
Q1623-BX453  &  2.1820  &  10.48$\pm$0.10  &  2.39$\pm$0.15  &  8.71$\pm$0.01  &  O3N2  &  10.88$\pm$0.06  &  0.40$\pm$0.12  &  1  &  5  \\
Q1623-BX528  &  2.2684  &  10.84$\pm$0.10  &  1.77$\pm$0.15  &  8.69$\pm$0.03  &  N2  &  $<$10.60  &  $<$-0.24  &  1  &  6  \\
Q1623-BX599  &  2.3312  &  10.75$\pm$0.10  &  2.12$\pm$0.15  &  8.58$\pm$0.02  &  N2  &  10.82$\pm$0.09  &  0.07$\pm$0.14  &  1  &  2  \\
Q1623-BX663  &  2.4335  &  10.81$\pm$0.10  &  2.16$\pm$0.15  &  8.63$\pm$0.07  &  N2  &  $<$10.97  &  $<$0.16  &  1  &  6  \\
Q1700-MD69  &  2.2881  &  11.21$\pm$0.10  &  2.03$\pm$0.15  &  8.71$\pm$0.01  &  O3N2  &  10.92$\pm$0.07  &  -0.29$\pm$0.12  &  1  &  5  \\
Q1700-BX691  &  2.1891  &  10.89$\pm$0.10  &  1.73$\pm$0.15  &  8.70$\pm$0.03  &  O3N2  &  10.55$\pm$0.12  &  -0.34$\pm$0.16  &  1  &  5  \\
Q2343-BX610  &  2.2107  &  11.00$\pm$0.10  &  2.33$\pm$0.15  &  8.72$\pm$0.01  &  N2  &  11.29$\pm$0.02  &  0.29$\pm$0.10  &  1  &  2  \\
Q2343-BX513  &  2.1082  &  10.43$\pm$0.10  &  1.45$\pm$0.15  &  8.60$\pm$0.02  &  N2  &  10.61$\pm$0.15  &  0.18$\pm$0.18  &  1  &  2  \\
Q2343-BX442  &  2.1752  &  11.12$\pm$0.10  &  1.91$\pm$0.15  &  8.85$\pm$0.03  &  O3N2  &  10.73$\pm$0.12  &  -0.39$\pm$0.15  &  1  &  5  \\
Q2343-BX389  &  2.1711  &  10.97$\pm$0.10  &  1.91$\pm$0.15  &  8.56$\pm$0.03  &  O3N2  &  $<$11.23  &  $<$0.26  &  1  &  5  \\
UDF3/3mm.01  &  2.5430  &  10.52$\pm$0.26  &  2.30$\pm$0.16  &  8.74$\pm$0.05  &  O3N2  &  11.43$\pm$0.02  &  0.91$\pm$0.26  &  7  &  8  \\
UDF4/3mm.03  &  2.4537  &  10.36$\pm$0.15  &  1.97$\pm$0.03  &  8.74$\pm$0.08  &  N2  &  11.00$\pm$0.04  &  0.64$\pm$0.16  &  7  &  8  \\
UDF11/9mm.5  &  1.9978  &  10.91$\pm$0.04  &  2.23$\pm$0.27  &  8.66$\pm$0.02  &  N2  &  10.80$\pm$0.12  &  -0.11$\pm$0.13  &  9  &  8  \\
Q1700-MD94  &  2.3330  &  11.18$\pm$0.10  &  2.43$\pm$0.15  &  ---  &  MZR  &  11.37$\pm$0.07  &  0.19$\pm$0.07  &  1  &  ---  \\
Q1700-MD174  &  2.3400  &  11.38$\pm$0.10  &  2.20$\pm$0.15  &  ---  &  MZR  &  11.05$\pm$0.05  &  -0.33$\pm$0.05  &  1  &  ---  \\
Q1700-BX561  &  2.4340  &  11.08$\pm$0.10  &  1.23$\pm$0.15  &  ---  &  MZR  &  10.41$\pm$0.15  &  -0.67$\pm$0.15  &  1  &  ---  \\
Q2343-MD59  &  2.0110  &  10.88$\pm$0.10  &  1.41$\pm$0.15  &  ---  &  MZR  &  10.88$\pm$0.08  &  -0.00$\pm$0.08  &  1  &  ---  \\
3mm.07  &  2.6961  &  11.10$\pm$0.10  &  2.27$\pm$0.05  &  ---  &  MZR  &  11.30$\pm$0.05  &  0.20$\pm$0.05  &  7  &  ---  \\
3mm.09  &  2.6977  &  11.10$\pm$0.10  &  2.50$\pm$0.05  &  ---  &  MZR  &  11.02$\pm$0.04  &  -0.08$\pm$0.04  &  7  &  ---  \\
3mm.12  &  2.5739  &  10.60$\pm$0.10  &  1.49$\pm$0.15  &  ---  &  MZR  &  10.68$\pm$0.06  &  0.08$\pm$0.06  &  7  &  ---  \\
   \hline\hline
 \end{tabular}
 \begin{tabular}{ l l l l l l }
 \multicolumn{6}{c}{Supplementary $z>1$ \aco\ sample} \\
\hline
ID & $z_{\text{spec}}$ & 12+log(O/H) & Type\tablenotemark{a} & CO Ref.\tablenotemark{b} & O/H Ref.\tablenotemark{b} \\
\hline
BzK-16000  &  1.5249  &  8.72$\pm$0.01  &  N2  &  10  &  3  \\
cB58  &  2.7293  &  8.43$\pm$0.04  &  O3N2  &  11  &  12  \\
CosmicEye  &  3.0740  &  8.62$\pm$0.02  &  N2  &  13  &  14  \\
Eyelash  &  2.3230  &  8.53$\pm$0.02  &  O2O3Ne3  &  15  &  16  \\
8oclock  &  2.7360  &  8.53$\pm$0.01  &  O2O3Ne3  &  17  &  14  \\
Horseshoe  &  2.3813  &  8.54$\pm$0.03  &  N2  &  17  &  18  \\
Clone  &  2.0026  &  8.41$\pm$0.02  &  N2  &  17  &  18  \\
HDF-M23  &  3.2159  &  8.51$\pm$0.08  &  O2O3Ne3  &  19  &  3  \\
UDF6/3mm.04  &  1.4152  &  8.75$\pm$0.06  &  O3N2  &  7  &  8  \\
3mm.06  &  1.0955  &  8.92$\pm$0.06  &  N2  &  7  &  7  \\
3mm.11  &  1.0964  &  8.70$\pm$0.03  &  N2  &  7  &  7  \\
3mm.14  &  1.0981  &  8.65$\pm$0.04  &  O2O3Ne3  &  7  &  7  \\
MP.3mm.02  &  1.0872  &  8.64$\pm$0.02  &  O2O3Ne3  &  7  &  7  \\
S15-830  &  1.4610  &  8.72$\pm$0.08  &  O2O3Ne3  &  20  &  21  \\
   \hline\hline
 \end{tabular}
\begin{flushleft}
\tablenotetext{a}{Line ratio used for metallicity derivation, described in Sec.~\ref{sec:metallicity}. Metallicities of sources with ``MZR'' are inferred from equation~\ref{eq:fitmzrz} for use in \mmol\ calculations.}
\tablenotetext{b}{References for CO and rest-optical spectroscopic observations.}
\tablerefs{1: \citet{tac13}; 2: \citet{for18}; 3: \citet{kri15}; 4: \citet{sha04}; 5: \citet{ste14}; 6: \citet{for06}; 7: \citet{boo19}; 8: \citet{sha20}; 9: \citet{rie20}; 10: \citet{dad10}; 11: \citet{bak04}; 12: \citet{tep00}; 13: \citet{rie10}; 14: \citet{ric11}; 15: \citet{dan11}; 16: \citet{oli16}; 17: \citet{sai13}; 18: \citet{hai09}; 19: \citet{mag12}; 20: \citet{sil15}; 21: \citet{kas19}}
\end{flushleft}
\end{table*}




\bibliography{ms}


\end{document}